\documentclass[10pt,journal,compsoc]{IEEEtran}

\ifCLASSOPTIONcompsoc
  \usepackage[nocompress]{cite}
\else
  \usepackage{cite}
\fi

\usepackage[usenames,dvipsnames,svgnames,table]{xcolor}
\usepackage{color}
\usepackage{framed}
\usepackage{amssymb}
\usepackage{soul}
\usepackage{textcomp}
\usepackage{multicol}  
\usepackage{multirow} 
\usepackage{xspace}
\usepackage{amsmath} 
\usepackage{graphicx}
\usepackage{breqn}
 \usepackage{hyperref}

\usepackage{stfloats}
\usepackage{subfig}

\usepackage{algorithm}
\usepackage{algorithmic}

\def\G{\mbox{$\mathcal{G}$}} %  
\def\V{\mbox{$\mathcal{V}$}} % 
\def\E{\mbox{$\mathcal{E}$}} %  
\def\d{\mbox{$d$}}  % damping factor

\newcommand{\argmax}{\operatornamewithlimits{argmax}}

\newcommand{\drawn}[1]{#1 \sim \mathcal{\uppercase{#1}}}

\newcommand{\teta}[1] {\vartheta_#1} % LT Model threshold
\newcommand{\iNeighbor}[1]{N^{in}(#1)} % in-neighbors 
\newcommand{\oNeighbor}[1]{N^{out}(#1)} % out-neighbors 
\newcommand{\olf}[1]{C(#1)} % credit 
\newcommand{\ActiveSet}[1]{\mu(#1)} % set of nodes active at the end of the process
\newcommand{\SetDiv}[1]{D(#1)} % set diversity
\def\TargetSet{TS} % target set

\def\LurkValue{L} % minimum threshold of nodes in the target set
\def\LurkValuePerc{L\textit{-perc}} % minimum threshold of nodes in the target set
\def\infp{inf}

\def\symbOF{DIC} %overall spread
\def\of{\textit{Diversity-sensitive Capital}} %objective function name

\def\formulaUno{\textit{global diversity}\xspace}
\def\formulaDue{\textit{local diversity}\xspace}
\def\cformulaUno{\textit{Global Diversity}\xspace}
\def\cformulaDue{\textit{Local Diversity}\xspace}

\newcommand{\data}[1]{{#1}}
\newcommand{\algo}[1]{{\textsf{#1}}}
\def\myalgo{\textsf{DTIM}\xspace}

\newtheorem{principle}{Principle}
\newtheorem{adefinition}{Definition}
\newtheorem{proposition}{Proposition}

\begin{document}

% Page heads
 \markboth{A. Cali\`{o} et al.}{Topology-driven Diversity for Targeted Influence Maximization}

% Title portion
\title{Topology-driven Diversity for Targeted Influence Maximization with Application to User  Engagement in Social Networks}

%\IEEEcompsocthanksitem J. Doe and J. Doe are with Anonymous University.}% <-this % stops an unwanted space

%\markboth{IEEE Transactions on Knowledge and Data Engineering}%
%{R. Interdonato et al.: Topology-driven Diversity for Targeted Influence Maximization}

\author{Antonio Cali\`{o}, Roberto~Interdonato,	Chiara~Pulice,  
	 and~Andrea~Tagarelli\thanks{Corresponding author: Andrea Tagarelli.}% <-this % stops a space
	\IEEEcompsocitemizethanks{\IEEEcompsocthanksitem A. Cali\`{o}  and  A. Tagarelli are with the DIMES Department, University of Calabria, Rende (CS), Italy.  %\protect\\
		E-mail: \{a.calio,tagarelli\}@dimes.unical.it 
		\IEEEcompsocthanksitem R. Interdonato is with Cirad, UMR Tetis, Montpellier,  France. E-mail: roberto.interdonato@cirad.fr (Work done at University of Calabria, prior to joining Cirad.)  
		\IEEEcompsocthanksitem C. Pulice is with   Dept. of Computer Science, Dartmouth College,
Hanover, NH, USA. %\protect\\ 
		E-mail:   chiara.pulice@dartmouth.edu 
	}% <-this % stops an unwanted space
}

\IEEEtitleabstractindextext{%
\begin{abstract}
Research on influence maximization has often to cope with marketing needs relating to the propagation of information towards specific users.
However, little attention has been paid to the fact that the success of an information diffusion campaign might depend not only on the number of the initial influencers to be detected but also on their \textit{diversity} w.r.t. the target of the campaign.   
Our main hypothesis is that  if we learn seeds that are not only capable of   influencing  but also are linked to more diverse (groups of)  users, then          the influence triggers will be diversified as well, and hence   the target users will get higher chance of being engaged. 
Upon this intuition, we define a novel problem, named \textit{Diversity-sensitive  Targeted Influence Maximization (DTIM)}, which assumes to model user diversity by exploiting only topological information within a social  graph.   
To the best of our knowledge, we are the first to bring the concept of topology-driven diversity into targeted IM problems, for which we define two alternative definitions. Accordingly, we propose approximate solutions of DTIM,  which detect a size-$k$  set of users that maximizes the diversity-sensitive capital objective function, for a given selection of target users. 
We evaluate our DTIM methods on a special case of user engagement in online social networks, which concerns users who are not actively involved in the  community life. Experimental evaluation on real networks has demonstrated the meaningfulness of our approach, also highlighting the opportunity of further development of solutions for DTIM applications. 
\end{abstract}

\begin{IEEEkeywords}
diversity-sensitive influence propagation, linear threshold diffusion model, social capital,   lurking behavior analysis.
\end{IEEEkeywords}}

 \maketitle
  \thispagestyle{empty}

%=====================================

%\section{Introduction}
\IEEEraisesectionheading{\section{Introduction}\label{sec:introduction}}
Online social networks (OSNs) are nowadays the preferred communication means for spreading information, generating and sharing  knowledge.   
 One central problem  is the identification of influential individuals in an OSN such that, starting with  them, one can trigger a chain reaction of influence driven by ``word-of-mouth'', which allows for reaching a large portion of the network   with a relatively little effort in terms of initial investment (budget). This is commonly referred to as \textit{viral marketing} principle, which is the underlying motivation for   a classic optimization problem in OSNs, namely  \textit{influence maximization} (IM).     The   general objective of an IM method is to find a set of initial influencers    which can maximize the spread of information through the network   (e.g.,~\cite{KempeKT03,GoyalLL11,TangXS14,GalhotraAVR15,LiBSC15,ZhouZZG15}).  
  
Most of existing works in IM and related applications focus on the entire social network through which the spread of influence is to be maximized. However, thinking in terms of viral marketing, an organization often wants to narrow the advertisement of its products to  users having certain needs or preferences, as opposed to targeting the whole crowd.  Also, in an OSN scenario, some  events or memes would be of interest only to users with certain tastes or social profiles. Our work fits into research on this problem, hereinafter referred to as \textit{targeted IM}.  

\textbf{Leveraging diversity for enhanced  IM.\ } 
While maximizing the  advertising of a product, an organization also needs to minimize the incentives offered to those users who will reach out the target ones.  This obviously raises  the necessity of choosing a proper number $k$ of seed users (i.e.,  initial influencers) to be detected, which corresponds to the budget constraint.    
Surprisingly, an important aspect that is often overlooked is   that the success of   a viral marketing process  might   depend not only on the size of the seed set but also on the \textit{diversity} that is reflected within, or in relation to, the seed set.  Intuitively, individuals that differ from each other in terms of kind (e.g., age, gender), 
 socio-cultural aspects (e.g., nationality, race)  
 or other characteristics, bring unique   opinions,  experiences, and perspectives to bear on the task at hand;  moreover, in an OSN context,   members naturally have different knowledge, community experience, participation motivation and shared information~\cite{santos08,PanLG14,RobertR15}. 
 It is worth noticing that diversity has been generally recognized as a key-enabling  dimension in data analysis, which is essential to enhance productivity, develop  wiser crowdsourcing processes, improve user satisfaction in content recommendation based on novelty and serendipity, avoid  information bubble effects, and ultimately have legal and ethical implications in  information processing~\cite{SciAme,Drosou+17}.   
     
Bringing this picture into  targeted IM scenarios, let us 
focus on the problem of \textit{user engagement}~\cite{MalliarosV13,leskovec13,ImlawiG14,PanLG14}.  
  Users that have not yet experienced community commitment (i.e., they are not actively involved in the community life)  often hail for different background and motivation, and communicate on diverse topics, which makes engaging them difficult.  
One effective strategy of user engagement should account for the support and guidance from  elder, active members of the community~\cite{Sun+14}.   
Therefore,    by identifying the most diverse, active members, the triggering stimuli will also be diversified. 
Since diverse individuals tend to connect to many different types of members,   the likelihood  of effective  engagement would be higher.  

\textbf{The challenge of diversity in targeted IM.\ } 
Existing targeted IM methods are not   designed to   embed a notion of diversity in their objective function.  
In this work, we aim to overcome this limitation,  using an  unsupervised approach. That is, our research relies on taking a  perspective that does not assume any side-information or a-priori knowledge on user attributes (e.g.,  personal profile, topical preference, community role)   that can enable diversification among users.   By contrast, we assume that  \textit{a user's diversity in a social graph can be determined  based on topological properties related to her/his  neighborhood}. Remarkably, this finds justifications from social science, particularly from  theories of  \textit{social embeddedness}~\cite{HarrisonK07}  
 and   \textit{boundary spanning}~\cite{ESNAM,SorokaR06}.  
 In particular, the latter   explains how  OSN users acquire  knowledge from some of their social contacts and then spread (part of) it   to other contacts that belong to  one or more components of the social graph, e.g.,    topically-induced communities, as found in~\cite{InterdonatoPT16}. 
 
Our main hypothesis is that  if we learn seeds that are not only capable of   influencing  but also are linked to more diverse (groups of)  users, then we would expect that  the influence triggers will be diversified as well, and hence   the target users will get higher chance of being engaged.

\begin{figure}[t!]
\centering
\includegraphics[width=0.65\columnwidth, height=0.5\columnwidth]{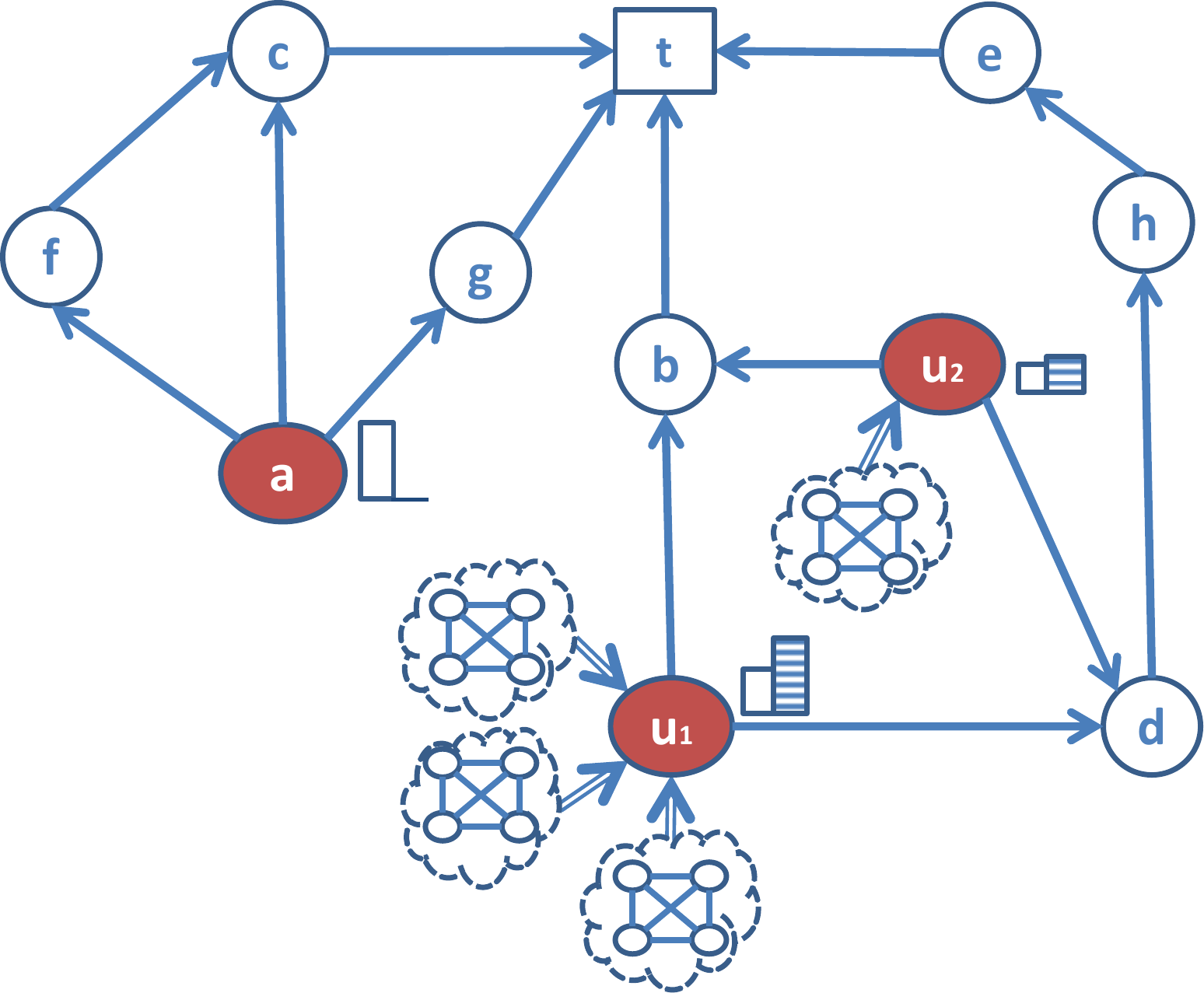}
\caption{Effect of topological diversity on the outcome of targeted IM.}
\label{fig:example}
\end{figure}

 %%%%%%%%%%%%%%%  EXAMPLE 1  %%%%%%
%\begin{aexample}\label{ex:ex1}
\textbf{\em Example 1.}
To advocate  the above hypothesis, consider the example social graph shown in Figure~\ref{fig:example}, where nodes represent individuals and  edges express influence relationships. 
%(The graph is directed as influence relationships are usually modeled as asymmetric.) 
 Suppose this graph corresponds to the context of a diffusion process, captured at a given time step, where for the sake of simplicity we omit to indicate both the influence probabilities as edge weights and the active/inactive nodes.  
Let us focus our attention on the square border node $t$, which represents  a target node, and assume that the colored nodes $a$, $u_1$, $u_2$  correspond to candidate seeds, for which we know  the individual cumulated  spreading influence towards $t$ and the individual topological diversity according to some diversity function; in the figure,  these scores are displayed by the leftmost bar and the rightmost bar, respectively, associated to each of the candidate seeds. \\
% %Assume also that the graph is being explored  backward, from $t$ to the other nodes.
%We annotated  each of the colored nodes  with two bars: the leftmost one  having height proportional to the cumulated  spreading influence towards $t$, and the rightmost one  having height proportional to the node's topological diversity.    \\ 
  A conventional targeted IM method would add   node $a$ to  the seed set, since 
 it has the highest capability of spread among the candidate seeds;  however, $a$'s location has two characteristics that, as we shall explain later,  would imply poor topological diversity: it does not receive any incoming connections from other components in the graph, and it diffuses towards nodes that are all in the same subgraph having  $t$ as sink. 
 By contrast, the location of nodes $u$  is strategical   in terms of topological diversity, since they  could be  influenced by one or more groups of nodes (in the figure indicated as components enclosed within dashed clouds), thus potentially acquiring a wider spectrum of varied information and perspectives. Selecting nodes $u$   would hence be favored by a diversity-aware targeted IM method  as they might be more effective in   increasing node $t$'s engagement. 
  %  \end{aexample}

Two main research questions here arise concerning  how to  leverage   users' social diversity  in order to enhance the performance of a targeted IM task:  \textbf{(R1)} how to determine diversity at a large-scale, when we have no a-priori knowledge on user attributes; and  
 \textbf{(R2)}  %how diversity relates to the influence spread and, consequently, 
how the seed users should be learned by also considering  %both aspects of spreading capability and 
diversity w.r.t. a target set.

 \vspace{1mm}
\textbf{Contributions.}  
% In this work we aim at  addressing the above stated research questions, and propose novel solutions to the problem of  \textit{targeted IM with diversity related to the seeds} to be detected.  
In this work we contribute with the definition of a novel problem, named {\em \textbf{D}iversity-sensitive \textbf{T}argeted \textbf{I}nfluence \textbf{M}aximization} (\myalgo). To the best of our knowledge, we are the first to bring the concept of topology-driven diversity into targeted IM problems. 
%In particular,   we will focus on the LT diffusion model, as we believe that a natural motivation to use LT (rather than IC or IC-based models) stands in the ability of this model to reflect the cumulative effect of exposure to multiple sources of influence.
%
%Pursuing our goal of modeling  topology-driven diversity to discover a ``better'' seed set for targeted IM, we define user diversity in relation to the structural context of the information diffusion graph.  
More specifically,   to answer \textbf{R1},  we provide two alternative ways of modeling topology-driven diversity for targeted IM,  
% \barrato{which impact on  the unfolding of the information diffusion process.}   
   which depend on the approach  adopted  to exploit structural information from the diffusion subgraph specific to a given target node.     
 (Loosely speaking, a target-specific diffusion subgraph corresponds the portion of the diffusion graph involved, at a given time step, in the unfolding of the diffusion towards a particular target node.)  
 The first method, dubbed \formulaDue, %features no requirement   to access the whole network information, while  it 
 is designed to  %\barrato{account for the progressive expansion of the information diffusion graph for a given target node.}  
  compute node diversity at each step of the  expansion of a target-specific diffusion subgraph. 
The \formulaDue of a  node captures  the likelihood of 
 %\barrato{exploring further the diffusion graph through that node,} 
   reaching it from nodes outside the currently unfolded target-specific diffusion subgraph.  %\barrato{, and is determined as proportional to  the increment contributed by that node to the number of incoming connections from outside the ``boundary'' of the target-specific diffusion graph being unfolded. }
Our second method of topology-driven diversity, dubbed \formulaUno,   
% relies on an opposite assumption w.r.t.   \formulaDue, i.e., 
exploits the  structural information of the fully unfolded  target-specific diffusion subgraph, and determines the diversity of nodes that  
 lay on the \textit{boundary} of the subgraph, 
   i.e., nodes that can receive influence links from  nodes external to the subgraph. 
  Intuitively, this would allow us to capture a boundary-spanning effect of external sources of influence coming from the rest of the social graph.

To address question \textbf{R2},  we capitalize on the \formulaDue and \formulaUno   definitions      to develop  alternative   algorithms for the \myalgo  problem, dubbed L-\myalgo and G-\myalgo. Both   algorithms follow a greedy approach that exploits  the search for shortest paths in the diffusion graph,   in a backward  fashion from the selected target set. 

We evaluate our \myalgo methods  on a special case of  user engagement in OSNs, which concerns the crowd of users who do not actively  contribute to the production of social content.  Such silent users, a.k.a. \textit{lurkers},    might have great potential in terms of  \textit{social capital}, i.e.,  acquired knowledge   through  the observation of  user-generated communications.   Therefore, it is highly desirable  to encourage (a portion of) silent users to more actively participate and give back to the community.     
Note that while we previously addressed this problem of user engagement in OSNs via a targeted IM approach in~\cite{InterdonatoPT15,InterdonatoPT16}, in this work we further delve into understanding such a challenging problem under the new perspective of diversity of the seeds to be identified for maximizing the engagement of silent users.  
%Note also that our proposed approach can be generalized not only to other cases of user engagement (for example, introducing newcomers to community life), but also to any other applications of targeted IM in which accounting for diversity of users based on their relationships/interactions with other users, is beneficial to the enrichment  of influence propagation outcome with effects of varied social capital. 

 Experimental evaluation using  three real-world OSN datasets was   conducted to assess the meaningfulness of our approach, mainly in terms of characteristics of the identified seeds and the activated target users, and how they are affected by tuning the input and model parameters of our methods. 
 We also included comparison with two of the most relevant existing IM methods, namely \algo{TIM+}~\cite{TangXS14} and \algo{KB-TIM}~\cite{LiZT15}, based on the state-of-the-art \textit{RIS} approach.     
  While this comparison has highlighted the uniqueness of our methods, it also suggested to improve their efficiency. In this respect, a further important contribution is the revisiting of RIS-based approximation theory to our diversity-sensitive  targeted IM problem.

\textbf{Plan of the paper.}
 The rest of the paper is organized as follows.  
Section 2 discusses related work, focusing on  diversity and   targeted IM.  
Sections 3 presents our diversity-sensitive targeted IM problem, defines two alternative formulations of topology-driven diversity, and presents the L-\myalgo and G-\myalgo algorithms.  In Section 4, we introduce  a  case study of user engagement  for the evaluation of  our proposed framework. Experimental evaluation methodology and results are reported in Section 5 and Section 6, respectively.  
 Section 7 describes a RIS-based formulation of \myalgo.    
  Section 8 draws   conclusions and provides pointers for future research.

%=======================================

\section{Related Work}
\label{sec:related} 
 
\textbf{Diversity in information spreading.\ }
\label{sec:related:diversitySN} 
Most existing  notions of   diversity   have been developed around  structural features of the network, or alternatively based on user profile attributes. This broad categorization applies to various contexts, such as,  e.g.,  web searching and recommendation~\cite{LiY13,SantosMO15,WuLCYGX16}, %BelemBSAG16,
  and information spreading.  
%The latter represents the focus of our discussion on related work we provide next.
Focusing on the latter aspect, the authors in~\cite{KunegisSSF12}  
%introduce several basic measures of diversity and show their evolution during the lifetime of the network. One of the measures   they 
propose a measure of controllability, defined as the number of nodes   %needed to control a network, i.e., nodes 
able to spread an opinion through the whole network. 
%The latter is estimated by finding a maximal matching in the bipartite cover of the network. 
%
In~\cite{BaoCZ13}, the       IC model is extended  to take into account the structural diversity of nodes' neighborhood. 
%For each node in the graph, they compute its neighborhood as the union of components, that in turn are defined as sets of its interconnected in-neighbors. 
%They define a component-based diffusion process and exploit the tightness of connections within the component to infer the model parameters. 
%They show that this model can  improve the accuracy of the diffusion process in social networks, making it closer to   real world dynamics.
%
%In~\cite{JohnsonPL15}, the authors study the complexity of computing diversity and clustering in social networks on a dot product graph model of the network. The dot product model assumes that two users are connected  if their attributes or opinions, represented through a  multidimensional vector, are similar to each other. The diversity considered in their work is measured by finding the size of the largest group of users with different opinions that, in that context, corresponds to solving an Independent Set problem. 
%
Main difference between the above mentioned approaches and our work,  relies on the fact that they do not take into account any optimization problem.
 %
%Diversity has also been introduced to efficiently identify influential users in social networks.  
%It is a matter of fact that, in order to maximize the social influence, we need to select nodes that could cooperatively trigger a big cascade of activations. However, in real-world marketing campaign, diversity of the seed nodes can play a key role as it contributes in reducing its risk of failure. Indeed, having a diverse seed set could bring many benefits. An obvious one is that if the seed set is diverse, we could expect the activated node set is diverse, too, reducing the risk of overlapping activations that prevent the influence to spread much farther. 
 Other works deal with the problem of estimating the spreading ability of a single node 
 %each node 
 in a network~\cite{FuHS15,HuangLCC13}.
%In~\cite{FuHS15}, the authors propose a   measure that combines global diversity and local feature (e.g., degree centrality) to identify the most influential spreaders. The diversity is determined by computing the neighbors'  position in the network, which can be obtained exploiting  community detection or k-shell decomposition. %~\cite{Seidman83}.  
%Experiments have been conducted on several networks to assess the validity of this measure w.r.t.   baselines (e.g. closeness and betweenness centrality), by modeling the diffusion process through the SIR epidemic model~\cite{Newman10}. %, as the choice of the top-k nodes in a multiple spreader scenario is not immune to the influence \disc{overlap} phenomenon.     
%However, in their analysis, the authors consider only a single initial spreader. 
% (the most influential node). 
%However, it is worth noting that  the problem of finding
%an optimal set of $k$ influential nodes is different from the problem of selecting $k$ individuals that are each individually optimal. Intuitively enough, if two top influential nodes both have strong influence on the same group of nodes, the influence exerted by this ensemble   will be quite a bit less than the sum of the influence of each of the nodes.  As a result,   this measure, as well as any naive centrality measures, will likely fail to find  the optimal set of influential nodes.   
 %
   Node diversity into the IM task has been introduced in~\cite{TangLZCZ14}. %Given a set of categories and the category distribution of each node, the authors propose a set of diversity measures that satisfy monotonicity and submodularity.  
%The overall objective function, to be maximized w.r.t. both influence and diversity, is also submodular under the IC and LT models. 
%, thus enabling the design of a greedy-based solution.    
%
This work shares with ours the linear combination of spread and diversity in the definition of objective function. However, our approach does not depend on user characterization based on topic-biased or categorical distributions.
%\disc{DIfferentemente da Zhou (ASONAM 2014) il target set che lor ovogliono differnziare (e lo fanno infine diversificndo i seed), non è predeterminato come nel nostro caso}

\textbf{Targeted influence maximization.}
\label{sec:related:TIM}
Research on targeted IM has gained attention in recent years.  
%that involves some notion of \textit{target} of the diffusion process.   
%
A few studies have assumed that the target is unique and a-priori specified.  
In~\cite{GuoZZCG13}, the authors address the problem of finding the top-$k$ most influential nodes for a specific target user, under the IC model. 
%They   show that the  problem has monotonicity and submodularity properties,  and design greedy algorithms based on repeated Monte Carlo simulations.  
%
In~\cite{GulerVTNZYO14}, the authors investigate optimal propagation  policies to influence a target user. % positively, i.e., in favor of a given idea or message.  %Given a  signed SN, where positive and negative signs correspond to friends and foes, and a selected pair of source and target nodes, the goal  is to find a path between the two nodes which is optimal in terms of   end-to-end propagation delay. 
% (i.e., the fastest way to positively influence the target). 
%The problem is solved via backward-induction dynamic programming. 
%
In~\cite{YangHLC13}, the authors consider the problem of acceptance probability maximization, whereby  a selected user (called initiator) wants to send a friendship invitation to a selected target which is not socially close to the initiator (i.e., the two nodes have no common friends). The goal is to find a set of nodes through which the initiator can best approach the target.   
%Two main issues are how  to model the invitation acceptance of users and how to compute the acceptance probabilities; the former is addressed through  a topology-based social influence factor, while the latter is addressed using  an extended IC model.  
   % 
% It should be noted that  the  above single-target IM problems  stem  from   perspectives different from ours. %:    a unique initiator is identified such that s/he has willingness to establish a connection with a unique target, so that the process starts from a fixed seed and ends in a fixed target.  
  Unlike the above single-target IM methods, our \myalgo   approach   aims at maximizing the probability of activating a target set which can be arbitrarily large, by discovering a seed set which is neither fixed and singleton %(\cite{GuoZZCG13,GulerVTNZYO14,YangHLC13})  
  nor has  constraints related to the topological closeness to a fixed initiator. % (\cite{YangHLC13}).  
 
In~\cite{LiZT15}, the authors describe a keyword-based targeted IM  method, named \algo{KB-TIM}. This assumes that each user  is associated with a weighted term vector to capture her/his preference on advertisements.  
 A user with keywords in common with the advertisement will belong to the target set.    
\algo{KB-TIM} relies on a state-of-the-art approach for the classic IM problem, named \textit{reverse influence sampling} (RIS)~\cite{BorgsBCL14,TangXS14}, which provides theoretical guarantees on the solutions.   
RIS   consists of two main steps: (i) computing, for a fixed number $\theta$ of nodes selected uniformly at random, the \textit{reverse reachable} sets, i.e.,  the sets of nodes that can reach them, and (ii)  
 selecting $k$ nodes that cover the maximum number of reverse reachable sets.  
  In~\cite{TangXS14}, the authors show that, when $\theta$ is large enough, this set has high probability of being a near-optimal solution to IM. More in detail, they propose the \algo{TIM+} algorithm which derives the parameter $\theta$ as function of a lower bound of the maximum expected spread among all size-$k$ node sets.  
The steps of \algo{KB-TIM} are similar to \algo{TIM+}, but as  the former   takes into account only users relevant to an advertisement, it defines a different lower bound for $\theta$. Moreover, while in \cite{BorgsBCL14,TangXS14}  the random reverse reachable sets are sampled online, \algo{KB-TIM} allows     the sampling procedure to be performed offline by building a disk-based reverse reachable index for each keyword.  
Other targeted IM approaches for target-aware viral marketing purposes are described in~\cite{LiLS11,LiuDLXCX14,LeeC15,MaLWC15}. 
 
It is worth emphasizing that, except \algo{KB-TIM} and \algo{TIM+}, \textit{all the above works focus on the IC diffusion model}. Note also that the study in~\cite{LiuDLXCX14}, which is in principle suited to any diffusion model, actually does not take into account the effect of multiple spreaders (i.e.,    the diffusion process is considered only for computing the potential influence of each node at a time).

%=====================================

\section{Targeted Influence Maximization with Topology-driven Diversity}

\subsection{Problem statement}
\label{sec:problemstatement}

Let $\G = \G_0(b, \ell) = \langle \V, \E, b, \ell \rangle$ be a directed weighted graph representing the information diffusion graph associated with the social network $\G_0=\langle \V, \E \rangle$, where $\V$ is the set of nodes, $\E$ is the set of edges, $b: \E \rightarrow \mathbb{R}^*$ is an edge weighting function, and $\ell: \V \rightarrow \mathbb{R}^*$ is a node weighting function. 
 The edge weighting function $b$ corresponds to the parameter of the \textit{Linear Threshold} (LT) model~\cite{Watts02,KempeKT03}, which we adopt as information diffusion model in this  work.  Under the LT model, each node can  be  ``activated''   by its active neighbors if their total influence weight  exceeds the threshold associated to that node. 
More formally, for any edge $(u,v)$, the weight $b(u,v)$ resembles a measure of ``influence'' produced by $u$ to $v$ and it is such that  
%  $$\sum_{u \in \iNeighbor{v}} b(u,v) \leq 1,$$ 
$\sum_{u \in \iNeighbor{v}} b(u,v) \leq 1,$
where $\iNeighbor{v}$ is the in-neighbor set of node $v$. At the   beginning of the diffusion process, each node $v$ is assigned a threshold uniformly at random from $\left[ 0, 1\right]$. Given a set $S \subseteq \V$ of initial active nodes, an inactive node $v$ becomes influenced or active at time $\tau\geq 1$, if the total weight of its active neighbors is greater than its threshold. The process runs until no more activations are possible. 
We denote with $\ActiveSet{S}$  the \textit{final active set}, i.e., the set of nodes that are active at the end of the diffusion process starting from $S$.

%The LT model deals with situations in which exposure to multiple sources is needed for a user before taking a decision. Therefore,  we  can profitably exploit it to pursue the goal of maximizing the likelihood of activating users that have been identified and scored as belonging to a target set. 

Given $\G = \langle \V, \E, b, \ell \rangle$, the node weighting function  $\ell$ determines  the status of each node as a \textit{target}, i.e.,   a node toward which the information diffusion process is directed. More specifically,  for any user-specified threshold $\LurkValue \in  [0,1]$, we define the \textit{target set}  $\TargetSet$ for $\G$    as: 
\begin{equation}\label{eq:TS}
\TargetSet  = \{ v \in \V   \ | \  \ell(v)\geq  \LurkValue\}.
\end{equation}
 
 The objective function of our targeted IM problem is comprised  of two functions. 
The first one, we call  \textit{capital}, is determined as proportional  to the cumulative status of the target nodes that are activated by the seed set $S$. 

\begin{adefinition}[Capital]
Given     $S \subseteq \V$,  the \textit{capital} $\olf{\ActiveSet{S}}$ associated with the final active set $\ActiveSet{S}$  is defined as:
\begin{equation}\label{eq:DC}
\olf{\ActiveSet{S}}=\sum \limits_{v \in (\ActiveSet{S}\cap \TargetSet)\setminus S} \ell(v) 
\end{equation}
\end{adefinition}

The capital function corresponds to the cumulative amount of the scores associated with the activated (target) nodes, i.e., $\olf{\ActiveSet{S}}$. 
Remarkably,   in Eq.~(\ref{eq:DC}) we do not consider nodes that belong to  the seed set $S$, in order to avoid biasing the seed set by nodes with highest scores.  
%The rationality behind this choice is that the selection of the seed set should not be biased by nodes with highest scores.   

The second measure  is   introduced to capture  the overall \textit{diversity} of the nodes in set $S$ w.r.t. the target set. We define it in terms of a function $div_t$ that   is in turn  designed to measure the diversity of a node with respect to each of the target nodes separately. 

\begin{adefinition}[Diversity]
Given  $S \subseteq \V$,  the \textit{diversity} $\SetDiv{S}$ associated with the target set $\TargetSet \subseteq \V$  is defined as:
\begin{equation}\label{eq:Div}
\SetDiv{S}=\sum \limits_{s \in S} \sum_{t \in \TargetSet} div_t(s)   
\end{equation}
\end{adefinition}
As previously mentioned, our approach is to measure node diversity in relation to the structural context of the information diffusion graph.  In Section~\ref{sec:diversity} we shall elaborate on different ways of computing \textit{topology-driven diversity}, and  provide alternative formulations for the $div_t$ function. 

We now formally define our proposed problem of targeted IM, named {\em \textbf{D}iversity-sensitive \textbf{T}argeted \textbf{I}nfluence \textbf{M}aximization} (\myalgo).   
% Afterwards, we discuss the properties of the proposed objective function, and we end this section by presenting our developed algorithms. 

\begin{adefinition}[Diversity-sensitive Targeted Influence Maximization]\label{def:tim} 
Given a diffusion graph $\G = \langle \V, \E, b, \ell \rangle$,   a budget $k$, and a threshold  $\LurkValue$, find a seed set $S \subseteq \V$ with $|S| \leq k$ of nodes (users) such that, by activating them, we maximize the $\of$ ($\symbOF$): 
\begin{equation}\label{eq:problem}
\begin{split}
S & = \argmax \limits_{S^' \subseteq \mathcal{V} \ s.t. \ \vert S^'\vert \leq k}  \symbOF  \\ 
&  = \argmax \limits_{S^' \subseteq \mathcal{V} \ s.t. \ \vert S^'\vert \leq k} \alpha  \olf{\ActiveSet{S^'}} + (1-\alpha)  \SetDiv{S^'}
\end{split}
\end{equation}
where $\alpha \in [0,1]$ is a smoothing parameter that controls the weight of capital $C$ with respect to diversity $D$. 
\end{adefinition} 

The objective function of the problem in Eq.~\ref{eq:problem} is defined in terms of linear combination of the two functions,  capital and diversity.  
The problem in Def.~\ref{def:tim} preserves the complexity of the IM problem and, as a result, it is computationally intractable, i.e., it is still NP-hard. However, as for the classic IM problem, a   greedy solution can be designed  since that  the natural diminishing property holds for   the considered problem, 
as stated in the following.  

\begin{proposition}
 The capital function 
	defined in Eq.~(\ref{eq:DC})  is monotone and submodular under the LT model.
\end{proposition}
 
\begin{proposition}
	 The diversity function  
	defined in Eq.~(\ref{eq:Div})  is monotone and submodular.
\end{proposition}

Proofs of  the above propositions can be found  in the \textit{Appendix}. 
In light of these theoretical results,   $\symbOF$ is also monotone and submodular as it corresponds to a non-negative linear combination of monotone and submodular functions.

 In the next  section, we conceptualize our notion of  user's \textit{topology-driven diversity},   which allows us  to completely specify the objective function $\symbOF$ in our \myalgo problem. 
 %, to completely specify the objective function of our proposed diversity-integrated  targeted IM problem.  

%=====================================

\subsection{Topology-driven Diversity}
\label{sec:diversity}

Our perspective in modeling user diversity is to utilize only structural information given by the topology of a social network graph. Therefore, we take the advantage of a completely \textit{unsupervised} process to avoid  requiring  any side-information or a-priori knowledge on user  attributes that can enable diversification among users.   Instead, we draw inspiration from   social science, in that the way  a user is connected to others within the OSN (a.k.a. \textit{social embeddedness}) is recognized as a  manifestation of diversity of the individual in that online social environment~\cite{HarrisonK07}.   This is also strictly related to the theory of \textit{boundary spanning}~\cite{ESNAM}, 
 %~\cite{CranefieldYH11}, 
  which essentially states that  OSN users may naturally  get knowledge from some of their social contacts and then spread (part of) it  to other contacts through one or more components of the social graph (e.g.,   topically induced communities). 
Boundary spanning has also been recognized as an important aspect to consider in order to adequately characterize those users that can show different behaviors in terms of information-production and information-consumption when considering them laying on the boundary of graph components~\cite{SNAM14,ESNAM}. %CranefieldYH11}. 
Upon the above intuitions, we start from the following basic assumption: 

\begin{principle} 
The  diversity of a user in a social graph can be determined  based on topological properties of   her/his  neighborhood. 
\end{principle}
 
%Since we want to account for topological diversity to discover a ``better'' seed set for targeted IM, 
%we define  user diversity in relation to the structural context of the information diffusion graph. 
%In the following, we  assume that  the information diffusion process unfolds \textit{backward} from a target node. Note that this does not cause loss of generality in both the capital and diversity   definitions we provide here, while it is useful for the purpose of visiting the information diffusion graph as it will be made clear in the next section.   

 \begin{adefinition}[Target-specific information diffusion subgraph]
% \barrato{Let $\G_0=\langle \V, \E \rangle$ denote  the social network graph,}   
Given the diffusion graph  $\G = \langle \V, \E, b, \ell \rangle$, defined over the social graph $\G_0=\langle \V, \E \rangle$,   a target node  $t \in \TargetSet$, and a time step $\tau$, we define the \textit{target-specific   diffusion subgraph} as the directed acyclic graph  $G_t^{(\tau)}=\langle V_t, E_t \rangle \subseteq \G_0$,    rooted in $t$,   
that corresponds to the  portion of $\G$ involved in the unfolding of the diffusion towards $t$, at time $\tau$. 
 \end{adefinition}

% \barrato{ the unknown portion of $\G$ is obtained by visiting  the  \textit{in}-neighbors of a node in the \textit{boundary set} $B_t$ at a time, which is defined as follows.   }
  
\begin{adefinition}[Boundary set]
Given a  target-specific information diffusion subgraph $G_t^{(\tau)}$,     its     \textit{boundary set}  is defined as the set of nodes having  at least one incoming connection from nodes in $\G$ outside $G_t^{(\tau)}$: %in-link outside $G_t$:
\begin{equation}\label{eq:boundaryDue}
B_t^{(\tau)} = \{v \in V_t \ | \ \exists(u,v) \in \E \setminus E_t\}
\end{equation}
  \end{adefinition}

 It is worth noticing here that, while the diffusion starts from a set of seed nodes and follows the directed topology of $\G$, a     widely adopted way of modeling the search for nodes that could reach target ones is to use the \textit{backward} or \textit{reverse} depth-first search  (e.g.,~\cite{GoyalLL11,BorgsBCL14,TangXS14}).

\begin{adefinition}[Expansion of target-specific diffusion subgraph]
Given a  target-specific information diffusion subgraph $G_t^{(\tau)}$ at time $\tau$,   its      \textit{expansion} at time $\tau+1$ is defined as the graph $G_t^{(\tau+1)}$ resulting from the reverse   unfolding of $G_t^{(\tau)}$ such that  $G_t^{(\tau+1)}$ contains nodes in $\G$ that can reach nodes in the boundary set of $G_t^{(\tau)}$. 
Moreover, a target-specific diffusion subgraph is said \textit{fully expanded} if no further backward unfolding over $\G$ is possible.  
  \end{adefinition}

 For the sake of simplification, 
 we hereinafter use symbols  $G_t, B_t$ instead of $G_t^{(\tau)}, B_t^{(\tau)}$ as the association with a particular  time step $\tau$ is assumed to be clear from the context. 
  Moreover, for any $v \in B_t$, we  denote with 
 $N^{in}_{\neg E_t}(v) = N^{in}(v) \setminus \{u \ | \ \exists (u,v) \in E_t \}$   
the set of in-neighbors of $v$ that are not linked to $v$ in $G_t$. 
%, through which the backward visit of $G_t$ can continue.  
 
We provide two alternative ways of modeling topology-driven diversity for targeted IM, which depend  on the strategy adopted to construct $G_t$: 
%, in order to integrate diversity in the unfolding of the information diffusion process:  

\begin{itemize}
\item the first method is designed to   compute node diversity at each step of  the expansion of the information diffusion subgraph for a given target $t$.% (i.e., as each previously unexplored portion of the target-specific diffusion subgraph is being revealed). 
 Since the method does not require information 
  %\barrato{on the whole structure of}  
    on the fully expanded  diffusion subgraph for $t$, it is referred to as \formulaDue.   
\item the second method, named \formulaUno, is instead designed to compute node diversity on the fully expanded  target-specific   diffusion subgraph. 
\end{itemize} 

In the following, we will provide a complete specification of  each of the above introduced diversity methods.

 \vspace{2mm}
 \subsubsection{\bf \cformulaDue}
 \label{sec:diversityDue}
 
Our notion of \formulaDue  of node is designed to  account for the progressive expansion of the information diffusion graph for a given target node.   
 
Given the currently unfolded $G_t$ and a node $v \in B_t$ with $N^{in}_{\neg E_t}(v) \neq \emptyset$, our goal is to determine the \formulaDue for every node $u$ in $N^{in}(v)$ based on two main criteria:

\begin{principle}\label{princip2}
The diversity of  node $u$ should be proportional to the likelihood of reaching  it from nodes outside the currently unfolded target-specific diffusion subgraph $G_t$, i.e., proportional to the number of $u$'s in-neighbors in $\G$ not already in $G_t$. 
\end{principle}

\begin{principle}
The diversity of  node $u$ should be proportional to the increment contributed by that node to the number of incoming links not already included in $G_t$.  
\end{principle}

Accordingly, we   first      characterize  the diversity in the boundary set of $G_t$, and its incremental update due to the insertion of a new node to $G_t$, then we provide our definition of \formulaDue.

\begin{adefinition}[Boundary diversity of set]
Given the currently unfolded $G_t$, % at a given time of its exploration,   
  the \textit{boundary diversity} $\delta_t$ of $G_t$ is defined as the number of nodes %in-neighbors 
   in $N_{\neg E_t}^{in}(v)$ averaged over nodes $v$ in $B_t$:
  
\begin{equation}\label{eq:deltaB2}
\delta_t = \frac{1}{|B_t|} \sum_{v \in B_t} |N_{\neg E_t}^{in}(v)|
\end{equation}
\end{adefinition}
Note that the above definition is simple yet convenient to use in incremental computations. Moreover, it is directly related to the amount of possible paths to diffuse towards a particular target node. 
 The study of alternative  definitions of boundary diversity could be an interesting direction as future work.

For each $u \in N^{in}(v)$,  with $v \in B_t$, if  $u$    is inserted in $G_t$,  the   boundary diversity will change accordingly, since $B_t$ is  updated to contain $u$. The  boundary diversity w.r.t. $B_t$ being  updated with $u$, denoted with $\delta^{+u}_t$, is straightforwardly determined  as follows:  

\begin{equation}\label{eq:deltaBinc}
\delta^{+u}_t = \frac{|B_t|\delta_t + |N^{in}_{\neg E_t}(u)|}{|B_t| + 1}  
\end{equation}

\begin{adefinition}[Local diversity]
The \formulaDue of $u$ is defined as the ratio of the boundary diversity conditional on   inclusion of $u$ in $G_t$, to the actual boundary diversity:

\begin{equation}\label{eq:div2}
div_t(u) =   \frac{\delta^{+u}_t} {\delta_t} = \frac{|B_t|}{1+|B_t|} \left(1+\frac{|N^{in}_{\neg E_t}(u)|}{\sum_{v \in B_t} |N^{in}_{\neg E_t}(v)|}\right)
\end{equation}
\end{adefinition}

Intuitively, the \formulaDue applies to any node $u$ that  is in-neighbor of some   node that lays on the boundary of the currently unfolded $G_t$, and expresses the increment due to   node $u$ to the overall likelihood of  being reached from more different portions of the diffusion graph $\G$.

\vspace{2mm}
 \subsubsection{\bf \cformulaUno}
 \label{sec:diversityUno}

Our second method of topology-driven diversity computation  relies on %an opposite perspective with respect to the previously introduced \formulaDue: 
 the availability of structural information of the fully expanded target-specific diffusion subgraph. 
  While this solution loses the advantage of incremental computation,   
  it also opens to the opportunity of using   more structural features to measure the diversity of a node. 

Given a target node $t$,    $G_t$ is here meant as the fully expanded diffusion subgraph for $t$. 
%   the directed acyclic graph  rooted in $t$. However, differently from the previous definition of diversity,     we regard $G_t$ as the fully-explored diffusion subgraph that is specific  to $t$.   
%For each node $v \in V_t$, let us introduce symbol $E_t(v) \subseteq E_t$ as the set of edges between node $v$ and its neighbors within $G_t$, i.e., $E_t(v) = \{(u,v) \in E_t\} \cup \{(v,u) \in E_t \}$.  
 Moreover, the definition of \textit{boundary} given in Eq.~\ref{eq:boundaryDue} 
as well as the definition of \textit{boundary diversity} given in Eq.~\ref{eq:deltaB2}  do not change; however,   we will exploit them at a ``node level'' rather than a ``set-level'' as for the \formulaDue.

First, the  boundary diversity here assumes a slight  different meaning with respect to   the \formulaDue case. It still captures the strength of the flow potentially spanning  over  
 %\barrato{unexplored   portions  of the social graph, } 
 portions of the diffusion graph not already unfolded,  
which makes Principle~\ref{princip2} hold; however, since   the target-specific diffusion subgraph $G_t$ is   considered as definitively unfolded, we conceptualize that:

\begin{principle}
The boundary spanning should  be regarded as \textit{exogenous} to the diffusion process  for a specific target, and hence intuitively associated to external sources of influence coming from the rest of the social graph. 
\end{principle}

\begin{adefinition}[Boundary diversity of node]
Given a node $v \in B_t$,  the \textit{boundary diversity} of $v$  is defined as  the contribution of $v$ to the   boundary diversity $\delta_t$: 

\begin{equation}\label{eq:divB1}
div_t^{B}(v) =  \frac{|N^{in}_{\neg E_t}(v)|}{|B_t|}
\end{equation}
Boundary diversity is set to zero for any $v \in V_t \setminus B_t$. 
\end{adefinition}

While the concept of boundary diversity is essential to characterize the   connectivity of boundary nodes from outside $G_t$, we also consider here to    measure their  \textit{outward} connectivity within  $G_t$ as the contribution a node gives to the average number of out-neighbors of nodes in $B_t$ that belong to $G_t$. We denote the latter as $|N^{out}_{E_t}(v)| / |B_t|$. 
%
%\begin{equation}\label{eq:deltaC1}
%\delta_t^{C} = \frac{1}{|V_t|} \sum_{v \in V_t} |N^{out}_t(v)|
%\end{equation}
 %Accordingly, for any node $v \in V_t$, we define the \textit{core density} of $v$ and the \textit{boundary diversity} of $v$, respectively, as follows:
%
%\begin{equation}\label{eq:divC1}
%den_t^{C}(v) =  \frac{|N^{out}_t(v)|}{|V_t|}
%\end{equation}
% where  
%%$N^{in}_t(v) = \{(u,v) \in E_t\}$  
%%and 
%$N^{out}_t(v) = \{(v,u) \in E_t\}$ 
%denotes  
%%the set of $v$'s in-neighbors 
%%and 
%the set of $v$'s out-neighbors 
%in $G_t$. %, respectively.  
%
Moreover, we observe that, from the perspective of maximizing diversity of nodes that propagates towards a given target, the overall measure of diversity of node should be not only obviously proportional to its boundary diversity, but also proportional to its outward internal span. 
% core density: in fact,   the latter quantifies how the node contributes to the   connectivity internal to the target-specific subgraph, which is however already taken into account when determining the contribution of node to the overall capital.  
The above considerations lead to the following definition.

\begin{adefinition}[Global diversity] 
The \formulaUno  of node $v$ is defined as:
\begin{equation}\label{eq:div1}
div_t(v) =  div_t^{B}(v) \times \mathrm{f}\left(\frac{|N^{out}_{E_t}(v)|}{|B_t|} \right) 
\end{equation}
where $\mathrm{f}$ is a smoothing function to assign the outward internal span a weight at most equal to the boundary diversity term.  
\end{adefinition}

In the following, we will refer to a logarithmic smoothing, i.e.,   $\mathrm{f} = \log(1+|N^{out}_{E_t}(v)|/|B_t|)$,   since we want the outward internal span of node  has an impact lower than the boundary diversity  on the overall value of diversity.

%=====================================
 
\subsection{The \myalgo algorithms}
\label{sec:algo}
In this section, we show our algorithmic solutions to the proposed \textbf{D}iversity-sensitive \textbf{T}argeted \textbf{I}nfluence \textbf{M}aximization problem. According to the \formulaDue and \formulaUno criteria previously introduced in Section~\ref{sec:diversity}, 
we provide  two methods, named  L-\myalgo and  G-\myalgo, respectively; due to space limits of this paper, they are concisely   reported in Algorithm~\ref{alg:formulazione1}.  
 
Following the lead of the study in~\cite{GoyalLL11}, L-\myalgo and G-\myalgo exploit as well the search for shortest paths in the diffusion graph, however in a backward  fashion. 
Along with  the information diffusion graph $\G$,  the budget integer $k$, the minimum score $\LurkValue$ and a parameter $\alpha \in \left[ 0, 1\right]$ which controls the balance between capital and  diversity, L-\myalgo and G-\myalgo take in input a real-valued threshold $\eta$. This parameter is  used to control the size of the neighborhood within which paths are enumerated: in fact, the majority of influence can be captured by exploring the paths within a relatively small neighborhood; note that for higher $\eta$ values, less paths are explored (i.e., paths are pruned earlier) leading to smaller runtime but with decreased accuracy in spread estimation.

 		\renewcommand{\algorithmicrequire}{\textbf{Input:}}
 		\renewcommand{\algorithmicensure}{\textbf{Output:}}
 			\begin{algorithm}\caption{\myalgo - \textbf{D}iversity-sensitive \textbf{T}argeted \textbf{I}nfluence \textbf{M}aximization}
 			%  ---  {\footnotesize  \textit{Note that the instruction in the red line (resp. blue line) is performed only by L-\myalgo (resp. G-\myalgo)}}}
 			\label{alg:formulazione1}
 			\begin{scriptsize}
 				\begin{algorithmic}[1]
 					\REQUIRE A graph $\G=\langle \V, \E, b, \ell \rangle$, a budget (seed set size) $k$, a target selection threshold $\LurkValue \in [0,1]$, a path pruning threshold $\eta \in [0,1]$, a smoothing parameter $\alpha \in [0,1]$.
 					\ENSURE  Seed set $S$.
 					%\STATE $S \leftarrow \emptyset$ 
 					\STATE $T \leftarrow \V$  \hfill \COMMENT{\textit{nodes that can reach target nodes}}
 					%\STATE $\TargetSet \leftarrow   \emptyset $
 					\FOR{$u \in \V$}
 					\IF{$\ell(u)\geq \LurkValue$} 
 					\STATE $\TargetSet \leftarrow \TargetSet \cup \{u\}$ \hfill  \COMMENT{\textit{identifies the target nodes}}
 					\ENDIF
 					\STATE $u.Dset \leftarrow \{\}$ \hfill  \COMMENT{\textit{initializes a data structure that keeps track of node diversity w.r.t. any target}}
 					\ENDFOR
 					\WHILE{$\vert S \vert < k$}  
 					%\STATE $bestSeed \leftarrow   -1$ \hfill \COMMENT{\textit{keeps track of the node with the highest spread}}
 					\FOR{$u \in T\setminus S$}
 					\STATE $u.C, u.D  \leftarrow 0$  \hfill \COMMENT{\textit{initializes each node's capital and diversity to zero}}
 					\ENDFOR
 					\STATE $T \leftarrow \emptyset$
 					\FOR{$t \in \TargetSet \setminus S$}  
 					\STATE $G_t = \langle V_t, E_t \rangle \leftarrow \langle \{t\}, \emptyset \rangle$ \hfill \COMMENT{\textit{initializes DAG rooted in $t$}}
 					\STATE $\algo{backward}(\langle t \rangle,1,t)$
 					\IF{$\vert S \vert = 0$} 
 					\STATE $\algo{updateDiversity}(t)$
 					\ENDIF
 					\ENDFOR
 					%\IF{$bestSeed \neq -1$}
 					\STATE $S \leftarrow S \cup \{ bestSeed \}$
 					%\ELSE \STATE \textbf{break}
 					%\ENDIF
 					\ENDWHILE
 					\RETURN $S$
 					\newline
 					\STATE \textbf{procedure} $\algo{backward}(\mathcal{P},pp,t)$   
 					\STATE $v \leftarrow \mathcal{P}.last()$, \ $T \leftarrow T \cup \{u\}$
 					\WHILE{$u \in N^{in}(v) \ \wedge \ u \not \in S \cup \mathcal{P}.nodeSet()$} 
 					\STATE $pp \leftarrow pp \times  b(u,v)$  \hfill \COMMENT{\textit{updates the path probability}}
 					\IF{$pp \geq \eta$}
 					\STATE $u.C \leftarrow u.C + pp \times \ell(t)$      \hfill  \COMMENT{\textit{updates the overall node capital}} 
 					\IF{$\vert S \vert = 0$}
 					\STATE $u.\infp \leftarrow u.\infp + pp $ \hfill \COMMENT{\textit{increases the overall influence of node $u$ on the current target}}
 					\STATE $(^{*})$ $u.Dset(t) \leftarrow div_t(u)$ \hfill \COMMENT{\textit{computes the current node diversity w.r.t. the target by Eq.\ref{eq:div2}}}
 					\STATE $G_t = \langle V_t \cup \{u\}, E_t \cup \{(u,v)\}\rangle $ \hfill \COMMENT{\textit{adds the edge $(u,v)$ to the explored DAG}}
 					\ELSE
 					\STATE $u.D \leftarrow u.D + pp \times u.Dset(t)$
 					%\IF{$bestSeed = -1 \vee u.\symbOF > bestSeed.\symbOF$}
 					\IF{$u.\symbOF > bestSeed.\symbOF$} 
 					\STATE $bestSeed \leftarrow   u$ \hfill  \COMMENT{\textit{sets the current best seed node as $u$}}
 					\ENDIF
 					\ENDIF
 					\STATE $\algo{backward}(\mathcal{P}.append(u),pp,t)$ 
 					
 					\ENDIF
 					\ENDWHILE
 					\newline
 					\STATE \textbf{procedure} $\algo{updateDiversity}(t)$
 					\FOR{$v \in V_t$}   
 					 \STATE $(^{**})$ $v.Dset(t)  \leftarrow div_t(v)$  \hfill  \COMMENT{\textit{computes node diversity w.r.t. the target $t$ by Eq. \ref{eq:div1}}}
 					\STATE $v.D \leftarrow v.D +  v.\infp\times v.Dset(t)$    \hfill  \COMMENT{\textit{updates the overall node diversity}} 
 					\STATE $v.\infp \leftarrow 0$
 					%\IF{$bestSeed = -1 \vee v.DIC > bestSeed.\symbOF$} 
 					\IF{$v.DIC > bestSeed.\symbOF$} 
 					\STATE $bestSeed \leftarrow   v$ \hfill  \COMMENT{\textit{sets the current best seed node as $v$}} 
 					\ENDIF
 					\ENDFOR 
 				\end{algorithmic}
 			\end{scriptsize}
 			
  % BEGIN COLOR
\vspace{.4em}
\scriptsize
$(^*)$ Instruction at line 31 is performed   by L-\myalgo only.\\
$(^{**})$ Instruction at line 44 is performed   by G-\myalgo only.\\
 		\end{algorithm}
 
As previously mentioned, L-\myalgo and G-\myalgo share the idea of performing a backward visit of the diffusion graph starting from the nodes identified as target (i.e., the nodes $u$ with $\ell(u) \geq \LurkValue$). To this end, all nodes are initially examined to compute the target set $\TargetSet$ (lines 2-5). 
In order to yield a seed set $S$ of size at most $k$, during each iteration of the main loop (lines 8-21), both the variants of Algorithm~\ref{alg:formulazione1} compute the set $T$ of nodes that reach the target ones and keep track, into the variable $bestSeed$, of the node with the highest marginal gain (i.e., diversity-sensitive capital $\symbOF$).  %This  enables an efficient reset of the attributes of each non-seed node contained in $T$. 

The $bestSeed$ node is found at the end of each iteration upon calling the subroutine \algo{backward} over all nodes in $\TargetSet$ that do not belong to the current seed set $S$. This subroutine takes a path $\mathcal{P}$, its probability $pp$ and the target $t$ from which the visit has started, and extend  $\mathcal{P}$ as much as possible (i.e., as long as $pp$ is not lower than $\eta$). Initially, a path is formed by one target node, with probability 1 (line 15). Then, the path is extended by exploring the graph backward, adding to it one, unexplored in-neighbor $u$ at time, in a depth-first fashion. 
Path probability is updated (line 26) according to the LT-equivalent ``live-edge'' model \cite{GoyalLL11,KempeKT03}, and so the  capital (line 28).  
The process is continued until no more nodes can be added to the path.

Both G-\myalgo and L-\myalgo  compute the node diversity only at the first iteration of the main loop, i.e., when the seed set $S$ is empty.  Indeed,  for each node, 
we keep track of its diversity w.r.t. each target it can reach, by using data structure  $Dset$.
A major difference between the two variants is that in G-\myalgo the node diversity is computed (through the subroutine \algo{updateDiversity}) only when the whole subgraph rooted in $t$ has been completely built (line 44). In L-\myalgo, instead, the node diversity is updated every time the node has been reached (line 31). Note that the instruction at line 31 (resp. 44) is performed by L-\myalgo (resp. G-\myalgo) only.  
The value of diversity of a node $v$ is, in both the variants, smoothed with the influence that $v$ might exert on $t$, contributing to the overall diversity $D$ of $v$ (line 45).

Note that both the numerical values yielded by both global diversity and local diversity functions $div_t$ might be subject to scaling in order to enable a fair comparison with the numerical value yielded by the capital.

\parskip1pt
%											 

\begin{figure}[t]
	\centering
	\includegraphics[width=0.35\textwidth]{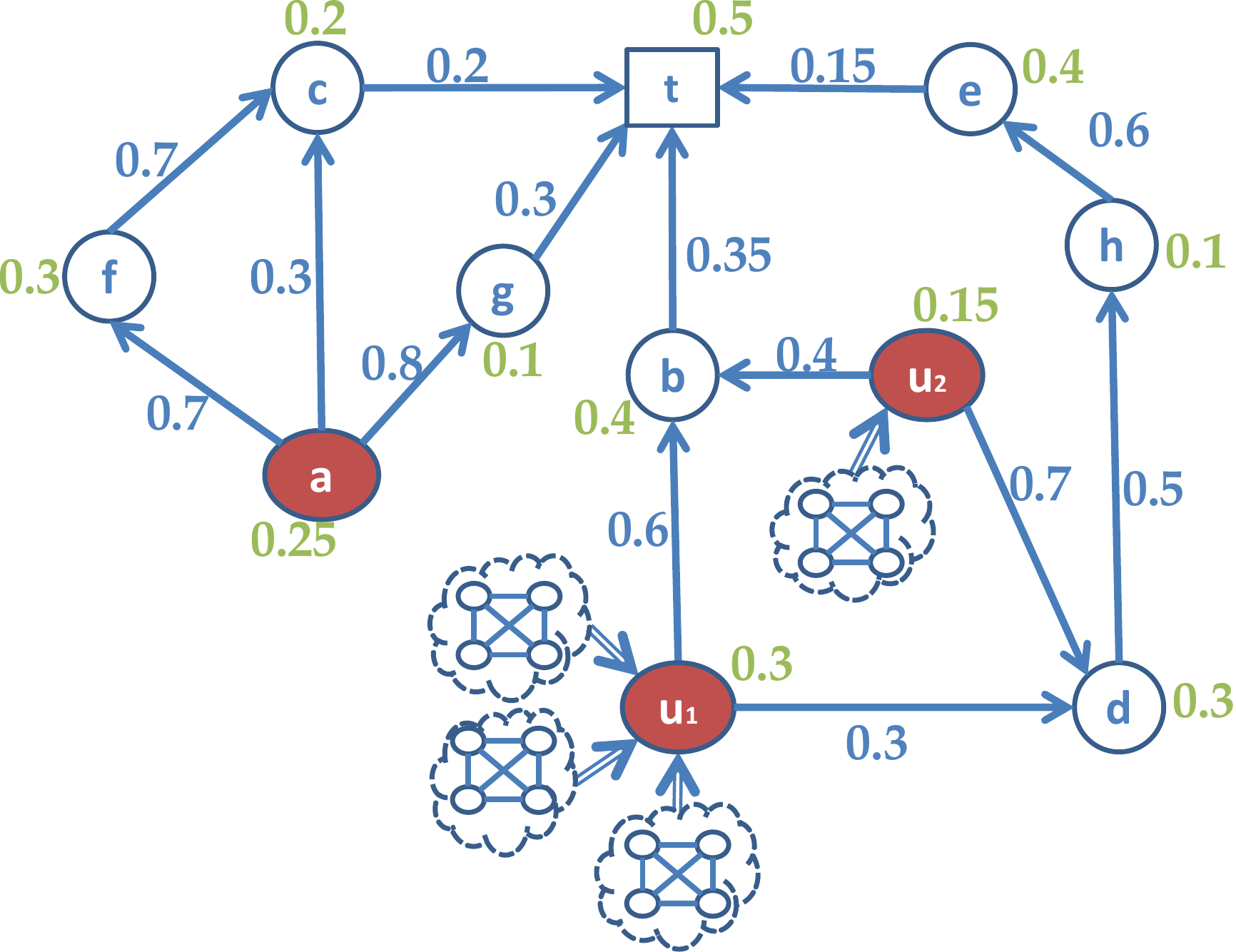}
	\caption{Targeted IM vs. diversity-sensitive targeted IM.  Edge weights (values in blue) and node weights (values in green) are computed by functions $b$ and $\ell$. To avoid cluttering of the figure, the node activation thresholds used by LT model here coincide with the node weights.}
	\label{fig:exampleDTIM}
\end{figure}

%\begin{aexample}
 %%%%%%%%%%%%%%%  EXAMPLE 2  %%%%%%
 \textbf{\em Example 2.}
Consider the example in Fig.~\ref{fig:exampleDTIM}, where the  target set includes the square border node $\{t\}$. Let's assume for simplicity we set $k=1$, $\alpha=0.5$, $\eta=0$ and we ignore the spread computation for nodes  inside the other components of $G_t$ (represented within clouds in the figure). Moreover, the double arrows connecting these components to nodes $u_1$ and $u_2$ count as two edges each. 
In the following, we denote with $pp\left[ x\rightarrow \cdots \rightarrow y\right] $ the probability of the path from $x$ to $y$, and with $x.\infp$ the overall influence exerted by node $x$ to the target.

The target node $t$ can be reached through $a$ (with $a.\infp = pp\left[ a\rightarrow f \rightarrow c \rightarrow t\right]  + pp\left[ a\rightarrow c \rightarrow t\right]  + pp\left[ a\rightarrow g \rightarrow t\right] = 0.098 + 0.06 + 0.24 $), $b$ (with $b.\infp = pp\left[ b\rightarrow t\right] = 0.35$), $c$ (with $c.\infp = pp\left[ c\rightarrow t\right] = 0.2$), $d$ (with $d.\infp = pp\left[d\rightarrow h \rightarrow e \rightarrow t\right] = 0.045$), $e$ (with $e.\infp = pp\left[ e \rightarrow t\right] = 0.15$), $f$ (with $f.\infp = pp\left[ f \rightarrow c \rightarrow t\right] = 0.14$), $g$ (with $g.\infp = pp\left[g \rightarrow t\right] = 0.3$), $h$ (with $h.\infp = pp\left[ h \rightarrow e \rightarrow t\right] = 0.09$), $u_1$ (with $u_1.\infp = pp\left[ u_1 \rightarrow d\rightarrow h \rightarrow e \rightarrow t\right] + pp\left[ u_1 \rightarrow b \rightarrow t\right] = 0.0135 + 0.21$), and $u_2$ (with $u_2.\infp = pp\left[ u_2 \rightarrow d\rightarrow h \rightarrow e \rightarrow t\right] + pp\left[u_2 \rightarrow b \rightarrow t\right] = 0.0315 + 0.14$).  Node $a$ has the largest chance of success in activating $t$, which results in the highest capital $C$. 
%($0.398\times 0.5$). 
However, since $a$ does not have in-neighbors, its diversity is equal to zero for both the diversity formulations. %Indeed, according to Eq. \ref{eq:boundaryDue}, the set of boundary nodes is  $B_t = \{ u_1, u_2\}$. 

Let us first focus on the behavior of G-\myalgo. According to Eq.~\ref{eq:boundaryDue}, the set of boundary nodes is  $B_t = \{ u_1, u_2\}$.
By definition of \formulaUno (Eq.~\ref{eq:div1}),  G-\myalgo computes the following values: 
$u_1.D = 2.08 $ (as  $div_t^{B}(u_1) = 6/2$ and $div_t(u_1) = 3 \times \log(1+2/2)$), 
$u_2.D = 0.69 $ (as $div_t^{B}(u_2) = 2/2$ and $div_t(u_1) = 1 \times \log(1+2/2)$). By applying the max-normalization to the node diversity, the final values are $u_1.D = 1 $ and $u_2.D = 0.33$.
As a result, for G-\myalgo node $u_1$ is chosen as seed node since it has    diversity-sensitive capital ($\symbOF = 0.22 \times 0.5 \times (0.5 + 1) = 0.165$) higher than that of $a$ ($\symbOF = 0.4 \times 0.5 \times (0.5 + 0) = 0.1$) and $u_2$ ($\symbOF = 0.13 \times 0.5 \times (0.5 + 0.33) = 0.05$).

The values of node diversity computed by L-\myalgo depend on   the order in which nodes are reached during the backward visit. Assume to visit first the branch starting from node $e$. According to Eq.~\ref{eq:div2}, L-\myalgo computes the following values of node diversity:
$e.D = 0.625 $ ($div_t(e) = 1/2 \times (1+1/4)$ as $B_t=\{t\}$), 
$h.D = 0.83 $ ($div_t(h) = 2/3 \times (1+1/4)$ as $B_t=\{t,e\}$), 
$d.D = 1 $ ($div_t(d) = 2/3 \times (1+2/4)$ as $B_t=\{t,h\}$), and, assuming to visit  $u_1$ before $u_2$,
$u_1.D = 1.47 $ ($div_t(u_1) = 2/3 \times (1+6/5)$ as $B_t=\{t,d\}$), 
$u_2.D = 0.9 $ ($div_t(u_2) = 3/4 \times (1+2/10)$ as $B_t=\{t,d,u_1\}$). 
Analogously, it proceeds in computing the node diversity through branches $c$ and $g$, whose values of diversity are lower than $0.9$ (not reported for the sake of readability). 
L-\myalgo eventually computes the following diversity:  
$b.D = 0.92 $ ($div_t(b) = 3/4 \times (1+2/9)$ as $B_t=\{t,u_1,u_2\}$), 
$u_1.D = 1.2 $ ($div_t(u_1) = 3/4 \times (1+6/10)$ as $B_t=\{b,u_1,u_2\}$), and 
$u_2.D = 0.92 $ ($div_t(u_2) = 3/4 \times (1+2/9)$ as $B_t=\{b,u_1,u_2\}$). 
Upon max-normalization to the values so obtained,   L-\myalgo  will choose  $b$ as seed node since it has diversity-sensitive capital ($\symbOF = 0.35 \times 0.5 \times (0.5 + 0.77) = 0.22$) higher than that of $u_1$ ($\symbOF = 0.22 \times 0.5 \times (0.5 + 1) = 0.165$).
%\end{aexample}

%=====================================

\section{Using \myalgo to engage silent users in social networks}

We evaluate our framework of targeted IM with topology-driven diversity on a special case of user engagement in OSNs, which refers to the problem of 
\textit{how to turn silent users into more active  contributors} in the community life. 

All large-scale OSNs are characterized by a participation inequality principle:   the crowd does not take an active role in the interaction with other members, rather it takes on a silent role. Silent users are also referred to as \textit{lurkers}, since they gain benefit from information produced by others, by observing the user-generated communications at all stages (e.g., reading posts, watching videos, etc.), but  without significantly giving back to the community~\cite{Edelmann13,Sun+14}.  
%Note that \textit{lurkers do not correspond to totally inactive users}, i.e.,   registered members who simply do not use their OSN account to log in and join the community life. 

Social science and human-computer interaction research communities have widely investigated   the main causes that explain  lurking behaviors, which include    subjective reticence (rather than malicious motivations) to contribute to the community wisdom, or a feeling that gathering information by browsing is enough without the need of being further involved in the community. Moreover, lurking can be expected or even encouraged because it allows users (especially newcomers) to learn or improve their understanding of the etiquette of an online community~\cite{Edelmann13}.   

Regardless of their  motivations, lurkers might have great potential in terms of \textit{social capital}, because they acquire knowledge from the OSN. They   can become aware of the existence of different perspectives and may make use of these perspectives in order to form their own opinions,   but they are unlikely to  let other people know their value. % (e.g., ideas, expertise, opinions, etc.). 
In this regard, it might be   desirable  to engage such users, or \textit{delurk} them, i.e., to develop a mix of strategies aimed at encouraging lurkers to return their acquired social capital, through a more active participation to the community life. %%This has the important long-term effect of helping sustain the OSN over time with fresh ideas and perspectives.

\textit{Engagement actions} towards silent users can be   categorized into four types~\cite{Sun+14}:   %  ~\cite{NonneckePAV04,Du06,LaiC14}:  
%\begin{itemize} 
%\itemsep1pt \parskip0pt \parsep0pt\leftmargin-.2in%\itemindent-.2in
reward-based external stimuli, % (e.g., tangible or intangible rewards); 
providing encouragement information, % (e.g., welcome statements, introduction to the netiquette rules);  
improvement of the usability and learnability of the system, % (to make it easier for users to participate);  
guidance from elders/master users to help lurkers become familiar with the system as quickly as possible.  
It is worth emphasizing   that \textit{our approach is  independent on the particular strategy of delurking  being adopted}.   
%Demonstrating the proposed approach on a real case study is beyond the scope of this paper;  nevertheless, as we believe  that the development of a  web-based system embedding our approach can really aid to support engagement of users, we envisage applications relevant in human-computer interaction and   marketing fields as well.  
The goal here is    how to    instantiate our \myalgo algorithms  in a user engagement scenario where \textit{lurkers are regarded as the target users} of the diffusion process. 
Therefore, our goal becomes:  
Given a budget $k$,  to  find a  set of $k$ nodes that are capable of maximizing the diversity-sensitive capital, i.e., the likelihood of activating the target silent users through diverse seed users.

A key aspect of our approach in this scenario is that  the selection of target users is  based on  the   solution produced by   a \textit{lurker ranking} algorithm~\cite{ASONAM,SNAM14,SNAM15} 
   applied to the social network graph $\G_0$. 
In Section~\ref{sec:lr} we provide a summary of the lurker ranking method we used in this work,  and in Section~\ref{sec:blfunctions}   we describe how the input  diffusion graph for \myalgo is  modeled, following our early work  in~\cite{InterdonatoPT15}.

\subsection{Identifying target users through LurkerRank}
\label{sec:lr}
Lurker ranking methods, originally proposed in~\cite{ASONAM,SNAM14},  are designed to mine silent user behaviors in the network, and hence to associate  users with a score  indicating her/his lurking status. 
 Lurker ranking methods  rely upon a
 \textit{topology-driven definition of  lurking} which is based    on the network structure only. 
 %, where    any edge $(v, u)$ means that   $u$  is ``consuming'' or ``receiving''  information from $v$.   %(e.g., $u$ follows $v$, or likes/rates/comments $v$'s contents).    
 Upon the assumption that   lurking behaviors build on the \textit{amount of information a node receives},  the key intuition is that  the strength of  a user's lurking status can be determined based on three basic principles:   overconsumption, authoritativeness of the information received, non-authoritativeness of the information   produced.

The above   principles form the basis for three ranking models that differently account for   the contributions of a node's in-neighborhood and out-neighbor\-hood. 
A complete specification of the lurker ranking models is provided in terms of PageRank  and Alpha\-Centrality based formulations.  
 For the sake of brevity here, we will refer to only one of the  formulations described in~\cite{ASONAM,SNAM14}, which is that based on the full \textit{in-out-neighbors-driven lurker ranking}, hereinafter dubbed  simply as \algo{LurkerRank} (LR). 

Given the directed social graph  $\G_0 = \langle \V, \E \rangle$, where any edge $(u, v)$ means that $v$ is is ``consuming'' or ``receiving'' information from $u$, the  LurkerRank $LR(v)$ score of  node  $v$ is defined as:
\begin{equation}\label{eq:LR}
LR(v) = \d  [\mathcal{L}_{\mathrm{in}}(v)  \ (1+\mathcal{L}_{\mathrm{out}}(v))]  +   (1-\d)p(v)  
\end{equation}
where $\mathcal{L}_{\mathrm{in}}(v)$ is the in-neighbors-driven lurking function:
\begin{equation}
 \mathcal{L}_{\mathrm{in}}(v) =  \frac{1}{out(v)} \sum_{u \in \iNeighbor{v}} \frac{out(u)}{in(u)} LR(u) 
\end{equation}  
and $\mathcal{L}_{\mathrm{out}}(v)$ is the out-neighbors-driven lurking function:
\begin{equation}
\mathcal{L}_{\mathrm{out}}(v) =  \frac{in(v)}{\sum_{u \in \oNeighbor{v}} in(u)} \sum_{u \in \oNeighbor{v}} \frac{in(u)}{out(u)} LR(u)  
\end{equation}  
where:
$in(v)$ (resp. $out(v)$) denotes the size of the set of in-neighbors (resp. out-neighbors) of $v$,  $\d$ is a damping factor ranging within [0,1] (usually set to 0.85), and $p(v)$ is the value of the personalization vector, which is set to $1/|\V|$ by default.   
To prevent zero or infinite ratios, the values of  $in(\cdot)$ and $out(\cdot)$ are Laplace add-one smoothed.

\subsection{Modeling the diffusion graph}
\label{sec:blfunctions}
In Section~\ref{sec:problemstatement}, we introduced symbol  $\ell(v)$ to denote the weight of node $v$   that quantifies  its status  as target. In this application scenario, the higher is the lurker ranking score of $v$  the higher should be $\ell(v)$.  

We define the node weighting function $\ell$ upon scaling and normalizing  the stationary distribution produced by the \algo{LurkerRank} algorithm over $\G_0$. 
The scaling compensates for the fact that  the lurking scores produced by   \algo{LurkerRank}, although distributed over a significantly wide range (as reported in~\cite{SNAM14}), might be numerically very low (e.g., order of  $1.0e\mbox{-}3$ or below). 
Moreover, we introduce a small smoothing constant in order to avoid that the highest lurking scores are mapped exactly to 1.  
  Formally, for each node $v \in \V$, we define the \textit{node lurking value}    $\ell(v) \in  [0,1)$ as follows:
\begin{equation}\label{eq:nodeweight}
\ell(v) = \frac{\widetilde{\pi_v}  - min_r }{(max_r - min_r)+\epsilon_r}
\end{equation}
\noindent where $\widetilde{\pi}$ denotes the stationary distribution of the lurker ranking scores ($\pi$) divided by the base-10 power of the order of magnitude of the minimum value in $\pi$, 
$\widetilde{\pi_v}$ is the value of $\widetilde{\pi}$ corresponding to node $v$, 
$max_r = \max_{u \in \V} \widetilde{\pi_u}$, $min_r = \min_{u \in \V} \widetilde{\pi_u}$,   and $\epsilon_r$ is a smoothing constant proportional to the order of magnitude of the $max_r$ value.  

In order to define the edge weights so that they express a notion of  strength of influence from a node to another (as normally required in an information diffusion model), 
we again exploit information derived from the ranking solution obtained by \algo{LurkerRank} as well as from the structural properties of the social graph. 
Our key idea is to calculate the  weight   on edge $(u,v) \in \E$ proportionally to the fraction of the original lurking score of $v$ given by its in-neighbor $u$:
\begin{equation}\label{eq:b0}
b_0(u,v) = \left[\sum_{w \in \iNeighbor{v}}  \frac{out(w)}{in(w)} \pi_w   \right]^{-1}      \frac{out(u)}{in(u)} \pi_u
\end{equation}

Using   Eq.~(\ref{eq:b0}), we finally define the   edge weight as:
\begin{equation}\label{eq:edgeweight}
b(u,v) = b_0(u,v)  \times  e^{\ell(v) -1} 
\end{equation}

\noindent 
Note that  Eq.~(\ref{eq:edgeweight})  meets the requirement $\sum_{u \in \iNeighbor{v}} b(u,v) \leq 1$, and  
accounts  for $\ell(v)$ such that the  resulting weight on $(u,v)$ is lowered   for higher $\ell(v)$, i.e., the more a node acts as a lurker, the more active in-neighbors are needed to activate that node.

%=====================================
\section{Evaluation methodology}
\label{sec:eval}

\subsection{\myalgo settings}
We experimentally varied the input and model parameters in \myalgo methods, namely: 
 the size of seed set ($k$),   the target selection threshold ($\LurkValue$),  the path pruning threshold ($\eta$), and  the   parameter $\alpha$ to control the contribution of diversity versus capital in the objective function of \myalgo methods.  
   Note that, to simplify the     interpretation of $\LurkValue$, we will instead use symbol  $\LurkValuePerc$ to denote a percentage value that determines the  setting of $\LurkValue$  such  that the selected target set   corresponds to the top-$\LurkValuePerc$ of the distribution of     scores yielded by function  $\ell$;  particularly, we set $\LurkValuePerc \in \{5\%, 10\%, 25\% \}$.  
    As concerns $\eta$,   though $\eta=1.0e\mbox{-}03$ is the  default   as used in other IM algorithms (e.g.,\cite{GoyalLL11}), 
 we set it to     a lower value,  $\eta = 1.0e\mbox{-}04$, 
to impact even less on the unfolding of the information diffusion process; moreover,  we will not present results corresponding to $\eta=0$ (i.e., no path-pruning), since we observed 
%in our early work on targeted IM~\cite{InterdonatoPT15}, evaluated on the same datasets (cf. Section~\ref{sec:data}), 
   this negatively affects the runtime by   several orders of magnitude while yielding nearly identical results to those corresponding to $\eta = 1.0e\mbox{-}04$.

\subsection{Competing methods}
We considered comparison with  \algo{TIM+}~\cite{TangXS14} and \algo{KB-TIM}~\cite{LiZT15}, which are   state-of-the-art solutions to the IM (resp. targeted IM) problem, based on the RIS approach (cf. Sect.~\ref{sec:related:TIM}).  

 Comparing \myalgo with a non-targeted  IM algorithm like  \algo{TIM+}   required to evaluate the quality of   seed sets produced by the    competing algorithm  under   a \textit{targeted} scenario.  To this purpose, we simply let \algo{TIM+} compute a size-$k$  seed set over the entire graph and then we estimated  the  capital  over different target sets  in accord  with the setting of   \myalgo.   
  We considered  two opposite settings for the main parameter ($\epsilon$) in \algo{TIM+}: (i)   the default  $\epsilon=0.1$, which provides strong theoretical guarantees yet is adversarial to the algorithm's memory consumption, and (ii) $\epsilon=1.0$, which conversely provides 
   no approximation guarantees but high empirical efficiency; note that the latter setting was also used by the \algo{TIM+}'s authors in~\cite{TangXS14} for the comparison with \algo{SimPath}.  
  We used default settings for the other parameters in \algo{TIM+}.  

 As concerns  \algo{KB-TIM}, we modified the keyword-based target selection stage to  make it equivalent to the target selection adopted in \myalgo.  
 \algo{KB-TIM} requires two main input files to drive the target selection: (i)   a sort of document-term sparse matrix, such that each node (document) in the graph is   assigned  a list of $keyword, \#occurrences$ pairs, and (ii) a list of keyword-queries, so that each query corresponds to the selection of a subset of nodes in the graph. 
To prepare these input files,  we defined three queries corresponding to the setting   $\LurkValuePerc \in \{5\%, 10\%, 25\% \}$, and accordingly created the sparse matrix so that each node was assigned a keyword for each of the top-ranked subsets   it belongs  to (e.g., a node in the top-10\% set of lurkers will be assigned two keywords, as it is also in the top-25\% set); moreover, the $\#occurrences$ associated with any keyword for a given node $v$ was calculated as   the  node lurking value     $\ell(v)$ suitably scaled  and truncated to its integer part.
 Also, we used the incremental reverse-reachable index ($IRR$) in  \algo{KB-TIM}.

 \subsection{Data} 
 \label{sec:data}
We used   FriendFeed~\cite{CelliLMPR10},  GooglePlus~\cite{McAuleyL12}, and Instagram~\cite{SNAM15}\footnote{Available at    http://people.dimes.unical.it/andreatagarelli/data/.} network datasets. Note that, for the sake of significance of the information diffusion process in latter network, we selected the induced  subgraph corresponding to the maximal strongly connected component of the original network graph,   hereinafter referred to   as \data{Instagram-LCC} (LCC stands for  largest connected component). 
 As major motivations underlying our data selection,  we  wanted to maintain continuity with our previous studies~\cite{SNAM14,SNAM15}  and     use publicly available datasets.   
 Table~\ref{tab:data} summarizes main structural characteristics of the evaluation network datasets.

\begin{table}[t]
	\centering
	\renewcommand\arraystretch{0.5}
%	\resizebox{\linewidth}{!}{%
\scalebox{0.9}{
\begin{tabular}{|l|c|c|c|c|c|c|}
\hline
\emph{data} & \textit{\# nodes} & \textit{\# links} &  \textit{avg}  &  \textit{avg} & \textit{clust.} & \textit{assorta-} \\% &  \textit{\# sources}    \\
  &   &  &  \textit{in-deg.}  &   \textit{path len.} & \textit{coeff.} & \textit{tivity} \\%& \textit{\# sinks}   \\
 \hline \hline
  \!\!\multirow{2}{*}{\textit{FriendFeed}}    &  \!\!\multirow{2}{*}{493,019}\!\! &   \!\!\multirow{2}{*}{19,153,367}\!\!  &  \multirow{2}{*}{38.85} & \multirow{2}{*}{3.82} &  \multirow{2}{*}{0.029} & \multirow{2}{*}{-0.128}  \\ %&  41,953    \\
 				& & & &  & & \\ %& 292,003     \\
 \hline 
 \!\!\multirow{2}{*}{\textit{GooglePlus}}    &  \!\!\multirow{2}{*}{107,612}\!\!  &   \!\!\multirow{2}{*}{13,673,251}\!\!  &  \multirow{2}{*}{127.06} & \multirow{2}{*}{3.32} &  \multirow{2}{*}{0.154} &  \multirow{2}{*}{-0.074}  \\ %&  35,341     \\
 				& & & &  & & \\ %&  22      \\
 \hline				
 \!\!\multirow{2}{*}{\textit{Instagram-LCC}}\!\!\!    &  \multirow{2}{*}{17,521}   &  \multirow{2}{*}{617,560}  &  \multirow{2}{*}{35.25} & \multirow{2}{*}{4.24} &  \multirow{2}{*}{0.089} &  \multirow{2}{*}{-0.012}  \\ % &  0     \\
 				& & & &  & & \\ % &  0      \\				
\hline   
\end{tabular}
}
\caption{Summary  of the evaluation network datasets\label{tab:data}}
\end{table}

%=======================================

\section{Results}

We present results of the evaluation of our proposed \myalgo algorithms according to three main objectives:     analysis of the identified seed nodes (Sect.~\ref{sec:seed-evaluation}),  analysis of the activated target nodes (Sect.~\ref{sec:target-evaluation}) and efficiency analysis (Sect.~\ref{sec:efficiency}).\footnote{All experiments were carried out on an Intel Core i7-3960X CPU @3.30GHz, 64GB RAM machine.   
 All algorithms were written in C++. All competing algorithms refer   to the original  source code provided by their authors.}

\subsection{Evaluation of identified seed nodes}
\label{sec:seed-evaluation} 

\bfseries
 \vspace{1mm}
\subsubsection{Seed set overlap}

\normalfont
In order to investigate the impact of taking into account diversity on the seed identification   process, we initially analyzed the matching  among seed sets produced by the two \myalgo methods with   varying $\alpha$. 

This analysis of seed sets  was twofold:  (i) pair-wise evaluation of the overlaps between seed sets produced by a particular  \myalgo  method  by  varying $\alpha$,  and (ii)  pair-wise evaluation of the overlaps between seed sets produced by G-\myalgo and L-\myalgo for particular values of $\alpha$.    
Unless otherwise specified, results correspond to  the largest sizes of target set and seed set we  considered  (i.e., $\LurkValuePerc=25\%$ and $k=50$), and express the \textit{normalized overlap} of any two seed sets, i.e., their intersection divided by the seed set size.

\begin{figure}
\centering
\begin{tabular}{cc} 
 \includegraphics[width=0.46\columnwidth, height=0.3\columnwidth]{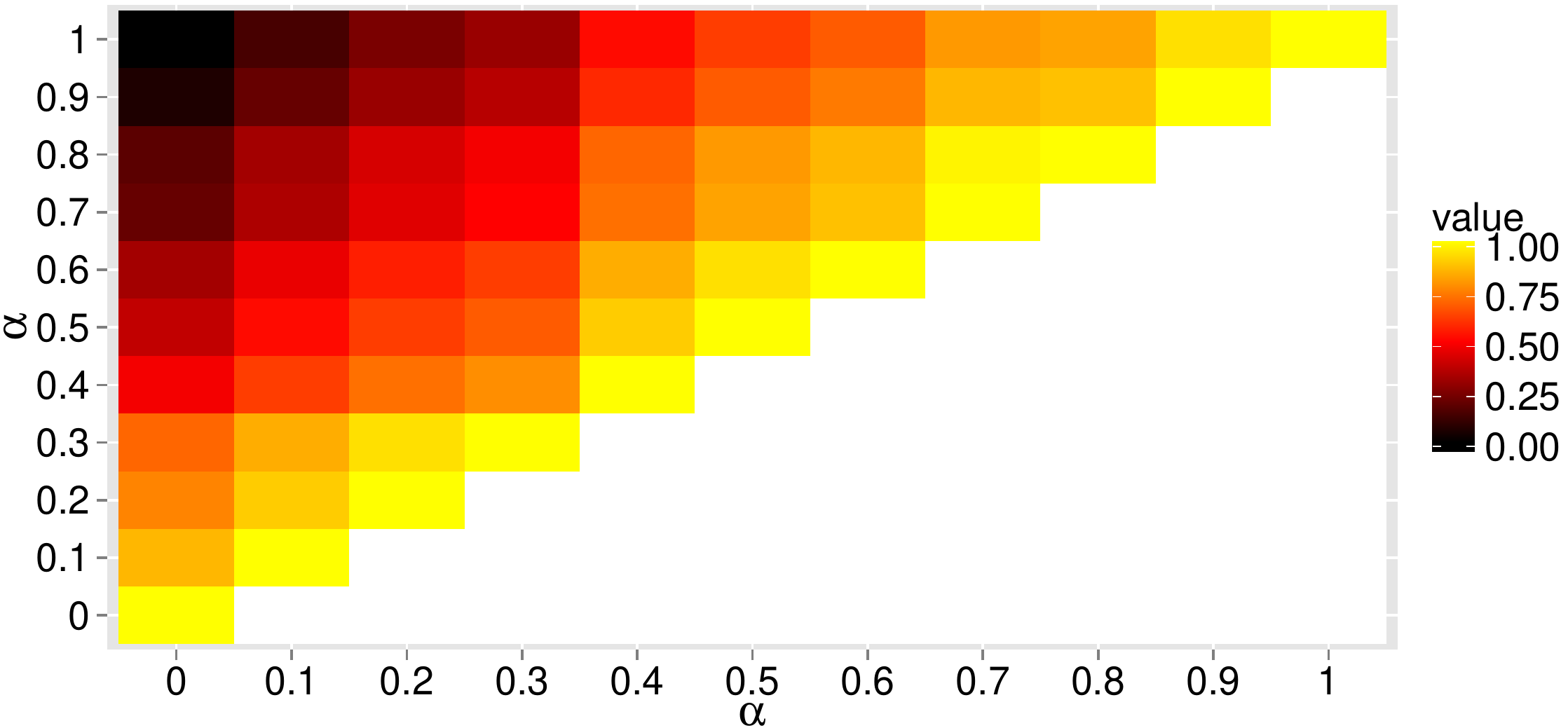} &  
 \includegraphics[width=0.46\columnwidth, height=0.3\columnwidth]{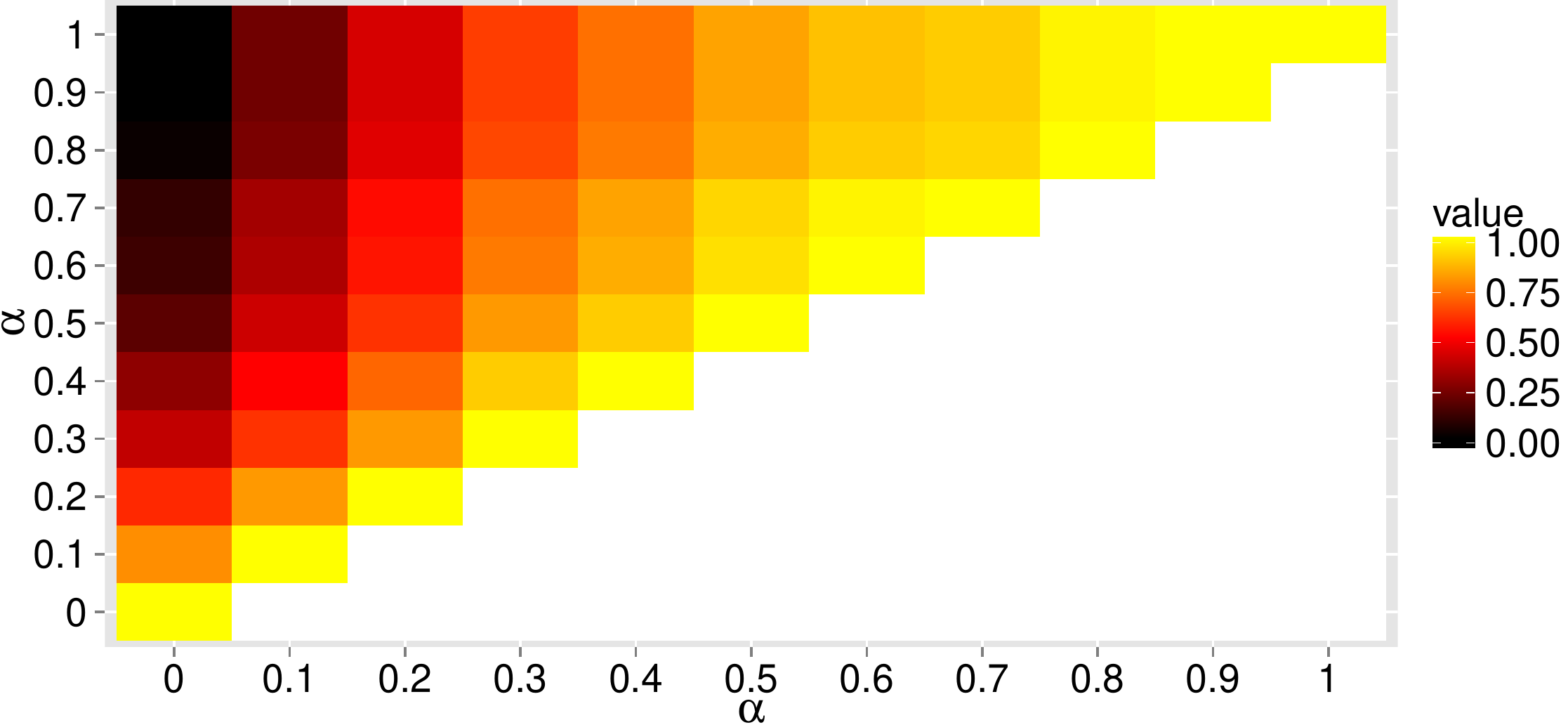}  \\
(a) G-\myalgo & (b) L-\myalgo   
\end{tabular}
\caption{Heatmaps of normalized overlap of seed sets,  for  varying $\alpha$, with $\LurkValuePerc=25\%$ and $k=50$, on 
\data{GooglePlus}.  
}  \label{fig:ssoverlap} 
\end{figure}

\vspace{1mm} 
\textit{Normalized seed set overlap.\ } 
 On \data{GooglePlus} (Fig.~\ref{fig:ssoverlap}), the normalized overlap values span over   the full range $[0.0,1.0]$, for both methods.   
In the heatmap corresponding to  G-\myalgo, an overlap above than   $50\%$ is observed    for  values of $\alpha$ in different subintervals, %the ranges $[0.0,0.3]$, $[0.4,0.6]$ and $[0.7,1.0]$.  
 while variations in the seed set are generally more uniform for L-\myalgo, whereby  the  normalized overlap increases   for higher values of $\alpha$. 
 %In fact, the largest overlaps are obtained for $\alpha$ in the ranges $[0.6,0.7]$ and $[0.8,1.0]$. 
   Also, for both methods  there is no overlap when comparing the seed set obtained for $\alpha=0$ (i.e., full contribution of diversity in the \myalgo objective function) with the seed set obtained for any       $\alpha > 0$.  
   These remarks generally hold regardless of the target set size when using L-\myalgo, while the contingencies of null overlap are more likely to occur for lower $\LurkValuePerc$ when using G-\myalgo.  
A  large  spectrum of normalized overlap values are observed on  \data{FriendFeed} as well (results not shown), %Fig.~\ref{fig:ssoverlap}(b)), 
 particularly at least   $0.25$ for G-\myalgo and   $0.4$ for L-\myalgo.  %Indeed,   variations in the seed set gradually occur with different  $\alpha$, though there is always a low, yet nonzero, overlap  with all settings (i.e., about $0.25$ for G-\myalgo and  $0.4$ for L-\myalgo). 
 Null overlap is mainly observed for low seed set size  ($k=5$ using L-\myalgo, and $k \leq 15$ using G-\myalgo). 
  By contrast, \data{Instagram-LCC} generally shows a quite higher overlap than in  the other networks (results not reported), which might be ascribed to the particular contingency of strong connectivity that  characterizes  \data{Instagram-LCC}.   
    
\begin{figure}
	\centering		
\begin{tabular}{cc} 
 \includegraphics[width=0.46\columnwidth, height=0.3\columnwidth]{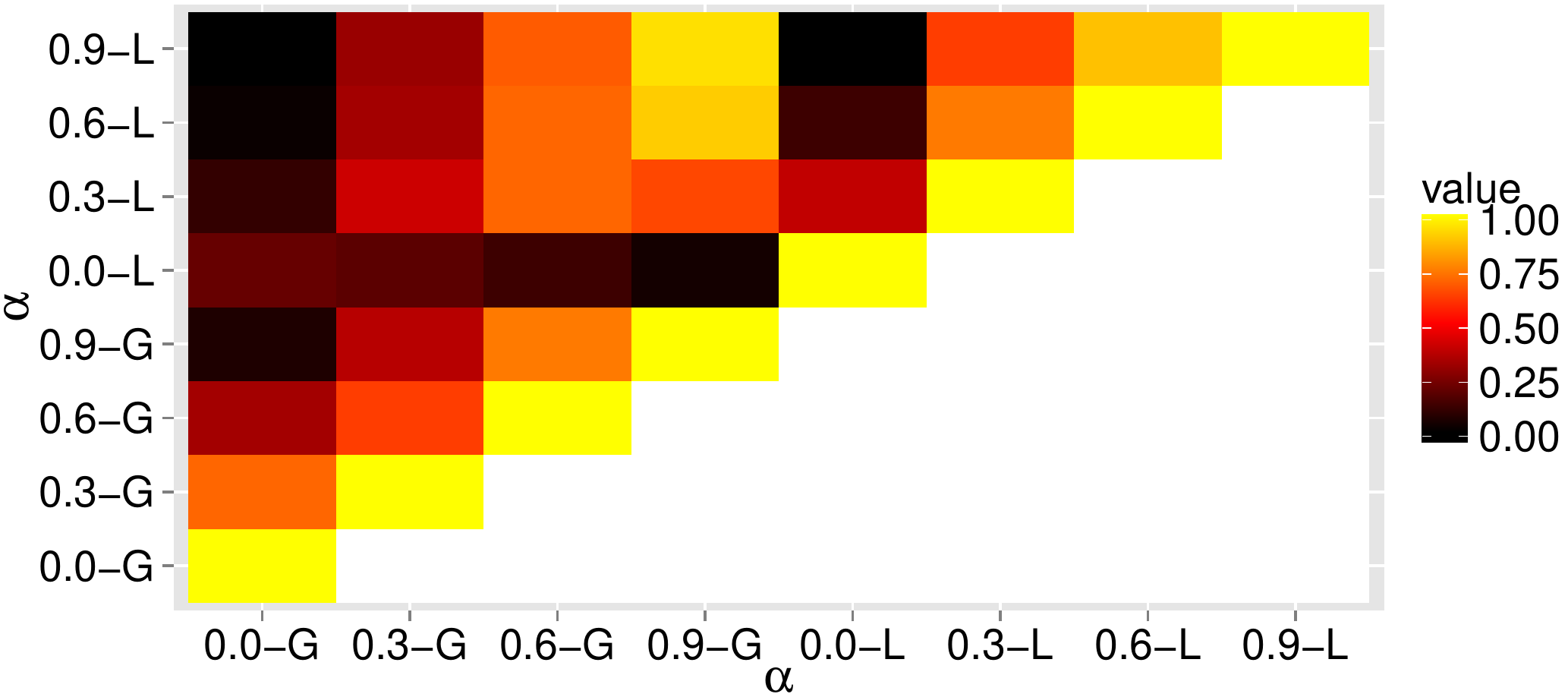}  & 
\includegraphics[width=0.46\columnwidth, height=0.3\columnwidth]{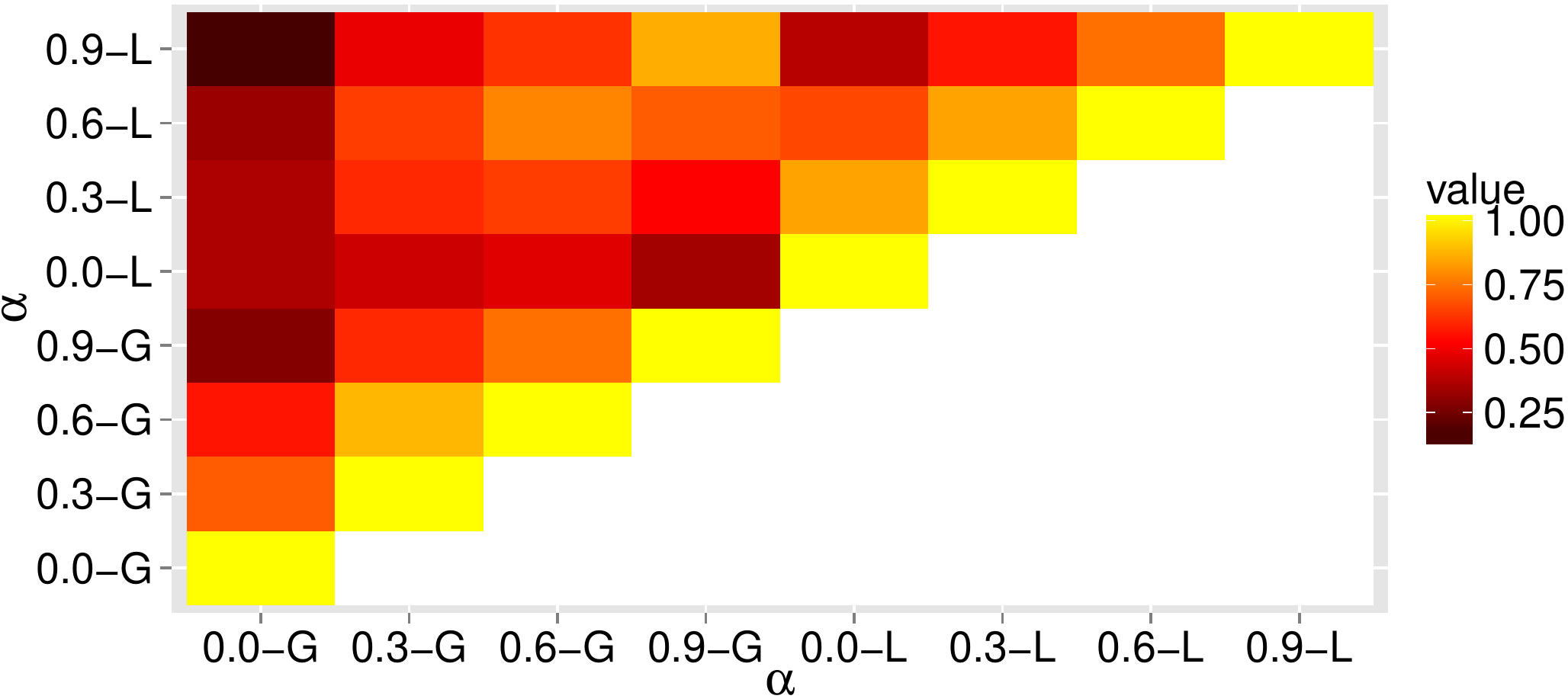} \\ 
(a) \data{GooglePlus} & (b) \data{FriendFeed}
\end{tabular}
\caption{Heatmaps of normalized overlap of seed sets between G-\myalgo and L-\myalgo, for $\alpha=\{0.0,0.3,0.6 ,0.9\}$, $\LurkValuePerc=25\%$ and $k=50$. (Suffix -L, resp. -G, denotes  a particular setting of $\alpha$ that refers to L-\myalgo, resp. G-\myalgo.)   
}
\label{fig:ssoverlap_gvsl}
 
\end{figure}

\vspace{1mm} 
\textit{Comparison between G-\myalgo and L-\myalgo seed sets.\ }    
Figure~\ref{fig:ssoverlap_gvsl} shows results on the comparison of seed sets identified by G-\myalgo and L-\myalgo, respectively, corresponding to   $\alpha=\{0.0,0.3,0.6,0.9\}$.  
On \data{GooglePlus} (Fig.~\ref{fig:ssoverlap_gvsl}(a)), the seed sets appear to be significantly different from each other for higher contributions of diversity in the objective function ($\alpha < 0.3$), while values of normalized overlap in the range $[0.5,1]$ are   observed for higher values of $\alpha$.  
Analogous observations can be drawn for \data{FriendFeed} (Fig.~\ref{fig:ssoverlap_gvsl}(b)),  yet with     lower overlap values also for values of $\alpha$ in the range $[0.6,0.9]$ (i.e., normalized overlap around $0.75$).

\vspace{1mm} 
\textit{Comparison with \algo{TIM+} and \algo{KB-TIM}.\ }
We also analyzed  the matching  between seed sets produced by \myalgo algorithms and   competing ones  (results not shown). Here we refer to the setting $\alpha=1.0$ (i.e., no diversity contribution), since \algo{TIM+} and \algo{KB-TIM} do not integrate any diversity notion in their formulations.  
 The minimum overlap of seed sets produced by \myalgo is reached against   \algo{KB-TIM} in all cases and on all datasets; in particular, with the setting $k=50, \LurkValuePerc=25\%$, 0.48  for \data{FriendFeed}, 0.46 for \data{GooglePlus}, 0.60 for \data{Instagram-LCC}.     
In general, for large $k$, the normalized overlap is within medium regimes, while it is close or equal to zero on FriendFeed. Only for $k=5$, the normalized overlap corresponds to mid-high values on GooglePlus and Instagram-LCC.   
 \myalgo with $\alpha=1$ can have relatively high overlap with \algo{TIM+} (about 0.75), especially for high $\LurkValuePerc$, on all datasets.  
 However, for lower $\LurkValuePerc$, the overlap is low (for smaller $k$) to medium (for higher $k$).

\vspace{1mm}
\textit{Discussion.\ }  
The seed set overlap analysis has revealed that accounting for diversity can yield  significant differences in the behavior of the \myalgo methods in terms of seed identification. Indeed, by varying $\alpha$ within its full regime of values  leads to a wide spectrum of  values of  normalized  seed set overlap. In particular, the changes in   overlap are more evident when varying $\alpha$ at lower regimes, thus indicating that higher contribution of diversity w.r.t. capital leads to more significantly diversified seed sets.  
  Remarkably,  the   overlap can be close to zero when comparing   two seed sets respectively obtained with   $\alpha=0$ and with $\alpha=1$,  i.e.,  completely different seed nodes can be identified when accounting for either diversity or capital only in the target IM objective function. 
%The latter is also confirmed by the analysis of variability (in terms of coefficient of variation) of the seed set overlaps conditionally to any given $\alpha$.  

The two   proposed notions of diversity turn out to be quite dissimilar to each other: indeed, the normalized overlap of seed sets yielded by  L-\myalgo and G-\myalgo, respectively, is generally below 50\%, which is   further reduced for low values of $\alpha$. 
 The local diversity notion appears to be less sensitive to $\alpha$ than   global diversity; however,   for low $\alpha$  and size of target set, L-\myalgo tends to produce more diverse seed sets than G-\myalgo, for any particular setting of $k$.

Our \myalgo  methods with $\alpha=1$  produce seed sets that have overlap with \algo{KB-TIM} ones   below $50\%$ on \data{FriendFeed} and \data{GooglePlus}, and $60\%$ on \data{Instagram} for $k=50$, $\LurkValuePerc=25\%$;  when compared to \algo{TIM+}, the seed set overlap can be relatively  higher.   

\bfseries
 
\subsubsection{Structural characteristics of seeds}

\normalfont

We analyzed  topological characteristics of the identified seeds, 
focusing on basic measures of node centrality, namely  \textit{outdegree},  \textit{betweenness},  and  \textit{coreness}. 
%, and \textit{closeness}. The latter is however left out of presentation of results, since it had   very negligible variations (order of $1.0e\mbox{-}05$) for all seeds and     methods on the various datasets.    
%
Due to space limits, we present here a summary of main findings, and refer the reader to the \textit{Appendix}  for detailed results.

One major remark that stands out is  that accounting for diversity in \myalgo methods produces the effect of choosing   seed nodes that can differ from those   that would be obtained otherwise (i.e., using only capital term in the objective function) according to  selected topological criteria.  This result, coupled with analogous considerations previously drawn about diversification in terms of set overlap, hence strengthens the significance of accounting for diversity in the targeted IM process.
 Structural characteristics tend  to be marginally affected by the setting of  $\LurkValuePerc$ when L-\myalgo is used, while the behavior with G-\myalgo is much more dependent on $\LurkValuePerc$, especially for smaller size of target set ($\LurkValuePerc=5\%$).
   Also, each of the  competing methods leads to the identification of seeds    that are   less different from each other  than \myalgo seeds being obtained   for most of the settings of $\alpha$, in terms of all the topological measures considered.

%========================================

\subsection{Evaluation of activated target nodes}
\label{sec:target-evaluation} 

\bfseries

\subsubsection{Capital}

\normalfont
We discuss  results on the expected capital   of the target users activated by a given set of seed users. 
 The estimation procedure is based on the results of $I_{MC}$ Monte Carlo simulations of the LT diffusion process, with $I_{MC}$ set to 10\,000.\footnote{A pseudo-code of the Monte Carlo based algorithm for   capital estimation can be found in the \textit{Appendix}. }   
  Note that while the identification of the seeds depends on the full \myalgo  objective function, here we     focus on the value of the  capital function $C$ only.

\begin{figure}[t!]
\centering
\begin{tabular}{cc}
\hspace{-4mm}
\includegraphics[width=0.5\columnwidth, height=0.33\columnwidth]{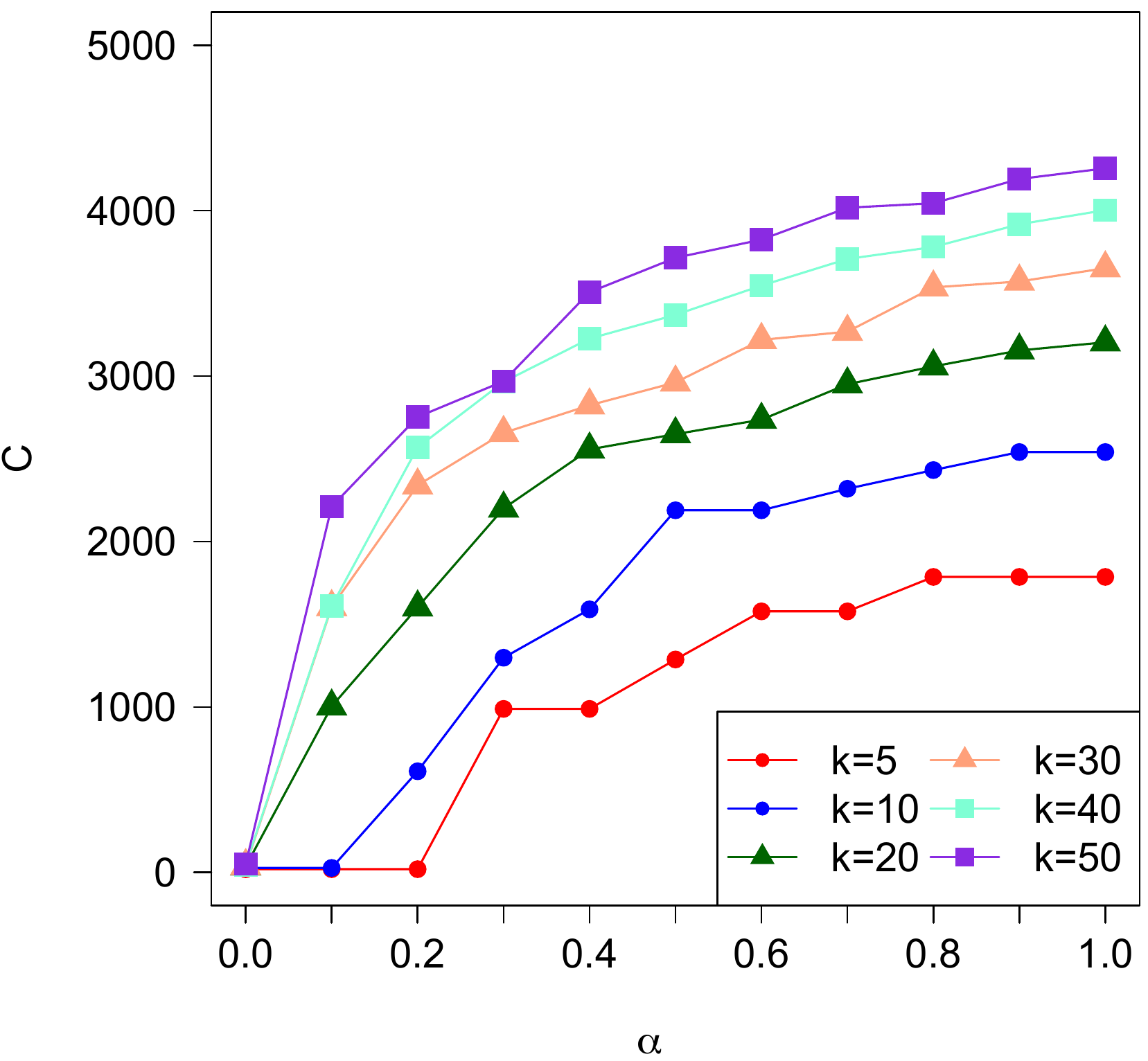} & 
\hspace{-3mm}
\includegraphics[width=0.5\columnwidth, height=0.33\columnwidth]{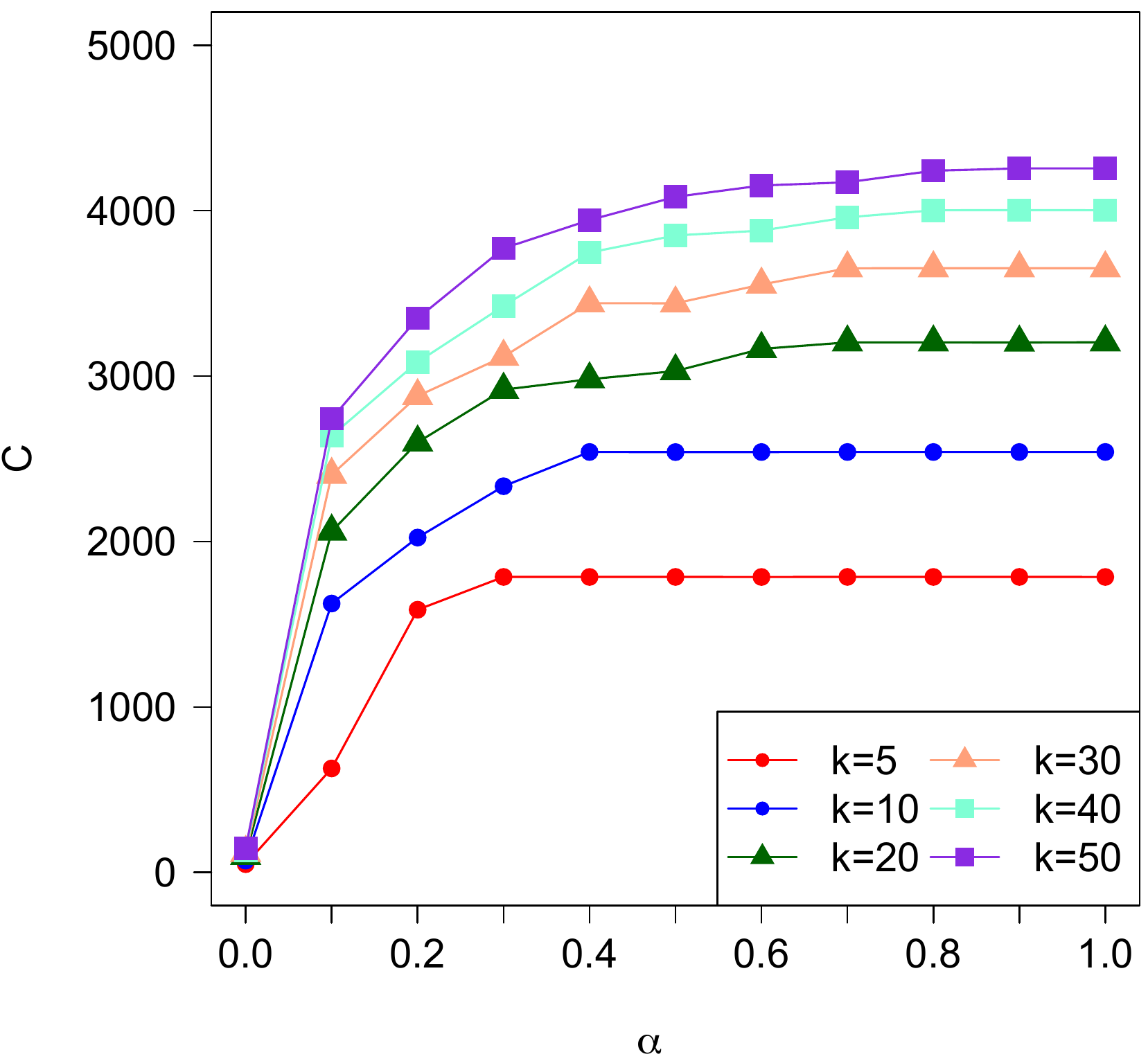} \\
\hspace{3mm} (a) G-\myalgo & \hspace{3mm} (b) L-\myalgo %&  \hspace{3mm} (c)
\end{tabular}
\vspace{-2mm}
\caption{Capital   in function of $\alpha$ and $k$, with $\LurkValuePerc$ set  to 25\%, on \data{GooglePlus}.  
}  
\label{fig:spread} 
\end{figure}

 Beyond the   expected increase in capital with $\alpha$ (which means weighting less  diversity  than  capital in the objective function), the  impact of $\alpha$ on the behavior of \myalgo algorithms is  evident, especially for $k>10$, with capital value that can vary up to three  orders of magnitude. 
 The generally upward trends  of $C$ are explained in function of both  $\alpha$ and $k$, particularly they are more rapidly increasing for mid-low $\alpha$ and  $k > 10$.  Also  on all datasets, L-\myalgo yields a higher average capital value, for every $k$, than that observed with G-\myalgo. 
  Similar overall behaviors are shown by the \myalgo algorithms for different sizes of target set. 

More in detail, 
on \data{GooglePlus} (Fig.~\ref{fig:spread}), 
when using G-\myalgo the capital value  increases rapidly, reaching around $80\%$ for $\alpha < 0.5$ and $k \geq 20$; 
%of the maximum $C$ increases for $0.0 \leq \alpha \leq 0.5$ and $k \geq 20$, reaching values around   $80\%$ of the maximum $C$,  then the increment becomes slower for $\alpha > 0.5$;  
%for lower $k$, the value of $C$  starts increase only for   $\alpha > 0.1$ (resp.  $\alpha > 0.2$) for $k=10$ (resp. $k=5$).  
%
for L-\myalgo,  we observe an even sharper increase in the value  of $C$ for small $\alpha$ (0.2), then the trends become nearly constant for higher $\alpha$.  
%For each $k$, values around the $80\%$ of the maximum $C$ can be obtained already for $\alpha=0.2$, while higher $\alpha$ corresponds to a plateau, which appears slightly flatter (showing nearly constant values of $C$) for lower $k$. 
  %
  Similar behaviors are shown on \data{FriendFeed}, though 
 the increasing trends are less monotone for $k<30$. 
 %  ttrends partially similar to the ones observed for \data{GooglePlus} characterizes   G-\myalgo (Fig.~\ref{fig:spread_f1}(b)), at least for relatively  high values of $k$ ($\geq 30$), % (i.e.,  $k=\{30,40,50\}$). while for lower values of $k$, the increasing trends are less monotone.  
  %
 On \data{Instagram-LCC}, the relatively small size and high connectivity of this  network makes capital values subject to an average variation of about $15\%$  over the full range of $\alpha$.  
%  Using  G-\myalgo, 
%%(Fig.~\ref{fig:spread_f1}(a)), 
%there is an average variation of $C$ of $15.32\%$ between the highest ($\alpha=1.0$) and lowest ($\alpha=0.0$) $C$ obtained; 
%%for each $k$, a slow decrease can be noted for $0.5 \leq \alpha \leq 1.0$, with  a plateau of  $C$ values for $\alpha < 0.5$. 
%   for L-\myalgo, %(Fig.~\ref{fig:spread_f2}(b)), 
%capital values obtained for each $k$ are even more similar when varying $\alpha$. %, with  an average variation of $0.03\%$. 
%This    is reflected  by the high overlap among the seed sets obtained when varying $\alpha$. 
% 

  \vspace{1mm}
\textit{Comparison with \algo{TIM+} and \algo{KB-TIM}.\ }   
Capital obtained by \myalgo methods is shown to be  much higher than that of competing methods, on all networks and for various   $k$ and $\LurkValuePerc$. The performance gain  is more significant on \data{FriendFeed}, with average percentage of increment from $9.85\%$ (for $\LurkValuePerc=5\%$) %, $8.82\%$ ($\LurkValuePerc=10\%$) and 
to $3.49\%$ ($\LurkValuePerc=25\%$) w.r.t. \algo{TIM+}, and even larger (from $35\%$ to $59\%$) w.r.t. \algo{KB-TIM}.  
 On the two largest networks, as the size of  target set increases, a general decreasing trend is  observed in the gap between \myalgo and \algo{TIM+} (resp. \algo{KB-TIM}) capital values, which  might be explained since a  larger target set implies that  a larger fraction of the entire node set could be reached.

\subsubsection{Target activation probabilities}
 
A further stage of evaluation was performed to understand how different settings of  $\alpha$ and  $k$ impact on the activation probability of nodes targeted by \myalgo methods.  We regard the activation probability of a node as the number of times it has been activated divided by the number of runs of   Monte Carlo simulation for the estimation of capital. 
Due to space limits, we present here  a summary of main findings concerning this evaluation, and refer the reader to   the \textit{Appendix}  for detailed results.

\begin{figure}
\centering
\begin{tabular}{cc}
 \includegraphics[width=0.45\columnwidth, height=0.3\columnwidth]{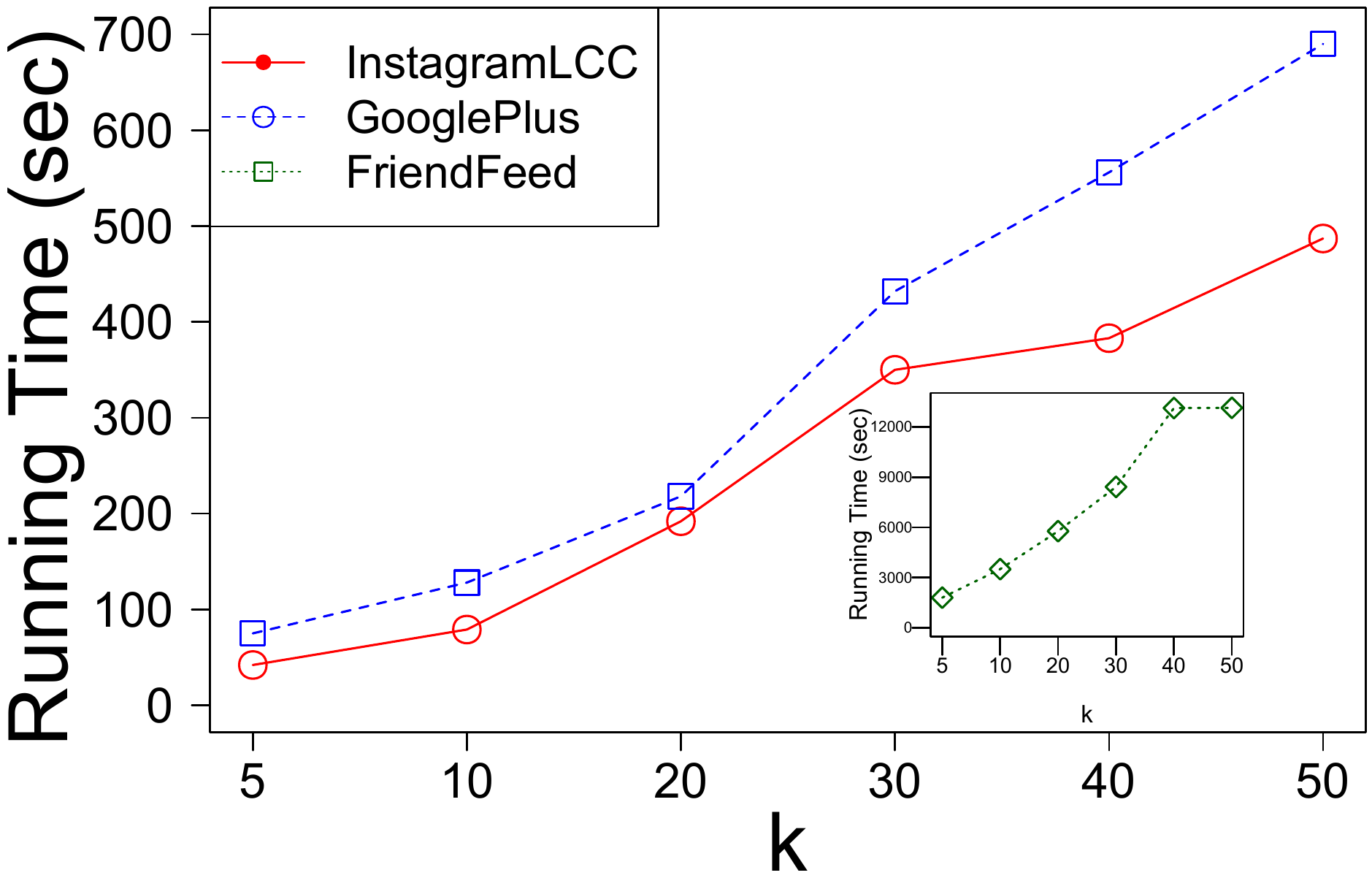} & 
\includegraphics[width=0.45\columnwidth, height=0.3\columnwidth]{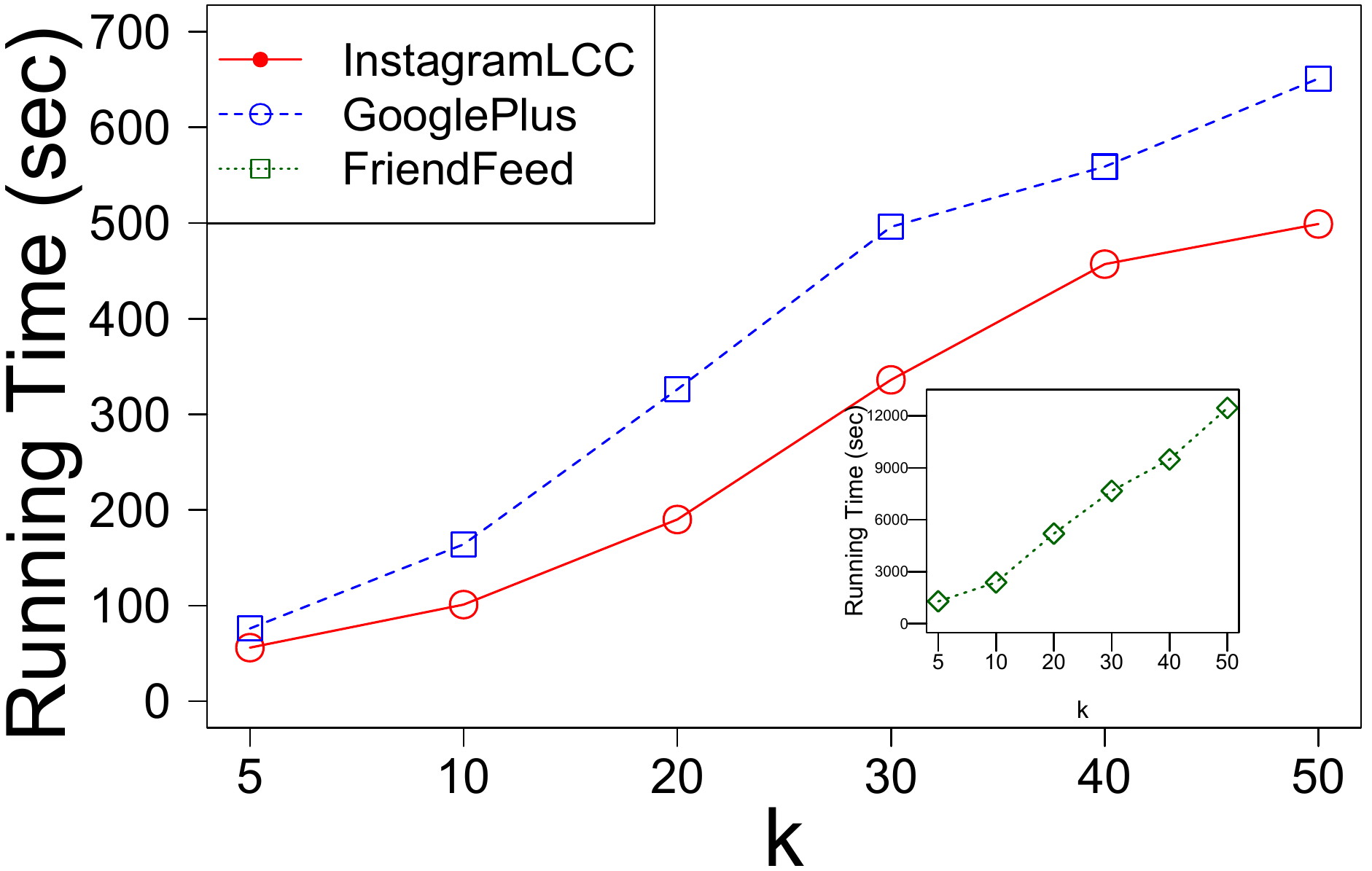} \\
(a) G-\myalgo & (b)  L-\myalgo
\end{tabular}
\caption{ 
Time performance (in seconds)   for varying $k$, with   $\alpha=0.5$ and $\LurkValuePerc=25\%$.
}  \label{fig:times_alpha05} 
\end{figure}

\begin{figure}
\centering
\begin{tabular}{cc}
 \includegraphics[width=0.45\columnwidth, height=0.3\columnwidth]{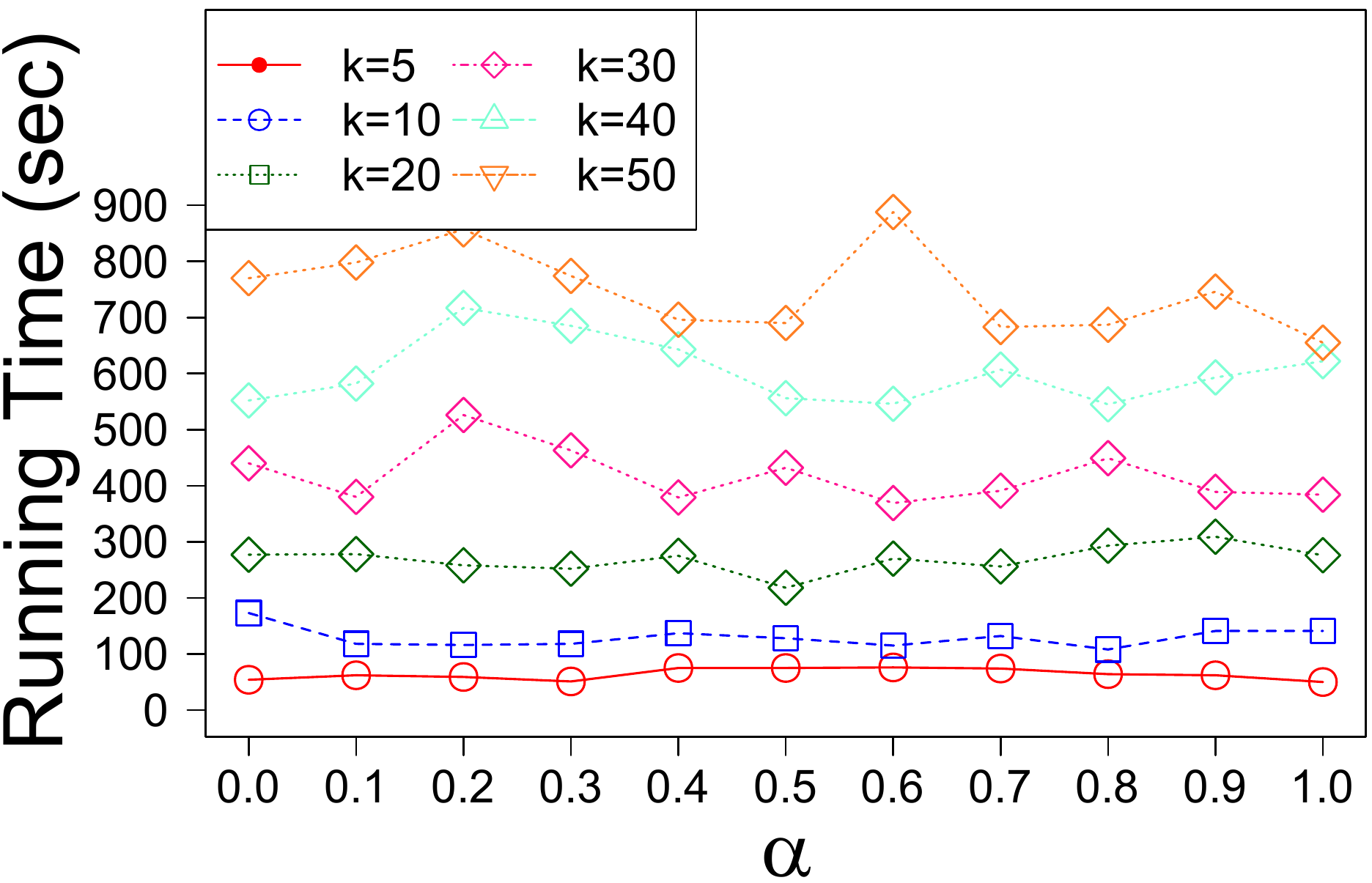} & 
\includegraphics[width=0.45\columnwidth, height=0.3\columnwidth]{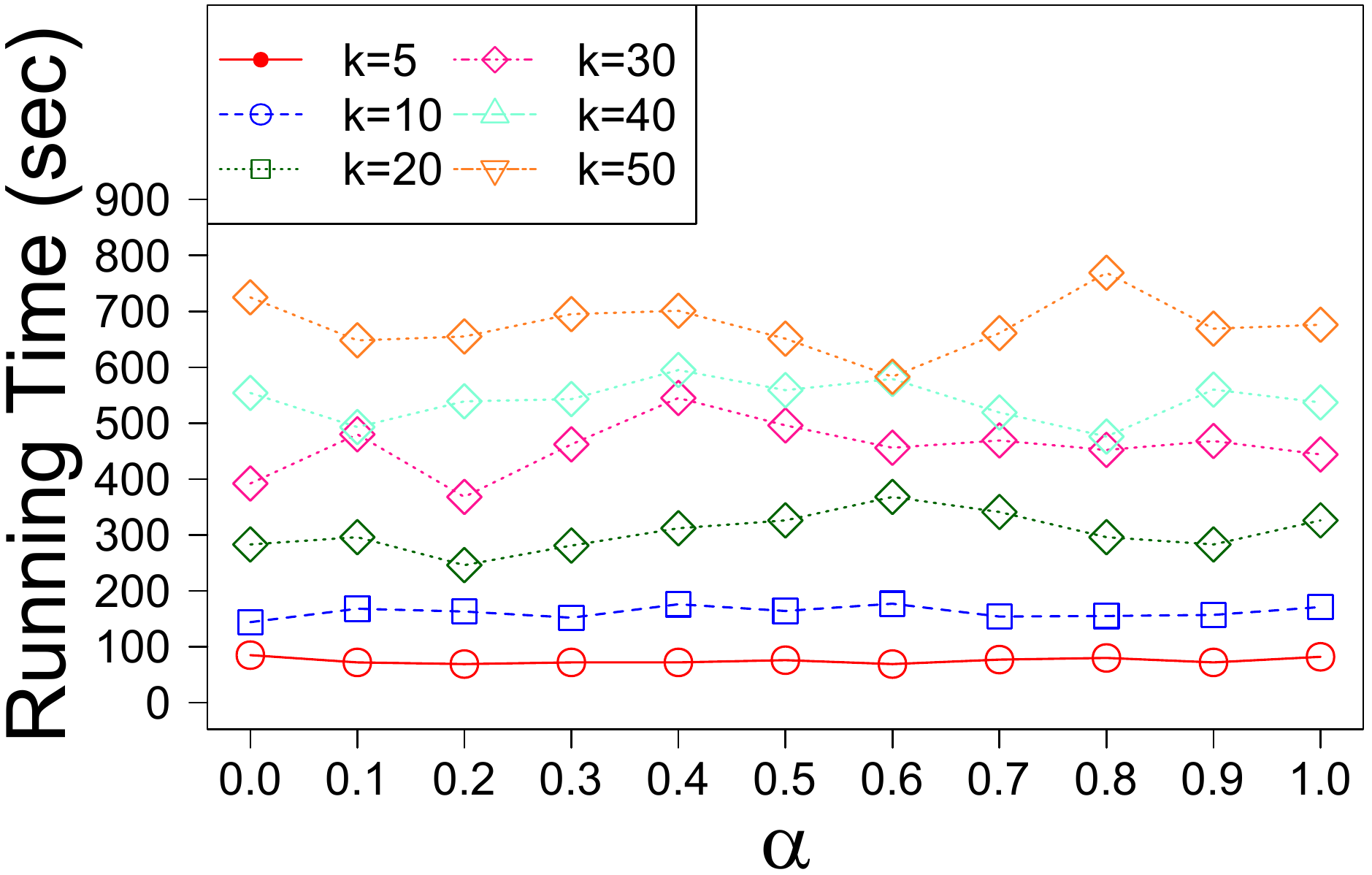} \\
(a) G-\myalgo & (b)   L-\myalgo
\end{tabular}
\caption{
Time performance (in seconds)   for varying $k$ and $\alpha$, with $\LurkValuePerc=25\%$, on 
\data{GooglePlus}. }  \label{fig:times_gplus} 
\end{figure}

For both \myalgo algorithms,   the activation probability follows a  non-decreasing trend as $\alpha$ increases. The likelihood of obtaining high  activation probability grows with $\alpha$, i.e., the amount of target nodes that have high probability of activation increases by increasing $\alpha$. 
The analysis of density distributions also puts in evidence that the density peak corresponding to   low activation probability is higher  for lower values of $\alpha$, whereas the density corresponding to high activation probability increases   for higher values of $\alpha$.  
Nevertheless, on the two largest datasets, we also observe that choosing a relatively large $k$ leads    a significant portion of target nodes to have  mid-high activation probabilities already for $\alpha=0.1$, thus  suggesting  that  target nodes can be activated even with strongly unbalancing capital with  diversity.   
By contrast, when choosing a small $k$, little changes in the value of $\alpha$ can significantly impact on the amount of more likely activated target nodes.

\subsection{Efficiency analysis}
\label{sec:efficiency} 

Figure~\ref{fig:times_alpha05} reports on time performance of  G-\myalgo and L-\myalgo on the various networks, for  $5 \leq k \leq 50$ and $\alpha=0.5$. The execution time of both methods  shows a roughly  linear increase with $k$, on all networks. (Note that the \data{FriendFeed} time series are shown in the figure insets, as they correspond to orders of magnitude higher than for the other networks, due to the larger size of \data{FriendFeed}).     
 Also,  G-\myalgo turns out to be slightly   faster than L-\myalgo, which might be ascribed to the fewer computations of node diversity needed by G-\myalgo w.r.t. L-\myalgo.  

As shown in Fig.~\ref{fig:times_gplus} for \data{GooglePlus} in particular (though similar behaviors   also characterize the other networks), varying $\alpha$ with fixed $k$ does not significantly impact on the time performance of both \myalgo methods. This would indicate that, for a given seed set size, the methods' effort in computing the global/local diversity as well as the capital contributions in the objective function is not greatly affected by the value of $\alpha$.
 Analogous remarks are also  drawn for the  other settings  of $\LurkValuePerc$.

As regards \algo{TIM+} and \algo{KB-TIM}  (results not shown), 
it comes without surprise that both  outperform     \myalgo methods. For instance,  on our largest network  (i.e.,  \data{FriendFeed}),  
  the execution times of \algo{TIM+} (with $\epsilon=0.1$)  are between  $6.3$ ($k=50$) and $11.9$ ($k=5$) seconds --- note that the increase in runtime   by decreasing $k$ is in line with the  theoretical and experimental results shown in~\cite{TangXS14};  
 yet, \algo{KB-TIM} execution times are always below 0.7 seconds regardless of $\LurkValuePerc$, which might  also  depend on the extremely low number of queries and keywords used by \algo{KB-TIM} in our setting.

%=======================================

\section{RIS-based formulation of \myalgo}
\label{sec:RIS-DTIM}

The gap in efficiency shown by our \myalgo algorithms w.r.t. the competing RIS-based ones, prompted us to investigate how to adapt RIS-based approximations to our diversity-sensitive,  targeted IM problem.   

\subsection{Revisiting RIS theory for the DTIM problem} 
The reverse influence sampling   (RIS)~\cite{BorgsBCL14} relies on the concept of \textit{reverse reachable} (RR) set.  
Intuitively, the random RR set generated from $\G$ for a randomly selected user $u$ (i.e., the \textit{root} of the RR set) contains the users who could influence $u$. By generating many random RR sets on different random users, if a user has high potential to influence other users, then s/he will likey appear in those random RR sets. Thus, if a seed set   covers most of the RR sets, it will likely maximize the expected spread. Upon this principle, Corollary 1 in~\cite{TangXS14} states that 
$\mathbb{E}[F(S) /\theta] =  \mathbb{E}[\mu(S)] / n$, 
 where $F(S)$ denotes the number of RR sets covered by the node set $S$,  $\mu(S)$ is the spread of $S$, $\theta$ is the number of RR-sets, and $n=|\V|$.\footnote{For the sake of simplicity of notation, we omit to declare random variable symbols when using the expected value operator $\mathbb{E}[\cdot]$.} 

 In our setting, every node $v \in \V$ is selected as root of an RR-set with probability proportional to its status as target node, i.e., $p(v) = \frac{\ell'(v)}{L_{TS}}$, 
where  
$\ell'(v) = \ell(v)$ if $v \in TS$, zero otherwise, and 
$L_{TS} = \sum_{v \in TS} \ell'(v)$.   
In the following, we state that for any set of nodes $S$, the expected value of the fraction of RR sets covered by $S$ is equal to the normalized expected value of the capital associated with the activation of target nodes due to $S$ as seed set. 
\begin{proposition}
\begin{equation}
\mathbb{E}\left[\frac{F(S)}{\theta}\right] = \frac{\mathbb{E}[C(\mu(S))]}{L_{TS}}
\end{equation}
\end{proposition}

The proofs of all propositions in this section are reported in the \textit{Appendix}.

\textbf{Estimation of the number of RR  sets.\ } 
In~\cite{TangXS14}, 
 the objective  is to find a number $\theta$ of RR sets such that 
  $\theta \geq \lambda / OPT$, 
where $OPT$ denotes the maximum expected spread  of any size-$k$ seed set, and $\lambda$ is determined as a function of the size of the graph, $k$ and the approximation factor $\epsilon$.  
 Since $OPT$ is unknown, a lower bound for it must be computed.    
 
Following from   Lemma 4 in~\cite{TangXS14}, 
% (and Claim 3.3 in~\cite{BorgsBCL12}),  
   the expected spread of a randomly sampled node can be expressed in terms of  the expected value $EPT$ of the number of edges pointing to nodes in an RR set (\textit{width}), such that  $EPT  \leq \frac{m}{n} OPT$ holds, with $m=|\E|$.  
We revise this result to state   that the expected value of the width of an RR set can be an accurate estimator of the capital associated with any  node when randomly selected  as a seed.  

\begin{proposition}
\begin{equation}
(L_{TS}/m) \ EPT = \mathbb{E}[C(\{v\})] \ \leq \ OPT 
\end{equation}
\end{proposition}

To avoid unnecessarily large values of $\theta$, it is desired to 
find a lower error bound in terms of the mean of the expected spread of a set $S$ (over the randomness in $S$ and the influence propagation process), denoted as $KPT$, such that  
$(n/m) EPT  \leq KPT \leq OPT$ holds. 
 To this aim, Lemma 5 in~\cite{TangXS14} estimates $KPT$ as  $KPT = n \mathbb{E}_{\drawn{R}}[\kappa(R)]$, taking the average over a set of random RR sets $R$ from the possible world  $\mathcal{R}$, where  $\kappa(R) = 1-(1-\frac{w(R)}{m})^k$ and $w(R)$ is the width of $R$. 
  Again, we revise  this result in our setting:
 
\begin{proposition}
Given a random RR set $R$, and denoted with $TS_R$ the set of target nodes in $R$, it  holds that 
\begin{equation}\label{eq:kappa}
\widehat{\kappa}(R) = \left[1 - \left( 1 - \frac{\vert TS_R \vert}{m}\right)^k \right]  \frac{\sum_{v \in R} \ell'(v)}{|TS_R|}.
\end{equation}
Therefore,  
\begin{equation}\label{eq:KPT}
KPT = n \mathbb{E}_{\drawn{R}}[\widehat{\kappa}(R)].
\end{equation}
\end{proposition}

\vspace{-3mm}
\subsection{Developing RIS-based \myalgo algorithms}  
We sketch here a reformulation of \myalgo  based on the   RIS approach.  To this purpose,  we start from \algo{TIM+} and adapt it to our DTIM problem.  This requires  four key modifications: 

 %
% \begin{description}
%\item[M1]
\noindent 
-~\textbf{M1}: Revise the sampling   over the nodes in $\G$.  

%\item[M2]
\noindent 
-~\textbf{M2}: Modify  the $KPT$ estimation procedure (i.e., \algo{TIM+}'s Algorithm 2).
 
%\item[M3]
\noindent 
-~\textbf{M3}: 
 Modify  the refinement of $KPT$ to obtain a potentially tighter lower-bound of $OPT$  (i.e., \algo{TIM+}'s Algorithm 3).  

%\item[M4]
\noindent 
-~\textbf{M4}:
Modify the node selection procedure (i.e., \algo{TIM+}'s Algorithm 1) for determining a size-$k$ seed set.  
%\end{description} 

 In the following, we elaborate on each of the above  points, which overall constitute a 4-stage workflow for the development of RIS-based \myalgo methods. 

\vspace{1mm}
\textbf{Sampling (M1).\ } 
As previously discussed, we define a probability distribution over the nodes in $\G$ such that the probability mass for each node $v$ is  non-zero and proportional to the value of $\ell(v)$ if $v \in TS$, and zero otherwise.  
 
\vspace{1mm}
\textbf{Parameter estimation (M2).\ } 
The  RR sets must be generated in such a way that   the roots are sampled from  the above defined probability distribution (i.e., the root of any RR set is a target node). Moreover, the   original function  $\kappa$ is replaced with  Eq.~(\ref{eq:kappa}).

\vspace{1mm}
\textbf{Parameter refinement (M3).\ }
 Starting from the set $\mathcal{R}'$ of all RR sets produced to estimate $KPT$,  the size-$k$ seed set $S'$ is generated by selecting those nodes that, while covering RR sets in $\mathcal{R}'$, maximize the capital  w.r.t. $\mathcal{R}'$. 
 More specifically,  each RR set in $\mathcal{R}'$  is associated with a score equal to the value of $\ell$ of its root node, and every node is associated with a score equal to the sum of RR-set-scores  the node belongs to.   
In the main loop, at each of the $k$ iterations, the node $v$ with maximum score is identified and added to $S'$, all RR sets covered by $v$ are removed from $\mathcal{R}'$, and the node scores are recomputed.  
 
Once computed $S'$, a new set  $\mathcal{R}''$ of RR sets is generated and used to derive  $\bar{\mathcal{F}}$, which contains the root nodes of all RR sets in $\mathcal{R}''$,   and $\mathcal{F}$, which is the subset of  root nodes of   RR sets that have non-empty overlap with $S'$. Next, we compute the fraction of capital associated with 
   $\mathcal{F}$, i.e., $f=\sum_{v \in \mathcal{F}} \ell'(v) / \sum_{v \in \bar{\mathcal{F}}} \ell'(v)$. Quantity $f$ is finally exploited to derive the new lower-bound analogously to the last two instructions in \algo{TIM+}'s Algorithm 3.

\vspace{1mm}
\textbf{Node selection (M4).\ } 
Let us first consider the case in which the diversity function is discarded from   the \myalgo objective function. The node selection procedure turns out to be   analogous to the first step described in \textbf{M3},  where the number $\theta$ of RR sets to generate is computed based on the refined  $KPT$.       
In the general case, the node selection procedure needs to also include  the global/local diversity values when scoring the nodes w.r.t. the RR sets they cover.  We provide here an informal description of the essential steps to perform. 

%To this purpose, we propose an index for every node appearing in an RR set 
Let $\mathcal{R}_v$ denote the set of RR sets rooted in $v$. Upon this, we build a tree index $\Lambda(v)$, with root $v$, by aggregating   all live-edge paths reaching $v$. Note that the tree is constructed in a backward fashion; also,   every node other than $v$ has at most one incoming edge, and it could appear in many paths and at different distance from $v$. 

Let us first consider the global diversity of a node in $\mathcal{R}_v$.   The boundary set of $\Lambda(v)$ is the multiset of all leaf nodes in the tree.  The \textit{RR-global-diversity} of a node $u$ in $\Lambda(v)$  is determined as the mean of  its global diversity values by possibly considering the multiple occurrences of $u$ as leaf. By averaging the RR- global-diversity values over all trees in which node $u$ appears, we compute the \textit{total RR-global-diversity} of $u$.   
To compute the \textit{RR-local-diversity}, we need to consider each \textit{level} of $\Lambda(v)$ at a time, and hence the boundary set of each subtree resulting from truncating $\Lambda(v)$  at a given distance from $v$.  
We then average the scores of a node $u$ over all trees in which $u$ appears to have the \textit{total RR-local-diversity} of $u$. 

Finally, the total   RR-diversity    of a node is    linearly combined with the corresponding capital score, in order to drive the search for the node with  maximum $DIC$   to be identified at the $k$-th iteration of  the node selection procedure.

%=======================================
\vspace{-1mm}
\section{Conclusions}
We presented a novel targeted IM problem in which the objective function  is defined in terms of  spreading capability and topology-based diversity  w.r.t. the target users. 
We proved that the proposed objective function is monotone and submodular,  and developed two alternative algorithms,  L-\myalgo and G-\myalgo, to solve the problem under consideration.   
Significance and effectiveness of our algorithms have been assessed, also in comparison with baselines and   state-of-the-art   IM methods, using publicly available, real-world  network graphs.  We have also provided theoretical foundations to develop RIS-based \myalgo methods. 
 
As future research, it would be interesting to investigate diversity notions  based on   boundary spanning principles that might rely on community  detection solutions; other  opportunities in this regard would certainly come from the integration of side information representing user profiles.  
 We also plan to evaluate the RIS-\myalgo method, which promises  to overcome the efficiency issues of the current \myalgo methods.  
 Finally, it is worth noting that our proposed approach is   versatile,  
 as it can easily be generalized  not only to other cases of user engagement (for example, introducing newcomers to a community), but also to any other application  of targeted IM in which accounting for diversity of users based on their relationships/interactions with other users, is beneficial to the enrichment  of influence propagation outcome with effects of varied  social capital.  In this respect,   we can  envisage further  developments from various perspectives, including human-computer interaction, marketing, and psychology.

\appendices

\section{Proofs} 

\begin{proposition}
	\em The capital function   $C$  (cf. Eq.~(2) in the main paper)  is monotone and submodular under the LT model.
\end{proposition}
\noindent
\textit{Proof sketch.} 
By exploiting the equivalence between   LT   and the live-edge model shown in~\cite{KempeKT03}, for any set $A \subseteq \V$ we can express the expected capital of the final active set $\ActiveSet{A}$ in terms of reachability under the live-edge graph:
\begin{equation}
\olf{\ActiveSet{A}} = \sum \limits_{\forall X} \Pr(X)\olf{R^X(A)}
\end{equation}
where $\Pr(X)$ is the probability that a hypothetical live-edge graph $X$ is selected from all possible live-edge graphs, and $R^X(A)$ is the set of nodes that are reachable in $X$ from $A$. 
Since for all $v \in \V$, $\ell(v)$ is a non-negative value, $\olf{R^X(A)}$ is clearly monotone and submodular. Thus, the expected capital under LT is a non-negative linear combination of monotone submodular functions, and hence it is monotone and submodular, which concludes the proof. ~\hfill$\Box$

\vspace{2mm}
\begin{proposition}
	\em The diversity function   $D$   (cf. Eq.~(3) in the main paper)  is monotone and submodular.
\end{proposition}
\noindent
\textit{Proof sketch.}  
As in both the formulations of topology-driven diversity provided above, $div_t(v)$ returns a non-negative value for all $v \in \V$, $\SetDiv{\cdot}$ is clearly monotone. To see that is also submodular, we have to verify that, $\forall S,T \subseteq \V$ with $S\subseteq T$ and $\forall v \in \V \setminus T$, $\SetDiv{S \cup \{v\}}-\SetDiv{S} \geq \SetDiv{T \cup \{v\}}-\SetDiv{T}$. For definition of diversity, the above expression can be written as $\SetDiv{S} + \SetDiv{\{v\}}-\SetDiv{S} \geq \SetDiv{T} - \SetDiv{\{v\}}-\SetDiv{T}$, hence it is nondecreasing submodular, which concludes the proof. ~\hfill$\Box$

\begin{proposition}
\begin{equation}
\mathbb{E}\left[\frac{F(S)}{\theta}\right] = \frac{\mathbb{E}[C(\mu(S))]}{L_{TS}}
\end{equation}
\end{proposition}

\noindent 
\textit{Proof sketch.\ }
Following notations used in~\cite{LiZT15}, let $p(S \to v)$ denote the probability that $v$ is activated by seed set $S$. Thus, the expected capital associated with $S$    can   be expressed as: % the sum of $p(S \rightarrow v)$ over all nodes in $\V$.  
\begin{equation}
 	\mathbb{E}[C(\mu(S))] =  \sum_{v \in \V} p(S \to v)   \ell'(v)
\end{equation}

By Lemma 2 in~\cite{TangXS14},  the probability that a set $S$ overlaps with an RR set $R_v$ rooted in a node $v$ is equal to the probability that $S$, when used as a seed set, can activate $v$, i.e., 
\begin{equation}
p(S \to v) = \Pr[S \cap R_v \neq \emptyset].
\end{equation}
 Therefore, it holds that
 \begin{equation}
\begin{split}
\mathbb{E}[F(S)/\theta] & = \sum_{v \in \V} p(v)  \Pr[S \cap R_v \neq \emptyset]  \\ 
 & = \sum_{v \in \V} \frac{\ell'(v)}{L_{TS}} p(S \to v)  \\
 & = \frac{\mathbb{E}[C(\mu(S))]}{L_{TS}}
\end{split}
\end{equation}
 ~\hfill$\Box$

\begin{proposition}
\begin{equation}
\frac{L_{TS}}{m} EPT = \mathbb{E}[C(\{v\})] \leq OPT 
\end{equation}
\end{proposition}

\noindent
\textit{Proof sketch.\ }  
Let $w(R_u)$ denote the width of an RR set rooted in node $u$, and 
 $R_u \sim \mathcal{R}$ denote  an RR set rooted in node $u$ sampled  from the distribution of all  RR sets.   
We have that:

\begin{equation}
\begin{split}
EPT & = \sum_{u \in \V} \frac{\ell'(u)}{L_{TS}} \mathbb{E}_{R_u \sim \mathcal{R}}[w(R_u)] \\
    &= \frac{1}{L_{TS}} \sum_{u \in \V} \ell'(u) \sum_{R_u \sim \mathcal{R}} \Pr[R_u] \sum_{v \in \V} \Pr[v \to u | R_u] \\
        &= \frac{1}{L_{TS}} \sum_{R_u \sim \mathcal{R}} \Pr[R_u] \sum_{ (v,u) \in \E} \ell'(u) \Pr[v \to u | R_u]  \\
    &= \frac{1}{L_{TS}} \sum_{(v,u) \in \E} \mathbb{E}[C(\mu(\{v\}))] \\
    &= \frac{m}{L_{TS}} \mathbb{E}[C(\mu(\{ v\}))] \\
  %
%EPT & = \sum_{u \in \V} \frac{\ell'(u)}{L_{TS}} \mathbb{E}_{\drawn{g}}[m_{g^{\mathrm{T}}}(u)] \\
%    &= \sum_{u \in \V} \frac{\ell'(u)}{L_{TS}} \mathbb{E}_{\drawn{g}}[\vert \{ v \vert v \in C_{g^{\mathrm{T}}}(u) \vert] \\
%    &= \frac{1}{L_{TS}} \sum_{u \in V} \ell'(u) \sum_{\drawn{g}} p(g) \sum_{v \in \V} p_{g^{\mathrm{T}}}(u \to v) \\
%    &= \frac{1}{L_{TS}} \sum_{\drawn{g}} p(g) \sum_{u \in V} \ell'(u) \sum_{v \in \V} p_{g}(v \to u) \\
%    &= \frac{1}{L_{TS}} \sum_{\drawn{g}} p(g) \sum_{ (v,u) \in \E} \ell'(u) p_{g}(v \to u)  \\
%    &= \frac{1}{L_{TS}} \sum_{(v,u) \in \E} \mathbb{E}_{\mathcal{G}}[C(\mu(\{v\}))] \\
%    &= \frac{m}{L_{TS}} \mathbb{E}_{\mathcal{G}}[C(\mu(\{ v\}))] \\
\end{split}
\end{equation}
~\hfill$\Box$

\begin{proposition}
Given a random RR set $R$, and denoted with $TS_R$ the set of target nodes in $R$, it  holds that 
\begin{equation}\label{eq:kappa}
\widehat{\kappa}(R) = \left[1 - \left( 1 - \frac{\vert TS_R \vert}{m}\right)^k \right]  \frac{\sum_{v \in R} \ell'(v)}{|TS_R|}.
\end{equation}
Therefore,  
\begin{equation}\label{eq:KPT}
KPT = n \mathbb{E}_{\drawn{R}}[\widehat{\kappa}(R)].
\end{equation}
\end{proposition}

\noindent
\textit{Proof sketch.\ }
Given an RR set $R$, let us denote with $A$ the event of selecting an edge in $\G$  that points to a target node, and with $B$ the event of selecting an edge in $\G$ that  points to a node in $R$. The probability of these events are $\Pr[A] =   |TS| / m$ and $\Pr[B]= w(R)/m$. The conditional probability of $A$ given $B$ is equal to 
$\Pr[A | B] = |TS_R|/w(R)$, where symbol  $TS_R$ is used to denote the set of    target nodes in $R$. Thus, the probability of selecting an edge pointing to a target node contained in $R$ is $\Pr[A \cap B] = \Pr[A | B] \Pr[B] = \frac{\vert TS_R \vert}{w(R)} \cdot \frac{w(R)}{m}  = \frac{\vert TS_R \vert}{m}$.     
 Given $k$ randomly selected edges, the probability that at least one of these points to a target node  in $R$ is $
 \widehat{\kappa}(R) = 1 - \left( 1 - \frac{\vert TS_R \vert}{m}\right)^k.$             
 This quantity is finally smoothed by $\frac{\sum_{v \in R} \ell'(v)}{|TS_R|}$, i.e., the average $\ell'$ value over the target nodes belonging to $R$.
 ~\hfill$\Box$

\section{Monte Carlo Estimation of Capital}
Algorithm~\ref{alg:MCEstimation} sketches the Monte Carlo  procedure of simulation of the LT diffusion process for estimating the capital associated with the target nodes that are finally activated by  a given seed set.

  \begin{algorithm} [t!]
\caption{Monte Carlo Estimation of Capital}\label{alg:MCEstimation}
\begin{small} 
\begin{algorithmic}[1]
\REQUIRE A graph $\G=(\V, \E, b, \ell)$, a target selection threshold $\LurkValue \in [0,1]$, seed set $S$, number of Monte Carlo iterations $I_{MC}$
\ENSURE Capital $\olf{\ActiveSet{S}}$ 
\STATE $curr\_C \leftarrow$ 0
\FOR{$u \in S$}
\STATE  $u.isActive \leftarrow$ \textbf{true} 
\ENDFOR
\FOR{$j=1 \ to \ I_{MC}$} 
\FOR{$v \in \V \setminus S$}
\STATE  $v.isActive \leftarrow$ \textbf{false} 
\STATE  $v.receivedInf \leftarrow 0$  
\STATE  $\teta{v} \leftarrow -1$  
\ENDFOR
\STATE $temp \leftarrow S$
\WHILE{$temp \neq \emptyset$}  
\STATE $u \leftarrow temp.remove(0)$
\FOR{$v \in \oNeighbor{u}  \ \wedge  \  v.isActive =\textbf{false}$}
\STATE  $v.receivedInf  \leftarrow v.receivedInf  + b(u,v)$
\STATE  \textbf{if} $\teta{v} = -1$ \textbf{then} \COMMENT{node $v$ has been reached for the first time during the current simulation}
\STATE  \qquad choose $\teta{v} \sim U\!\left[0,1\right]$ 
\STATE  \textbf{if} $v.receivedInf  \geq \teta{v}$ \textbf{then}
\STATE  \qquad $v.isActive \leftarrow \textbf{true}$
\STATE  \qquad $temp \leftarrow temp \cup \{v\}$
\STATE  \qquad \textbf{if} $\ell(u)\geq \LurkValue$ \textbf{then}
\STATE  \qquad \qquad   $curr\_C  \leftarrow curr\_C + \ell(v)$  
\ENDFOR
\ENDWHILE
\ENDFOR
\RETURN $curr\_C / I_{MC}$  
\end{algorithmic}
\end{small}
\end{algorithm}

%==============================================

\section{Note on LurkerRank for targeted IM}
 \algo{LurkerRank} does not require any information other than the network topology, in which node (user) relationships are asymmetric and indicate that one node receives information from another one. The actual meaning of  ``received information'' can depend on the specific context of network evaluation; in general, it refers to either a social graph (i.e., $(u,v) \in \E$ means that $v$ \textit{is follower} of $u$)  or an interaction graph (e.g., $v$ \textit{likes} or \textit{comments} $u$'s posts);   \algo{LurkerRank} has been indeed evaluated on both scenarios~\cite{SNAM14,SNAM15}. 
 
For purposes of targeted IM,  both social and interaction relations can be seen as indicator of user influence. However, we note that influence is normally produced regardless of actual, visible interaction between two users. Yet, information on  interaction data might be significantly sparse in real SNs, causing a flawed  setting for an IM task. 
Without any loss of generality, in the main paper, we have  assumed that the graph $\G_0$ (on which \algo{LurkerRank} is applied) is a followship graph.

%==============================================

\section{Additional Results}

\subsection{Structural characteristics of seeds}
In this section  we report details concerning   analysis of structural characteristics of the detected seeds (cf. Section 6.1.2 in the main paper)

 \begin{figure*}[t!]
\centering
\begin{tabular}{ccc}
\hspace{-5mm}
\includegraphics[width=0.3\textwidth]{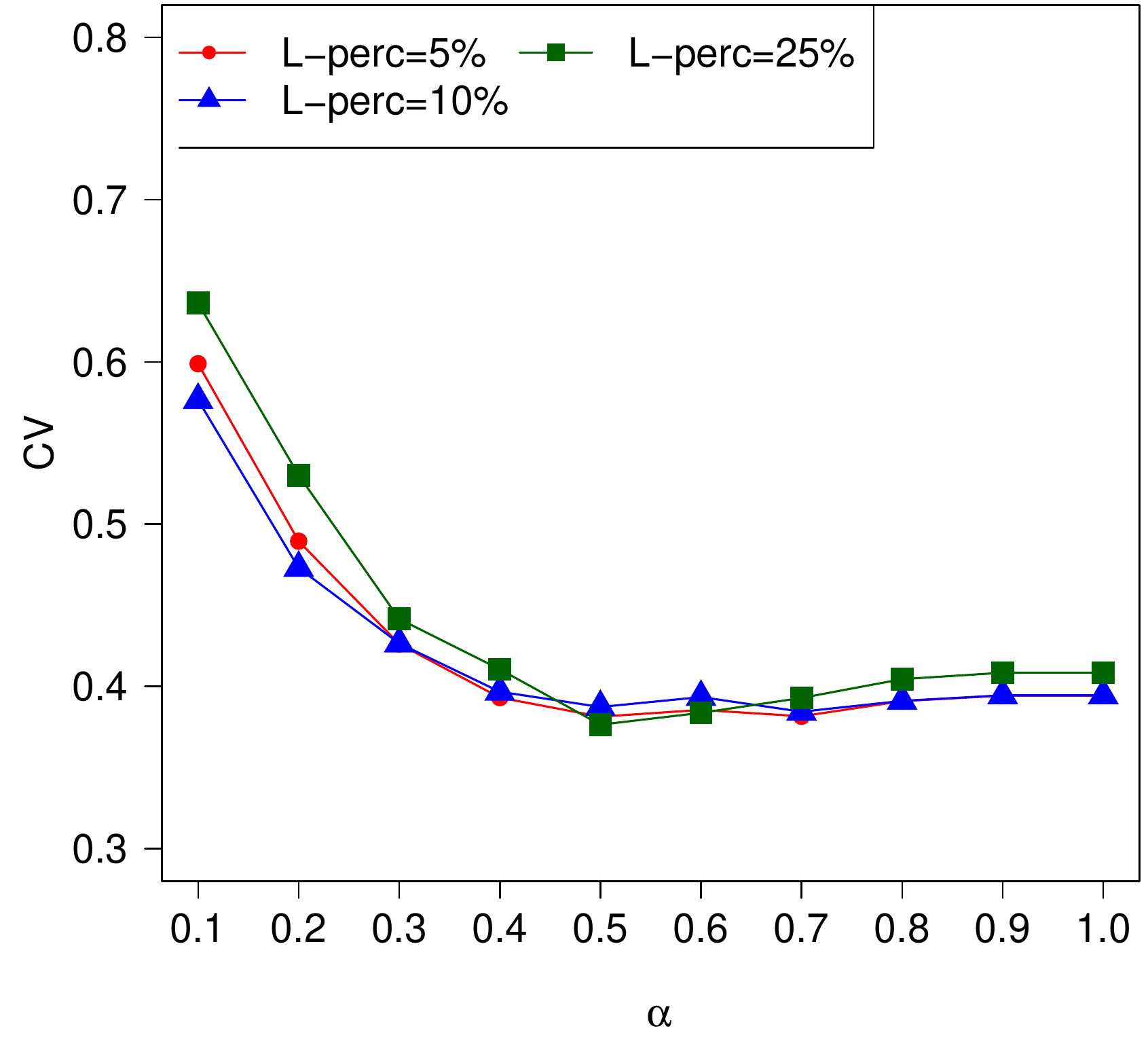}    &
\includegraphics[width=0.3\textwidth]{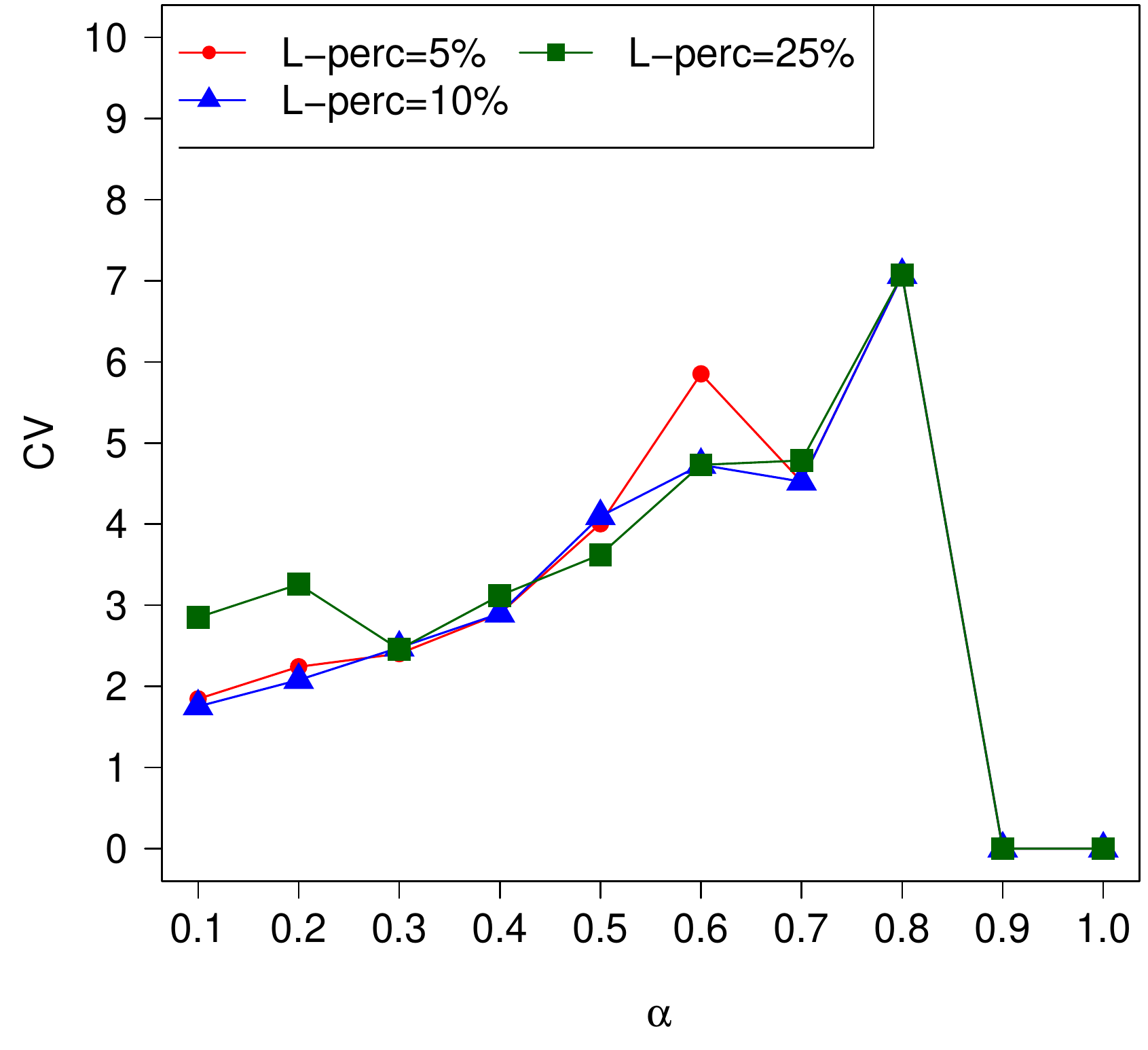}    &
\includegraphics[width=0.3\textwidth]{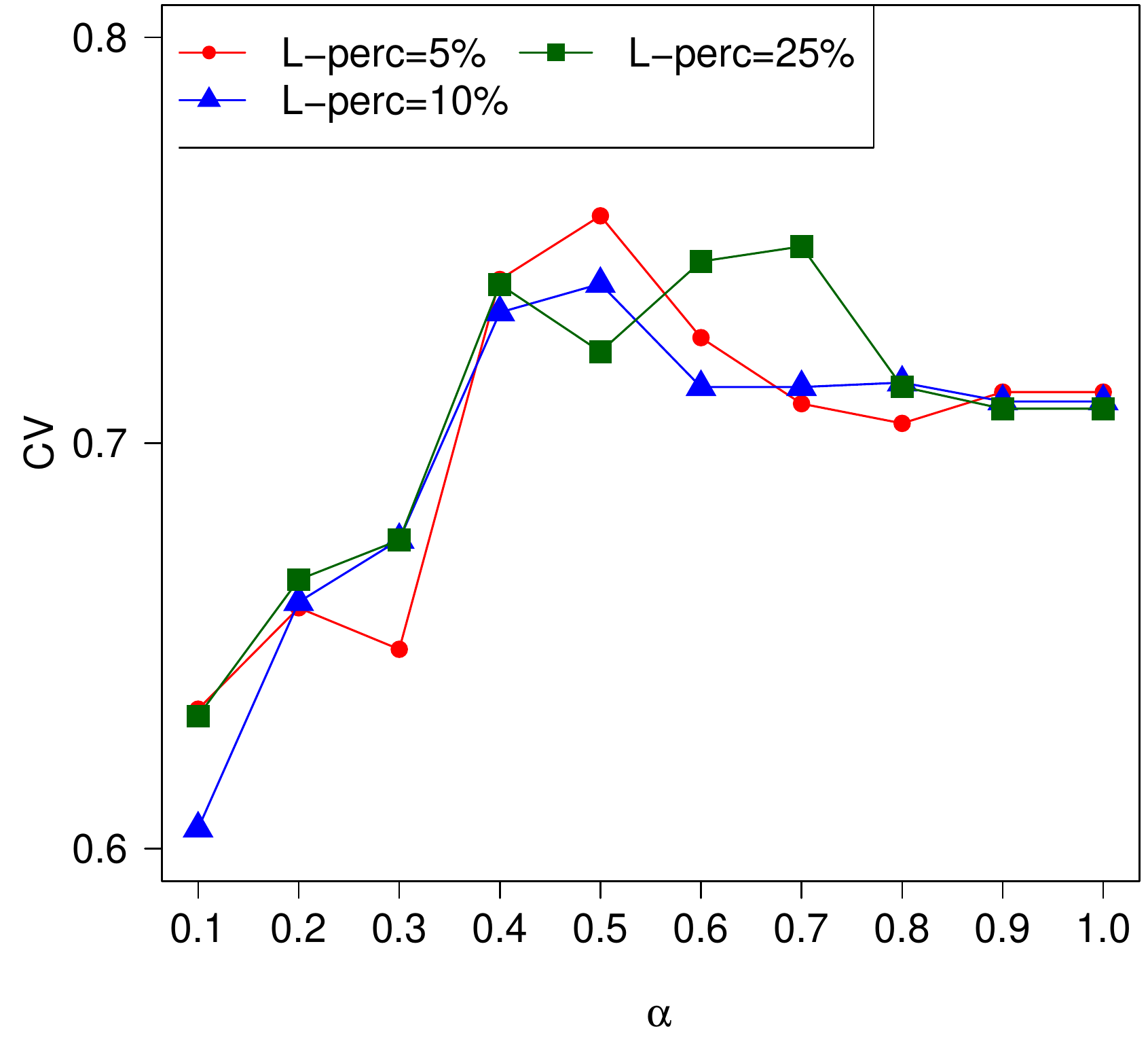}    \\
\hspace{-5mm} (a) outdegree & (b) betweenness & (c) coreness \\
\hspace{-5mm} \includegraphics[width=0.3\textwidth]{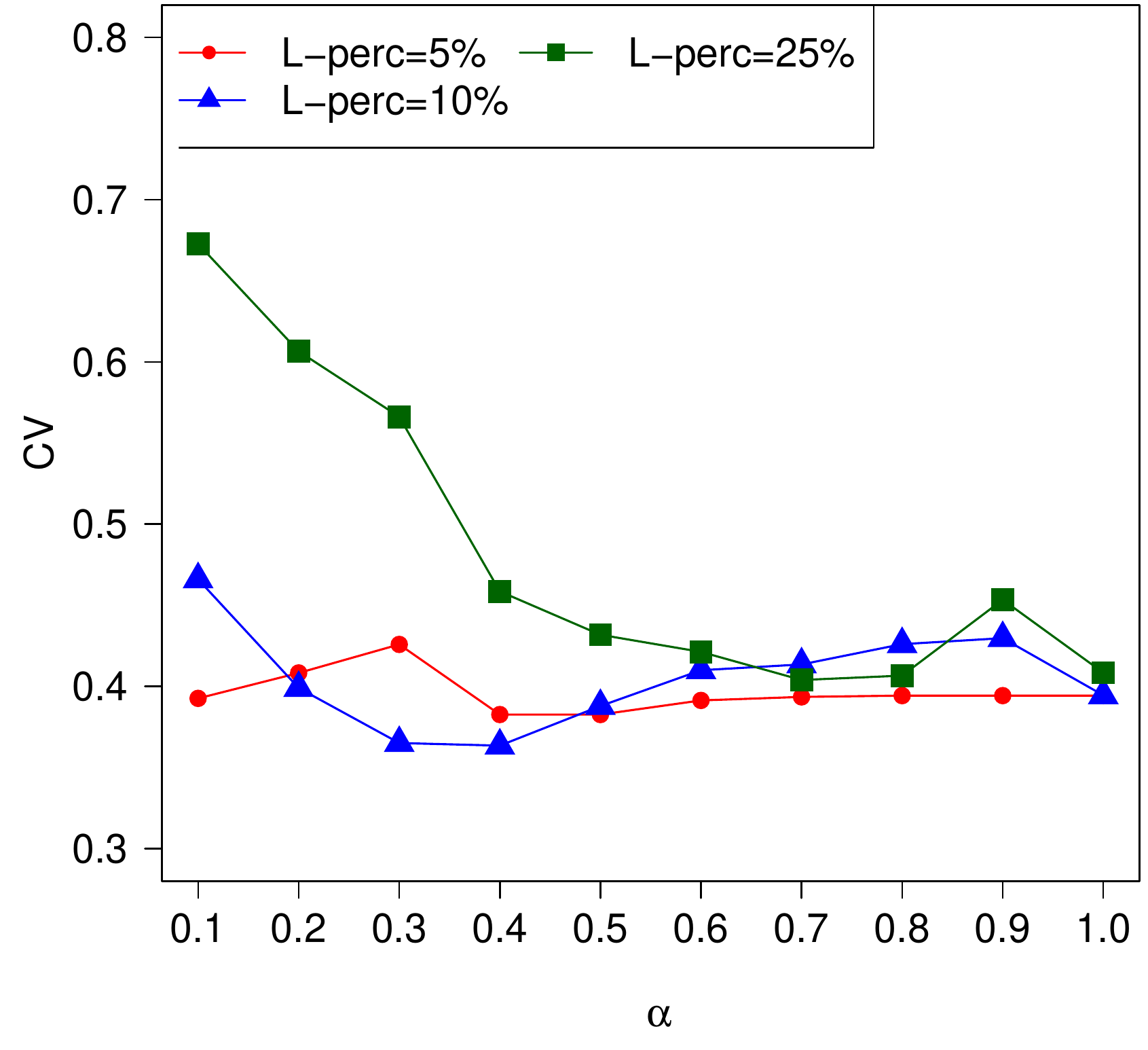}    &
\includegraphics[width=0.3\textwidth]{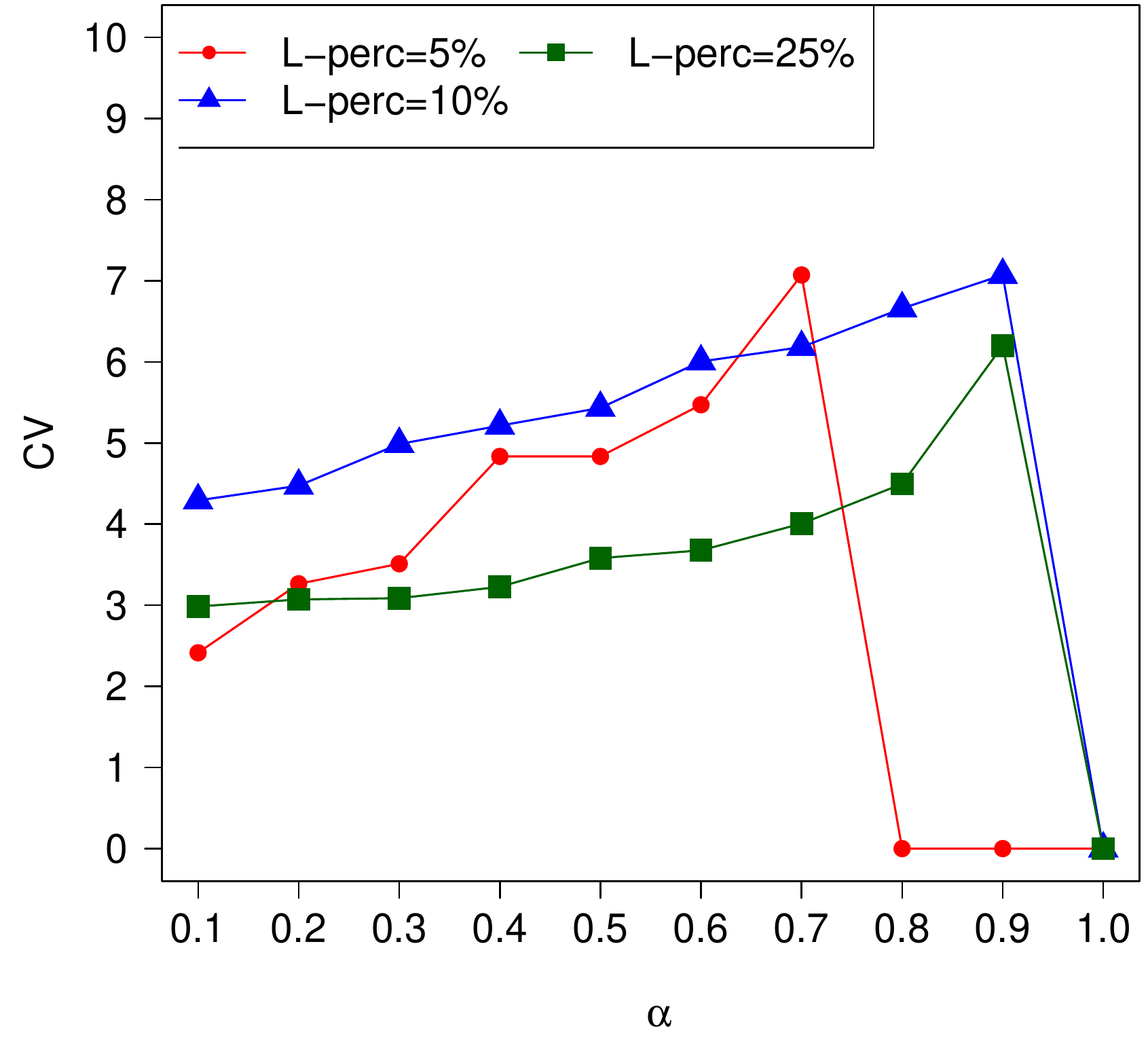}    &
\includegraphics[width=0.3\textwidth]{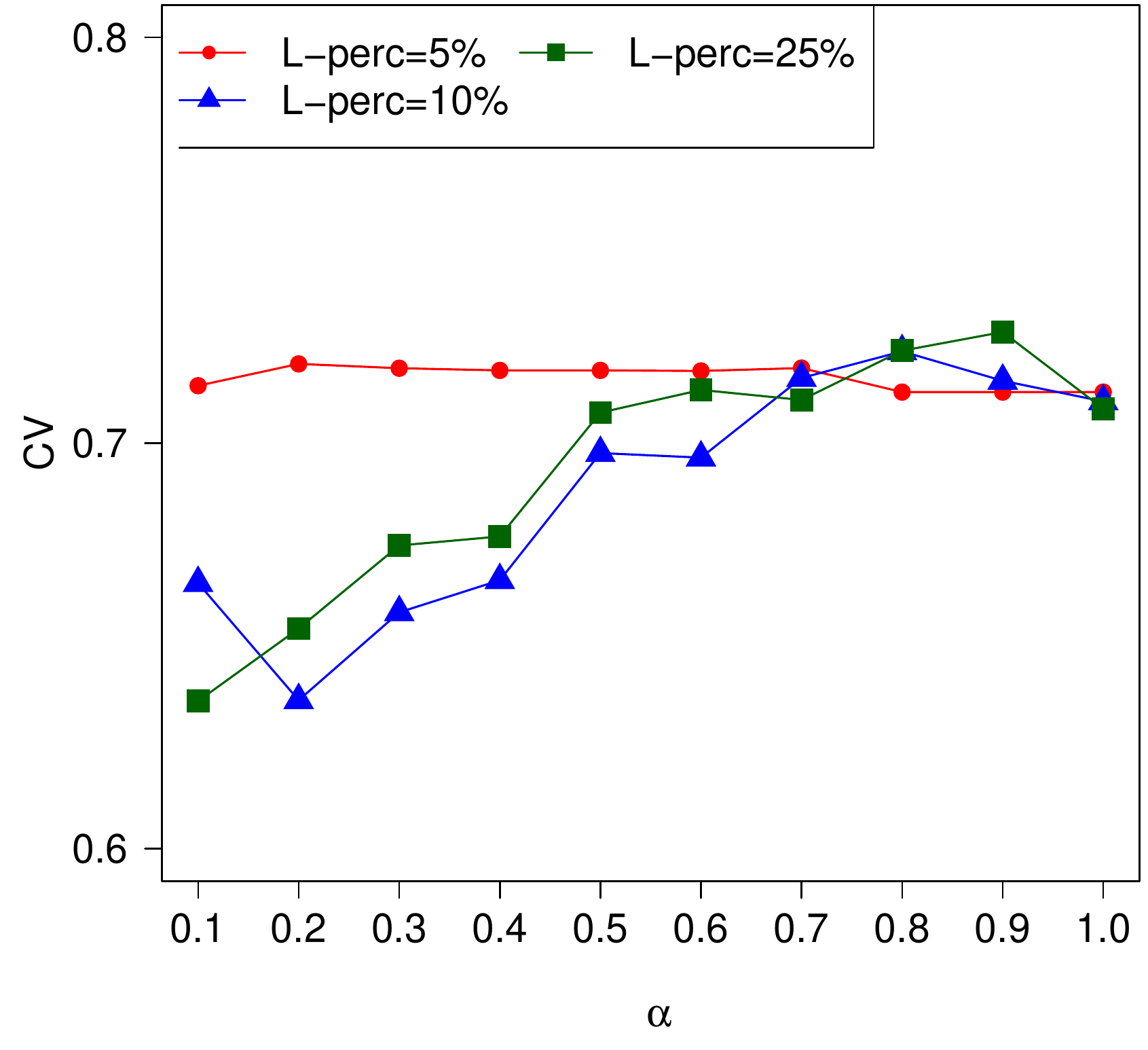}    \\
\hspace{-5mm} (d) outdegree & (e) betweenness & (f) coreness
\hspace{-1mm}
\end{tabular}
\caption{Coefficient of variation (CV) of topological properties of identified seed nodes, with $k=50$, by varying $\alpha$ and $\LurkValuePerc$, on \data{GooglePlus}: (a)--(c) L-\myalgo, (d)--(f) G-\myalgo.}
\label{fig:seed-topological-props}
\end{figure*} 

Figure~\ref{fig:seed-topological-props} shows the \textit{coefficient of variation} (hereinafter denoted as CV) of   selected topological measures over the seed nodes, by varying $\alpha$ and target set size ($\LurkValuePerc$).   
   Looking at results on the outdegree, we observe  decreasing trends for CV by increasing $\alpha$ up to 0.5, followed by   roughly constant trends set around 0.4, for both \myalgo methods.  Consistently with the analysis on seed set  overlap, L-\myalgo seeds tend to have   similar outdegree regardless of $\LurkValuePerc$, while in the case of  G-\myalgo,  relatively small variations occur for $\LurkValuePerc=\{5\%, 10\%\}$ by varying $\alpha$.  
   As concerns betweenness, CV generally increases with $\alpha$ up to high values (0.7, 0.9), then drastically reduces to zero; this indicates that when diversity is discarded, seeds tend to correspond to source nodes in the graph.   Analogously to the outdegree analysis, the trends for varying $\LurkValuePerc$ are quite similar   to each other in the L-\myalgo case. 
    Considering coreness, CV ranges within a much smaller interval than that corresponding to outdegree and betweenness, i.e., (0.6, 0.76) with L-\myalgo,  (0.64, 0.73) with G-\myalgo.      
    Again, the variability over the seeds computed by L-\myalgo is much less affected by the setting of $\LurkValuePerc$ than in the G-\myalgo case, with a general increasing trend up to mid-high values of $\alpha$. 
 
As concerns the competing methods, 
\algo{KB-TIM} identifies seed nodes having average CV that does  not   significantly change in terms of $\LurkValuePerc$, specifically: (0.42, 0.40) for outdegree, (3.41, 3.52) for betweenness, and 0.61 for coreness. 
\algo{TIM+} identifies seed nodes that have on average 0.45 CV of  outdegree, 0.0 CV of betweenness, and 0.70 CV of coreness.

\subsection{Target activation probabilities}
In this section  we report detailed results concerning the analysis of the target activation probabilities (cf. Section 6.2.2 in the main paper) with the aim of deepening our understanding of how different settings of $\alpha$ impact on the activation probability of nodes targeted by \myalgo.
We regard the activation probability of a node as the number of times the node has been activated divided by the number of Monte Carlo runs ($I_{MC}$, cf. Algorithm~\ref{alg:MCEstimation}).

\begin{figure*}[t!]
\centering
\begin{tabular}{ccc}
\hspace{-1mm}
\includegraphics[width=0.3\textwidth]{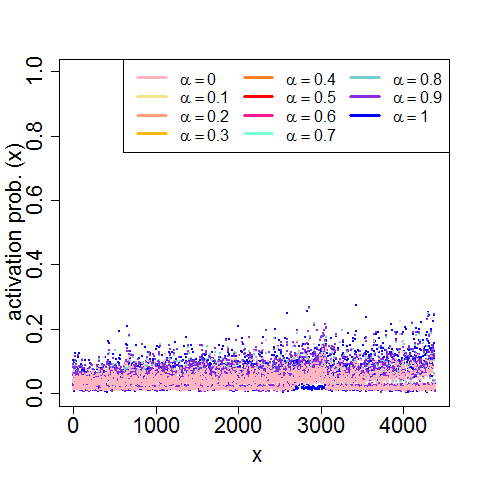} &
\hspace{-1mm}
 \includegraphics[width=0.3\textwidth]{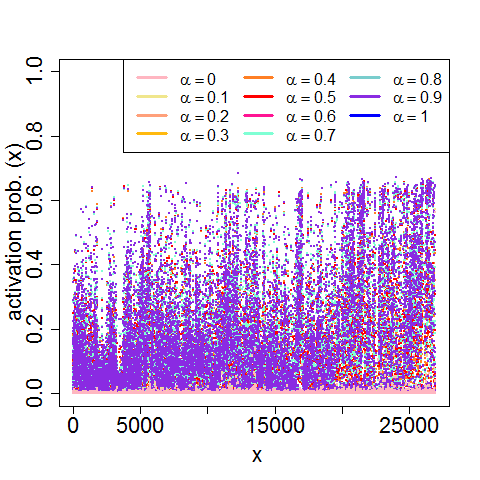} & 
\hspace{-1mm}
\includegraphics[width=0.3\textwidth]{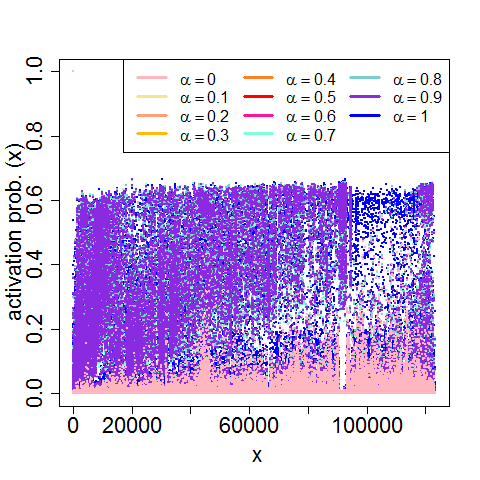} \\
\includegraphics[width=0.3\textwidth]{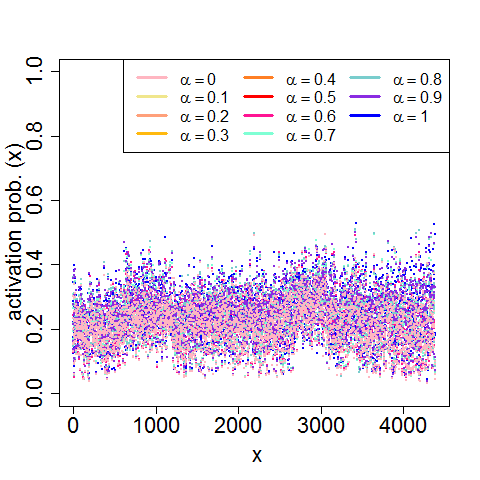} &
\hspace{-1mm}
 \includegraphics[width=0.3\textwidth]{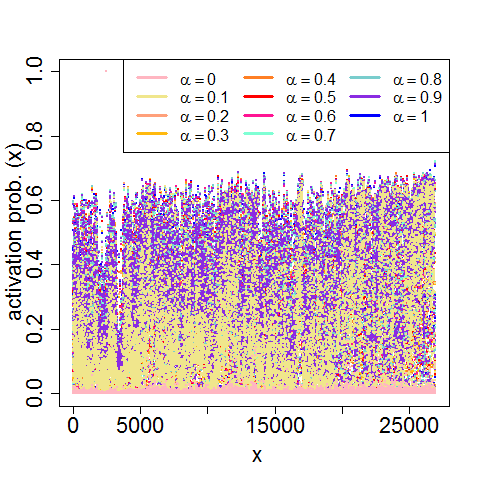} & 
\hspace{-1mm}
\includegraphics[width=0.3\textwidth]{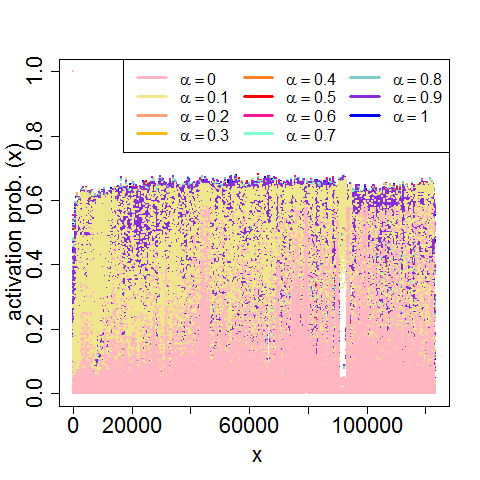} \\
\hspace{3mm} (a) & \hspace{3mm} (b) &  \hspace{3mm} (c)
\end{tabular}
\caption{Activation probabilities (y-axis)  for each target node (x-axis), obtained by G-\myalgo for varying $\alpha$. Results correspond to   $\LurkValuePerc=25\%$, $k$ set to $5$ (top) and $50$ (bottom), on (a) \data{Instagram-LCC}, (b) \data{GooglePlus}, and (c) \data{FriendFeed}. 
}  
\label{fig:ap_scat_f1} 
\end{figure*}

\begin{figure*}[t!]
\centering
\begin{tabular}{ccc}
\hspace{-1mm}
\includegraphics[width=0.3\textwidth]{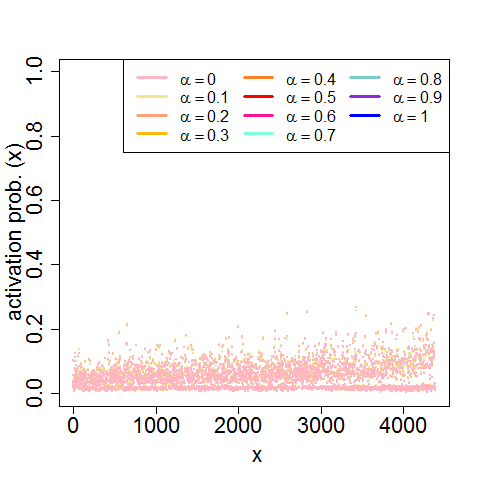} &
\hspace{-1mm}
 \includegraphics[width=0.3\textwidth]{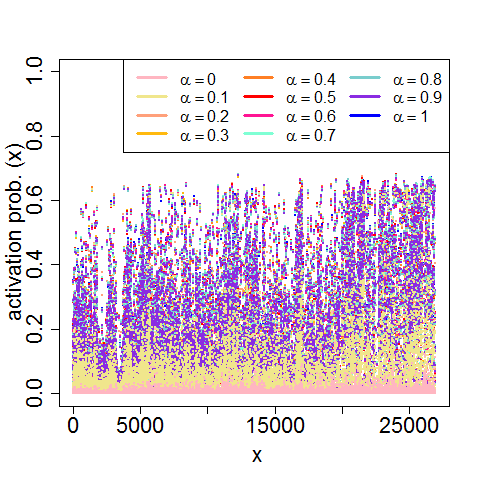} & 
\hspace{-1mm}
\includegraphics[width=0.3\textwidth]{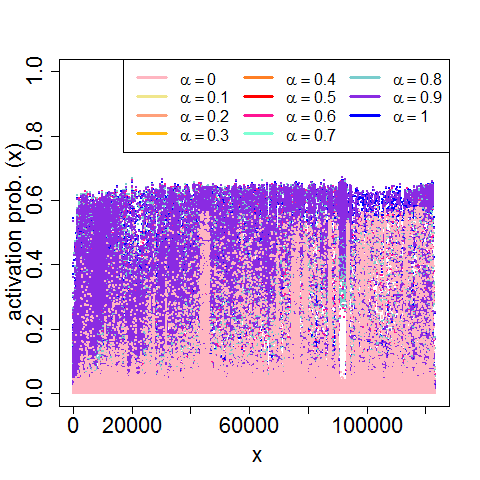} \\
\includegraphics[width=0.3\textwidth]{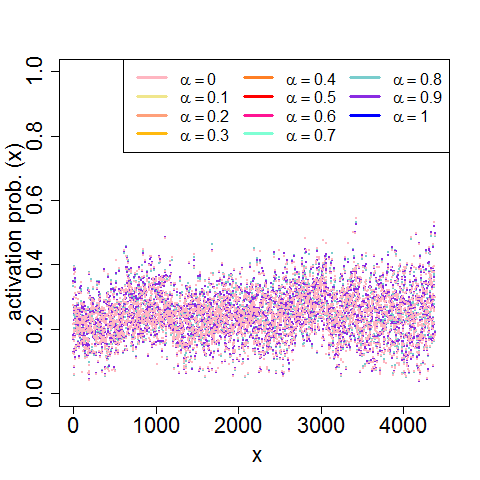} &
\hspace{-1mm}
 \includegraphics[width=0.3\textwidth]{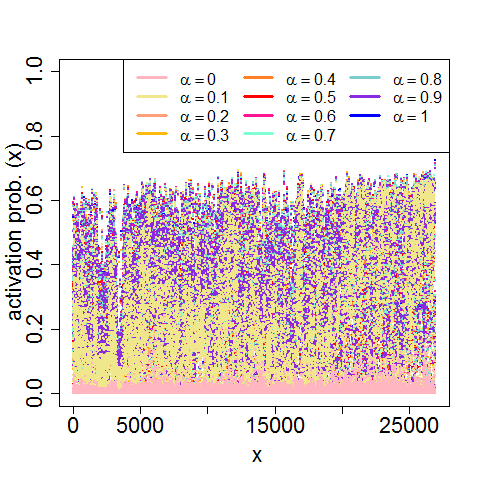} & 
\hspace{-1mm}
\includegraphics[width=0.3\textwidth]{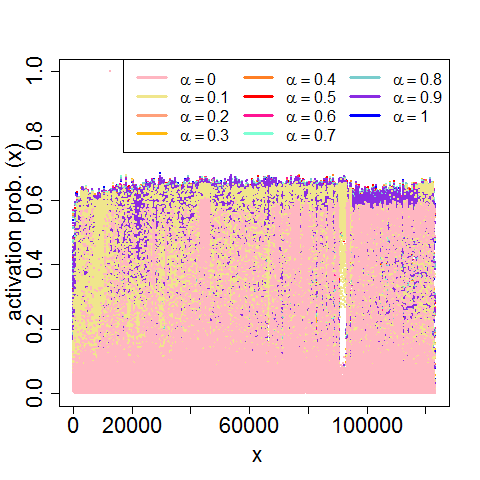} \\
\hspace{3mm} (a) & \hspace{3mm} (b) &  \hspace{3mm} (c)
\end{tabular}
\caption{Activation probabilities (y-axis)  for each target node (x-axis), obtained by L-\myalgo for varying $\alpha$. Results correspond to   $\LurkValuePerc=25\%$, $k$ set to $5$ (top) and $50$ (bottom), on (a) \data{Instagram-LCC}, (b) \data{GooglePlus}, and (c) \data{FriendFeed}.  
}  
\label{fig:ap_scat_f2} 
\end{figure*}

\begin{figure*}[t!]
	\centering
	\subfloat[]{
		\includegraphics[width=0.3\textwidth]{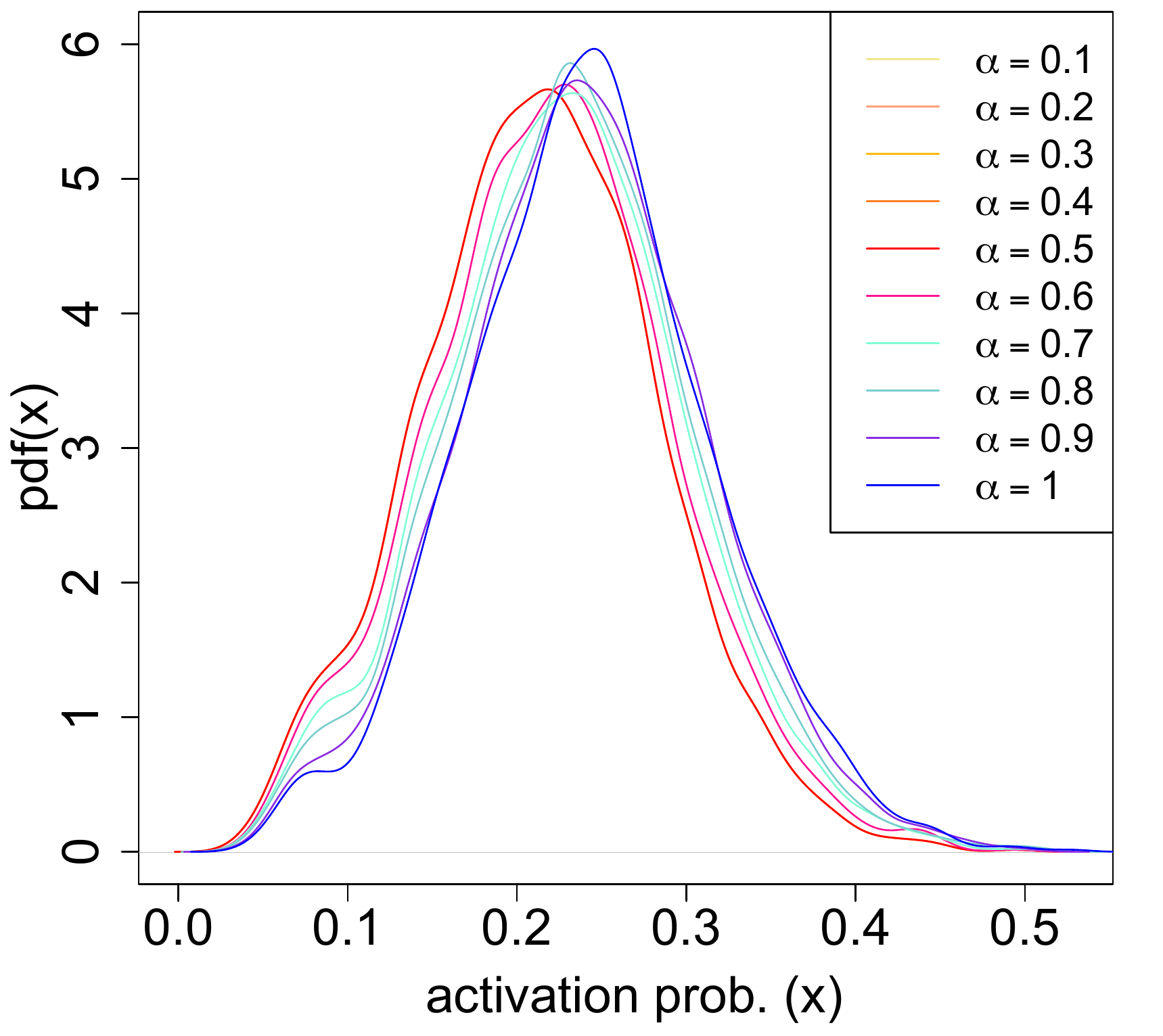} }
	\subfloat[]{
		\includegraphics[width=0.3\textwidth]{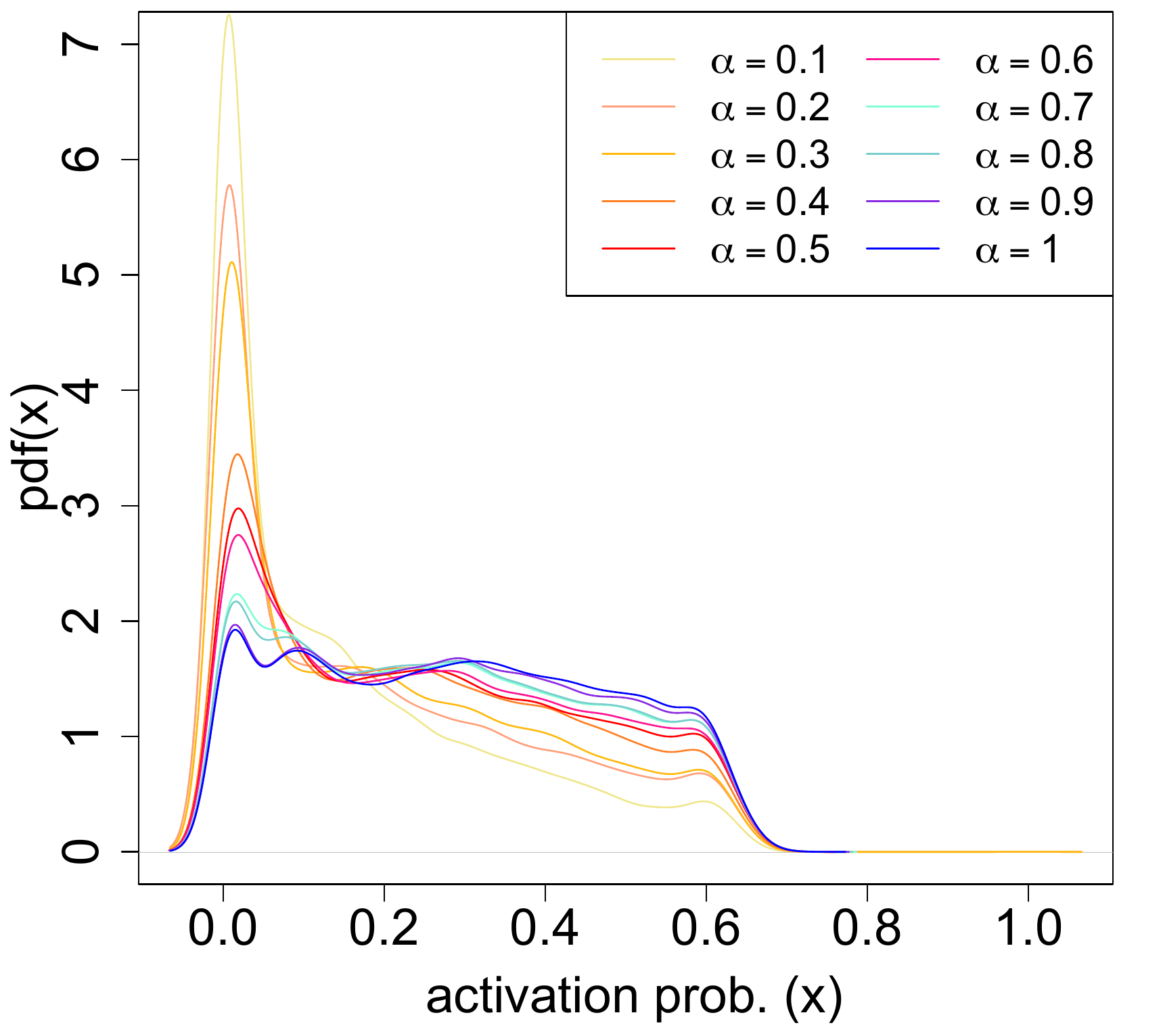}
	}
	\subfloat[]{
		\includegraphics[width=0.3\textwidth]{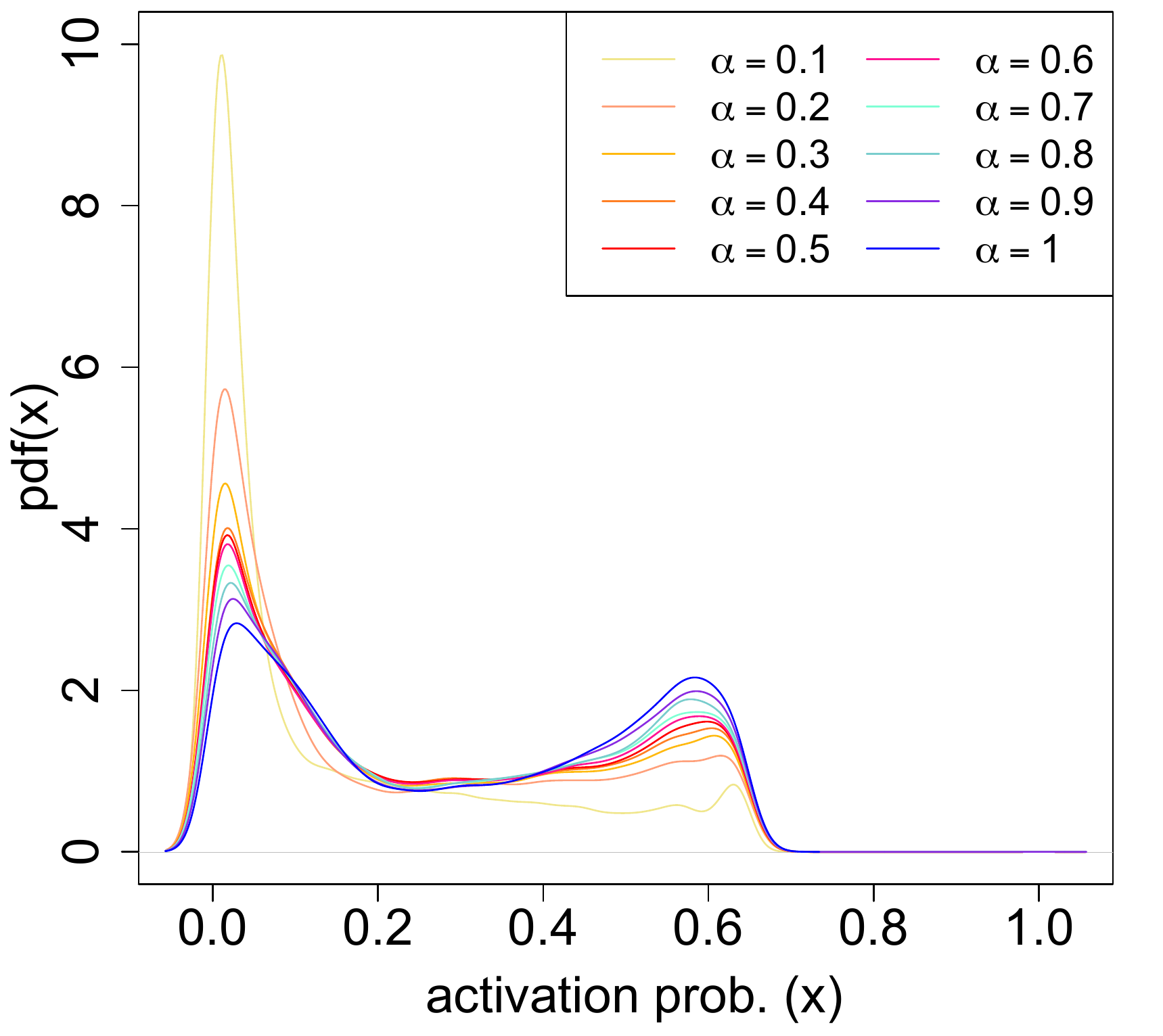} 
	}
	\caption{Density distributions of activation probabilities obtained by G-\myalgo, for varying $\alpha$, with $\LurkValuePerc$ set  to 25\%, $k=50$, on (a) \data{Instagram-LCC}, (b) \data{GooglePlus},   and (c) \data{FriendFeed}.  }
	\label{fig:ap_pdf_f1} 
\end{figure*}

\begin{figure*}[t!]
	\centering
	\subfloat[]{
		\includegraphics[width=0.3\textwidth]{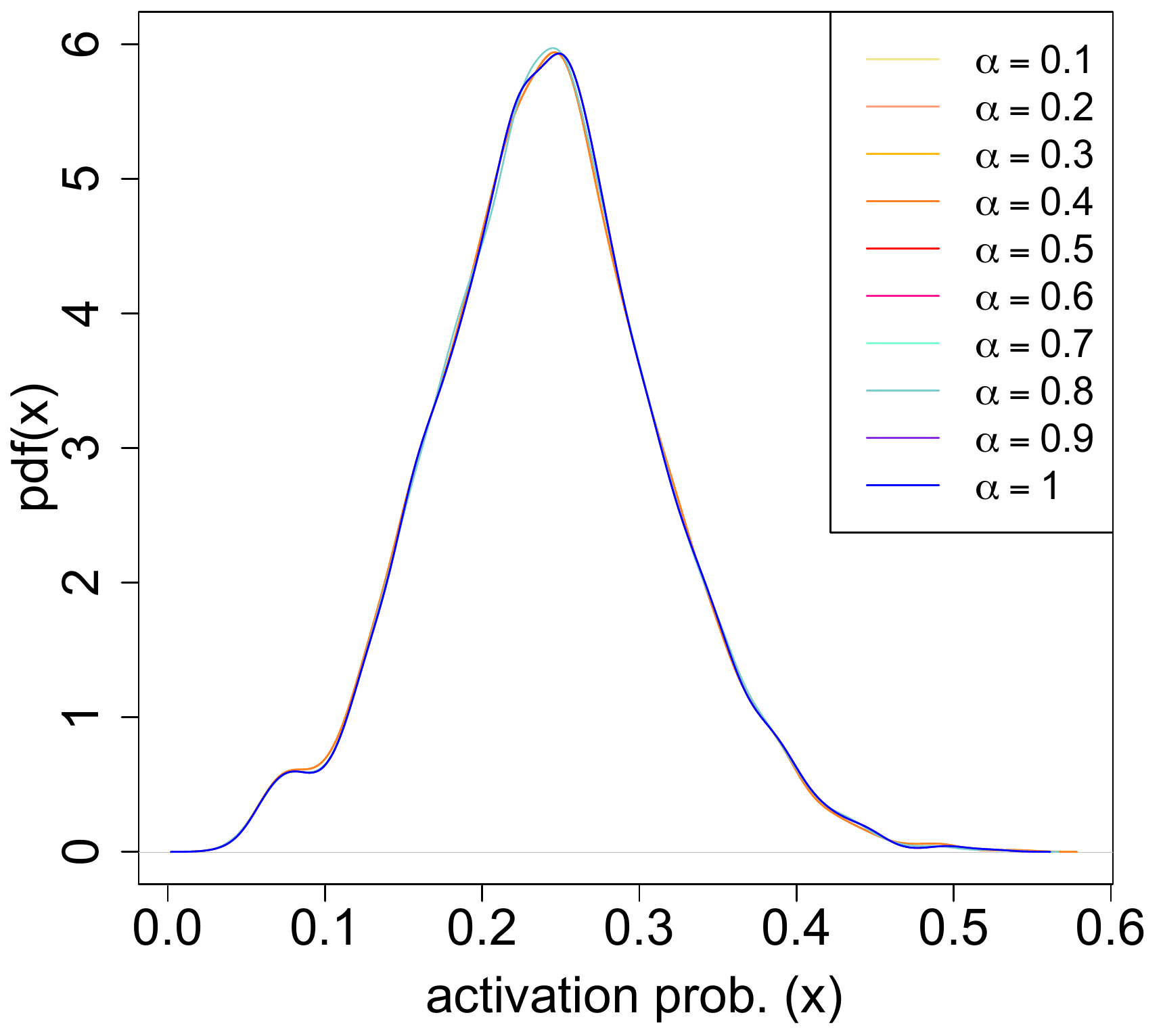} }
	\subfloat[]{
		\includegraphics[width=0.3\textwidth]{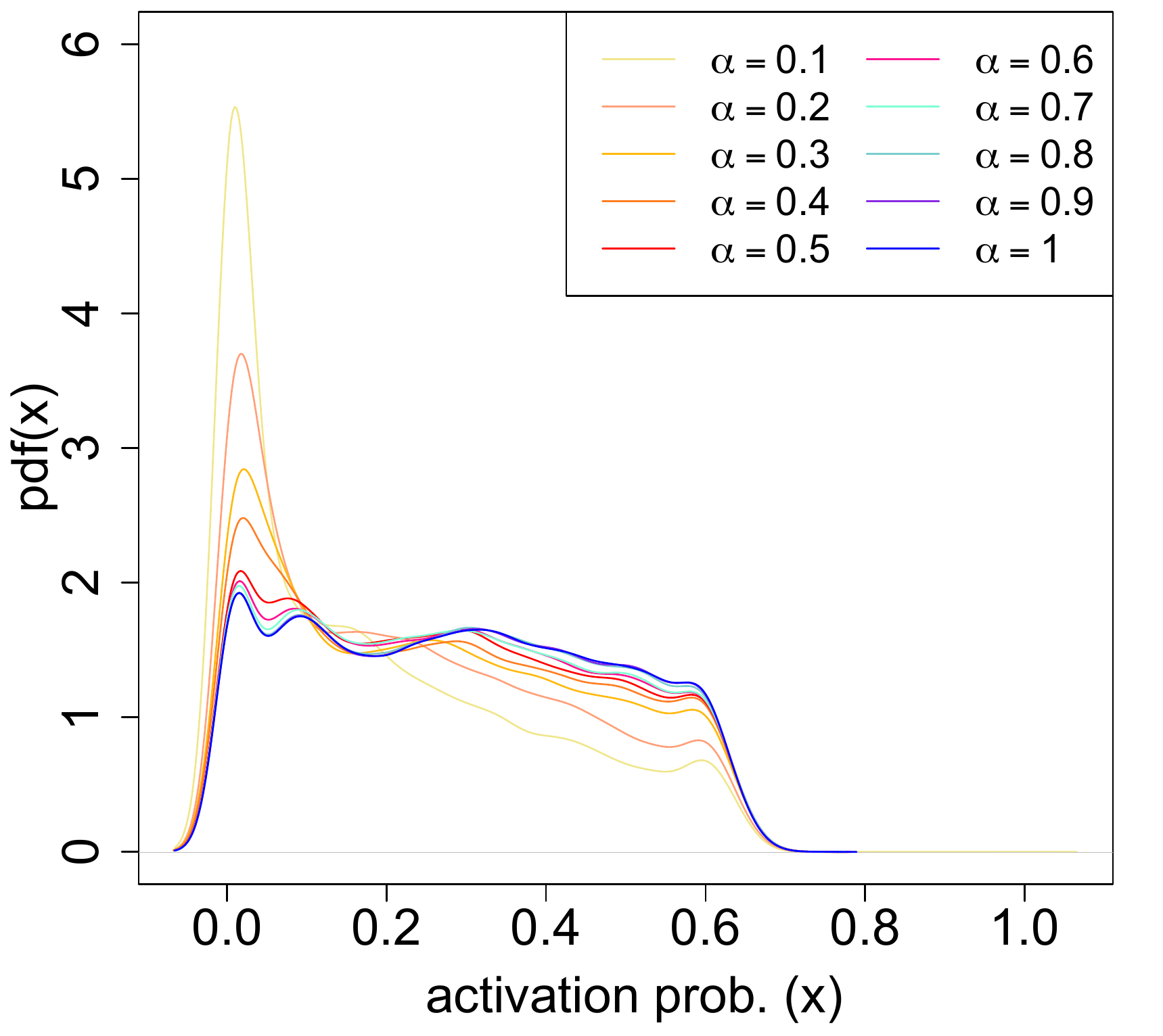}
	}
	\subfloat[]{
		\includegraphics[width=0.3\textwidth]{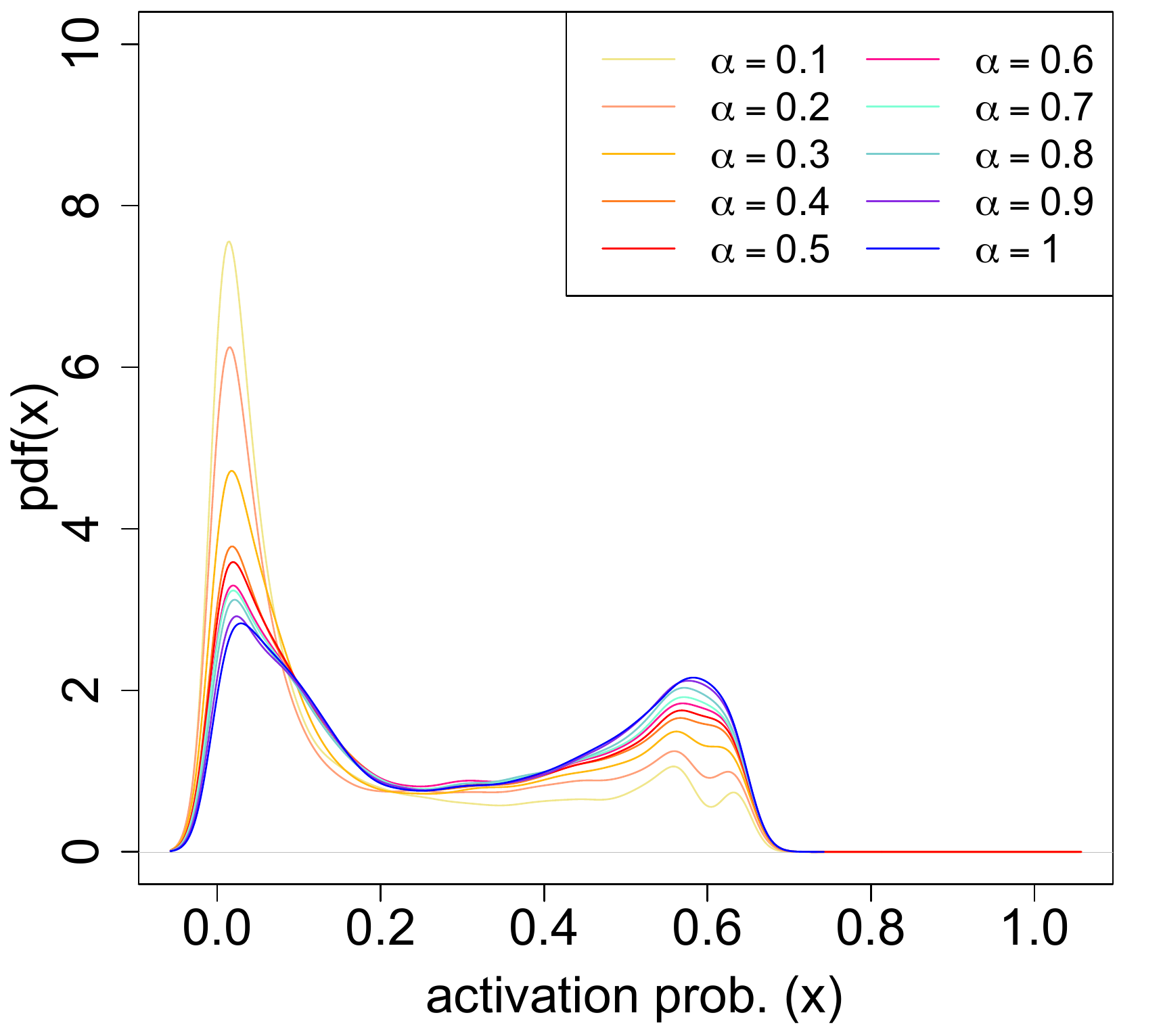} 
	}
	\caption{Density distributions of activation probabilities obtained by L-\myalgo, for varying $\alpha$, with $\LurkValuePerc$ set  to 25\%, $k=50$, on (a) \data{Instagram-LCC}, (b) \data{GooglePlus},   and (c) \data{FriendFeed}.     }  
	\label{fig:ap_pdf_f2} 
\end{figure*}

In order to analyze the above property of target nodes,   we present first the activation probability values of the nodes in the final active set, shown in  Figures~\ref{fig:ap_scat_f1} and \ref{fig:ap_scat_f2}. Next we discuss  
the density distributions $pdf(x)$ with variable $x$ modeling the vector of activation probabilities associated with the nodes in the final active set, reported in   Figures~\ref{fig:ap_pdf_f1} and  \ref{fig:ap_pdf_f2}.

\vspace{2mm}
\textbf{Plots of activation probability distributions.\ }
 Figures~\ref{fig:ap_scat_f1} and \ref{fig:ap_scat_f2} show the  activation probabilities versus the target nodes, by varying the values of $\alpha$  and $k$, for G-\myalgo and L-\myalgo.  
 
Considering first the performance of G-\myalgo (Fig.~\ref{fig:ap_scat_f1}),   there is an evident gap between the activation probabilities obtained for low     $\alpha$ (i.e., $\alpha \leq 0.4$), and higher values of the parameter, with the maximum activation probability values (and maximum coverage of the target set) generally obtained for $\alpha=0.9$ and $\alpha=1.0$.
On \data{Instagram-LCC} (Fig.~\ref{fig:ap_scat_f1}(a)), given the generally low values of activation probabilities, and the high overlap among the seed sets obtained when varying $\alpha$, the gap between minimum and maximum  values is strongly reduced w.r.t. other datasets, with $\alpha=1.0$ showing only small increase on the activation of targets w.r.t. $\alpha=0.0$. Note also that  for $k=50$, there is a very small number of  nodes showing activation probability within $[0.0, 0.1]$: this  would hint that, when estimating the activation probabilities,  the set of activated nodes remains almost unaffected in all the \textit{R} Monte Carlo runs (while in other cases there are a bunch of nodes which are reached by the influence diffusion process only for a small number of runs, resulting in near-zero   activation probabilities).   
More interesting behaviors are    observed for  \data{GooglePlus} (Fig.~\ref{fig:ap_scat_f1}(b)). For $k=5$ (upper plot), mid-high activation probabilities are reached for a small set of nodes starting from $\alpha=0.5$, but the   majority of target nodes is activated for $\alpha \geq 0.9$, with activation probabilities in the range $[0.0, 0.6]$.  
However, for $k=50$ (lower plot), a significant set of target nodes shows mid-high activation probabilities already for $\alpha=0.1$, indicating that, with a relatively large $k$, low values of $\alpha$ are sufficient to activate target nodes while taking into account diversity. 
As regards \data{FriendFeed} (Fig.~\ref{fig:ap_scat_f1}(c)), activation probabilities obtained for $0.0 \leq \alpha \leq 0.6$ are generally higher than the ones obtained for the other two datasets. Nevertheless, for $k=5$ (upper plot), a value of $\alpha=0.7$ is needed to reach significant activation probabilities on a vast portion of the target set. Most  target nodes are again reached for $\alpha=0.9$, but it can be noted that there is a large band of target nodes (on the right side of the plot) which reaches mid-high probabilities only for $\alpha=1.0$. This indicates that in large networks, when using low $k$, even small variations on the value of $\alpha$ can significantly impact on the effectiveness of the influence maximization process.
Looking at the results obtained for $k=50$ (lower plot), we observe that the set of target nodes obtaining a significant activation probability is relevant already for $\alpha=0.0$, with a coverage on a large portion  of the target set starting for $\alpha=0.1$.
 
Quite similar qualitative remarks can be drawn about the performance of L-\myalgo (Fig.~\ref{fig:ap_scat_f2}).
As regards \data{Instagram-LCC} (Fig.~\ref{fig:ap_scat_f2}(a)), for $k=5$ (upper plot) no visible improvement in the activation probabilities can be observed starting from $\alpha \geq 0.1$, while the results are similar to the ones discussed for G-\myalgo  for $k=50$ (lower plot).
On \data{GooglePlus} (Fig.~\ref{fig:ap_scat_f2}(b)), a general improvement of the performance obtained for $\alpha=0.1$ can be noted, while the results obtained for different $\alpha$ values are similar to the ones observed for G-\myalgo. The improvement is more evident for $k=5$ (upper plot), but remains significant also for $k=50$ (lower plot).
On \data{FriendFeed}, an increment in the activation probability values obtained for $0.0 \leq \alpha \leq 0.5$ can be noted for $k=5$ (upper plot), w.r.t. the situation described for G-\myalgo. With $k=50$ (lower plot), higher probabilities than the ones observed for G-\myalgo  are observed for $\alpha=0.0$.

\vspace{2mm}
\textbf{Density distributions of activation probability.\ } 
Figures~\ref{fig:ap_pdf_f1} and~\ref{fig:ap_pdf_f2} show density distributions of activation probability obtained for G-\myalgo and L-\myalgo, respectively.  

Focusing first on  \data{GooglePlus}, similar trends can be noted for both  G-\myalgo (Fig.~\ref{fig:ap_pdf_f1}(b)) and L-\myalgo (Fig.~\ref{fig:ap_pdf_f2}(b)). A density peak corresponding to   low activation probability values (close to $0.0$) can be noted for low values of $\alpha$ (i.e., $\alpha \leq 0.6$ for G-\myalgo and $\alpha \leq 0.4$ for L-\myalgo). This peak slightly decreases for increasing values of $\alpha$, yielding a relatively wide area of nearly constant density (e.g., around $2$) which covers a range of activation probabilities from $0.0$ up to about $0.6$.

A roughly bi-modal distribution can be observed for \data{FriendFeed}, for both G-\myalgo  (Fig.~\ref{fig:ap_pdf_f1}(c)) and L-\myalgo (Fig.~\ref{fig:ap_pdf_f2}(c)). It is easy to recognize a first peak corresponding to near-zero activation probability values, and a second one located around  $0.6$;  hence,  the first peak   becomes lower and the second peak higher by increasing   $\alpha$. 

Analogously to previous evaluation settings,  situation on   \data{Instagram-LCC} is drastically different from the other two datasets, which in this case corresponds to roughly Normal  distributions for varying $\alpha$.  
Using  G-\myalgo (Fig.~\ref{fig:ap_pdf_f1}(a)), the density distribution has a mean   activation probability which spans from approximately $0.2$ for low values of $\alpha$ to values close to $0.3$ for higher values of $\alpha$. Using L-\myalgo (Fig.~\ref{fig:ap_pdf_f1}(b)), due to  the high overlap of the seed sets obtained when varying $\alpha$, all distributions are nearly identical, and centered on an average value of activation probability around $0.25$.

It should be noted that the density distributions referring to the setting $\alpha=0.0$ are omitted from Figures~\ref{fig:ap_pdf_f1} and~\ref{fig:ap_pdf_f2}. The reason behind this choice is that, as discussed in the previous analysis, %of activation probability distributions
 in some cases there is a large gap between the activation probabilities obtained with  $\alpha=0.0$ and  $\alpha=0.1$. Here the entity of such a  gap  causes the curve of density distribution for $\alpha=0.0$ to have a peak corresponding to very high values of probability density function for near-zero values of activation probability (which, if showed, would force us to use a larger scale, making the other curves difficult to read). 
This contingency is observed on \data{GooglePlus} for both versions of \myalgo, and \data{FriendFeed} for G-\myalgo, while in other cases the density curve for $\alpha=0.0$ can be relatively close (\data{FriendFeed} with L-\myalgo) or nearly identical (\data{Instagram-LCC} for G-\myalgo and L-\myalgo) to the curve shown for $\alpha=0.1$.

\subsection{Correlation analysis between capital and diversity measurements}
Tables~\ref{tab:correlation:a} and \ref{tab:correlation:b} summarize results of  correlation analysis between the sequence of capital values and the sequence of diversity values associated to the nodes at convergence of the diffusion process, for each of the \myalgo methods   and for selected settings of parameters.  

\begin{table}[t!]
\centering
\caption{Correlation analysis between capital and diversity measurements: G-\myalgo}
\label{tab:correlation:a}
\scalebox{0.8}{
\begin{tabular}{|l|c|c|c|c|}
\hline
network	 & $\alpha$ &	$L\textit{-perc}$	 & 	$k$	 &  correlation \\
 & & $(\%)$ & & \\
\hline \hline
GooglePlus & 0.1 & 10 & 5 & -0.001 \\
GooglePlus & 0.5 & 10 & 5 & -0.004 \\
GooglePlus & 0.9 & 10 & 5 & -0.005 \\
GooglePlus & 0.1 & 25 & 5 &  0.006 \\
GooglePlus & 0.5 & 25 & 5 & -0.001 \\
GooglePlus & 0.9 & 25 & 5 & -0.006 \\

\hline
FriendFeed & 0.1 & 10 & 5 & -4.4e-05 \\
FriendFeed & 0.5 & 10 & 5 & -7.8e-05 \\
FriendFeed & 0.9 & 10 & 5 & -8.1e-05 \\
FriendFeed & 0.1 & 25 & 5 & 0.004 \\
FriendFeed & 0.5 & 25 & 5 & 0.003 \\
FriendFeed & 0.9 & 25 & 5 & 0.001 \\

\hline \hline
GooglePlus & 0.1 & 10 & 50 & -0.008 \\
GooglePlus & 0.5 & 10 & 50 & -0.008 \\
GooglePlus & 0.9 & 10 & 50 & -0.007 \\
GooglePlus & 0.1 & 25 & 50 & -0.008 \\
GooglePlus & 0.5 & 25 & 50 & -0.006 \\
GooglePlus & 0.9 & 25 & 50 & -0.011 \\

\hline  
FriendFeed & 0.1 & 10 & 50 & -1.6e-04 \\
FriendFeed & 0.5 & 10 & 50 & -2.3e-04 \\
FriendFeed & 0.9 & 10 & 50 & -2.7e-04 \\
FriendFeed & 0.1 & 25 & 50 & 5.5e-04 \\
FriendFeed & 0.5 & 25 & 50 & 3.0e-04 \\
FriendFeed & 0.9 & 25 & 50 & 3.3e-04 \\

\hline
\end{tabular} 
}
\end{table}

\begin{table}[t!]
\centering
\caption{Correlation analysis between capital and diversity measurements: L-\myalgo}
\label{tab:correlation:b}
\scalebox{0.8}{
\begin{tabular}{|l|c|c|c|c|}
\hline
network	 & $\alpha$ &	$L\textit{-perc}$	 & 	$k$	 &  correlation \\
 & & $(\%)$ & & \\
\hline
GooglePlus & 0.1 & 10 & 5 & 0.169 \\
GooglePlus & 0.5 & 10 & 5 & 0.059\\
GooglePlus & 0.9 & 10 & 5 & 0.008 \\
GooglePlus & 0.1 & 25 & 5 & 0.148\\
GooglePlus & 0.5 & 25 & 5 & 0.054 \\ 
GooglePlus & 0.9 & 25 & 5 & 0.004 \\
\hline
FriendFeed & 0.1 & 10 & 5 & 0.085 \\
FriendFeed & 0.5 & 10 & 5 & 0.046 \\
FriendFeed & 0.9 & 10 & 5 & 0.018 \\
FriendFeed & 0.1 & 25 & 5 & 0.076 \\
FriendFeed & 0.5 & 25 & 5 & 0.052 \\
FriendFeed & 0.9 & 25 & 5 & 0.020 \\
\hline \hline
GooglePlus & 0.1 & 10 & 50 & 0.225 \\
GooglePlus & 0.5 & 10 & 50 & 0.088 \\
GooglePlus & 0.9 & 10 & 50 & 0.025 \\
GooglePlus & 0.1 & 25 & 50 & 0.229 \\
GooglePlus & 0.5 & 25 & 50 & 0.097 \\
GooglePlus & 0.9 & 25 & 50 & 0.020 \\
\hline
FriendFeed & 0.1 & 10 & 50 & 0.164 \\
FriendFeed & 0.5 & 10 & 50 & 0.126 \\
FriendFeed & 0.9 & 10 & 50 & 0.069 \\
FriendFeed & 0.1 & 25 & 50 & 0.180 \\
FriendFeed & 0.5 & 25 & 50 & 0.131 \\
FriendFeed & 0.9 & 25 & 50 & 0.064 \\
\hline
\end{tabular} 
}
\end{table}

% Acknowledgments
%\begin{acks}
%\verde{The authors} \blu{would like} to thank ................................
%\end{acks}

% Bibliography

% \bibliographystyle{IEEEtran}
% \bibliography{LRrefs}

                             % Sample .bib file with references that match those in
                             % the 'Specifications Document (V1.5)' as well containing
                             % 'legacy' bibs and bibs with 'alternate codings'.
                             % Gerry Murray - March 2012

\vspace{-1mm}


\vspace*{-12mm}

\begin{IEEEbiography}[{\includegraphics[width=1in,height=1.2in]{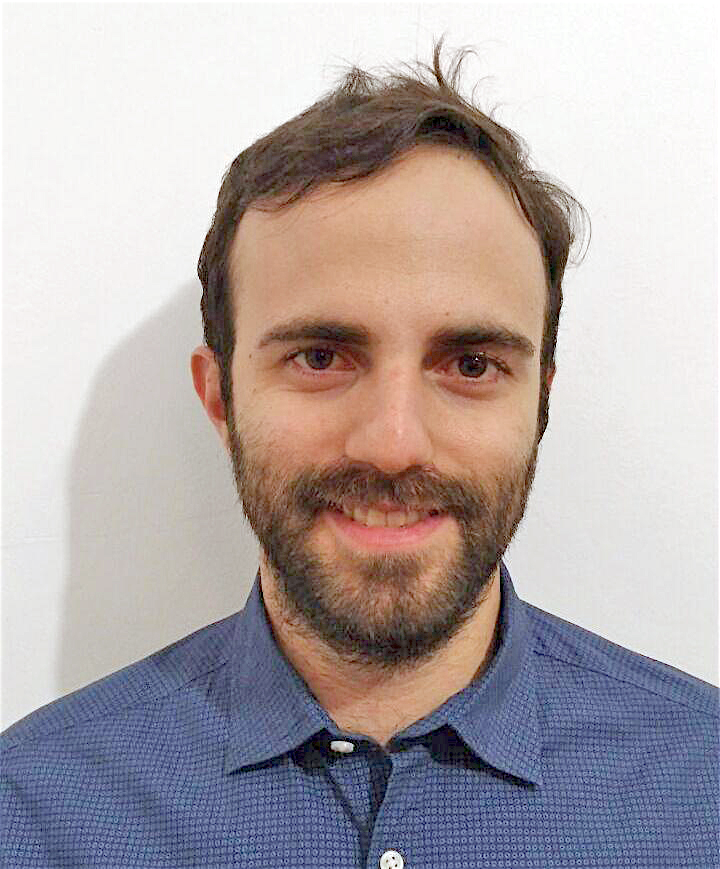}}]{Antonio Cali\`{o}}
	is a  first-year PhD student in information and communication  technologies with the DIMES Department, University of Calabria,  Italy.  	His research interests are focused on  information diffusion models,  influence propagation, and related computational problems in network and data  science.
\end{IEEEbiography}

 \vspace*{-12mm}
 \begin{IEEEbiography}[{\includegraphics[width=1in,height=1.2in]{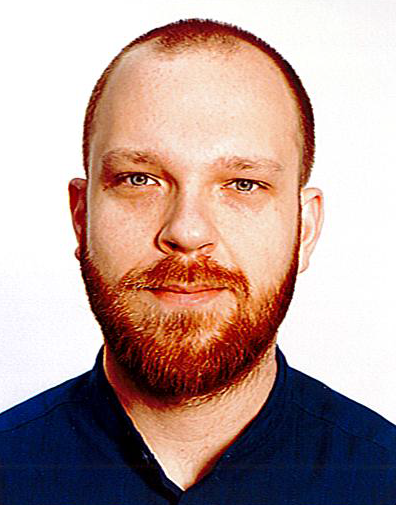}}]{Roberto Interdonato} is a research scientist at Cirad, UMR TETIS, Montpellier, France. He was  a postdoctoral  researcher with the University of La Rochelle (France), Uppsala University (Sweden), and   University of Calabria (Italy), where he received his PhD degree in computer and systems engineering in 2015.   His research interests include topics in data mining and machine learning applied to complex networks analysis   and to remote sensing analysis. 
 %On these topics he has coauthored journal articles and conference papers, organized workshops, presented tutorials at international conferences and developed practical software tools.
 \end{IEEEbiography}

\vspace*{-12mm}

\begin{IEEEbiography}[{\includegraphics[width=1in,height=1.2in]{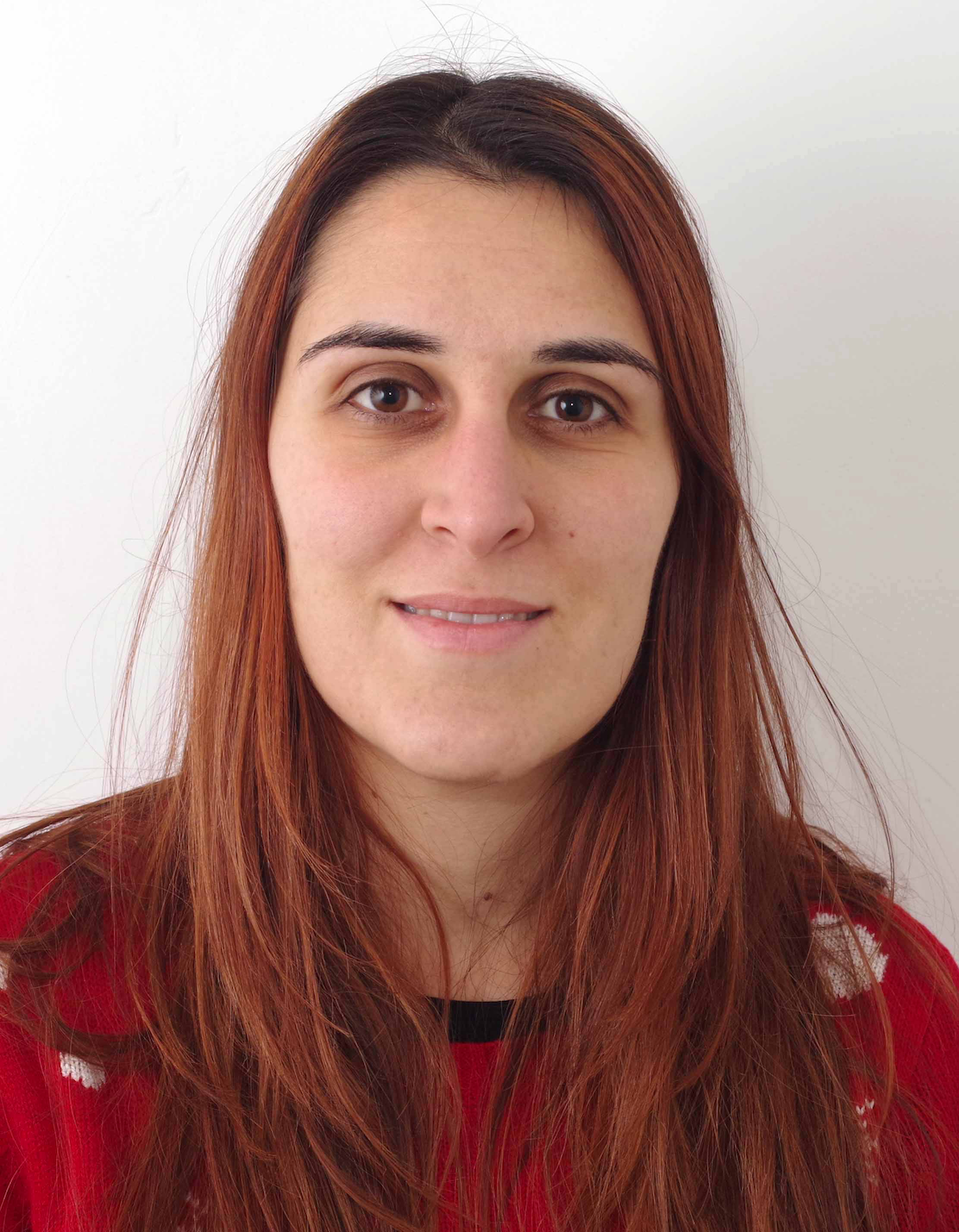}}]{Chiara Pulice}
	received the PhD degree in computer and systems engineering from the University of Calabria, Italy. 
	She was a visiting scholar at the Department of Computer Science of the University of British Columbia  (2013-2014), and a
postdoctoral researcher with the University of
Maryland’s Institute for Advanced Computer Studies
(2016-2017).   Currently, she is a postdoctoral researcher with the Dartmouth College Department of Computer Science in Hanover, New Hampshire.  Her research interests include data integration, inconsistent databases, data mining, machine learning, and social network analysis.
\end{IEEEbiography}

\vspace*{-12mm}

\begin{IEEEbiography}[{\includegraphics[width=1in,height=1.2in]{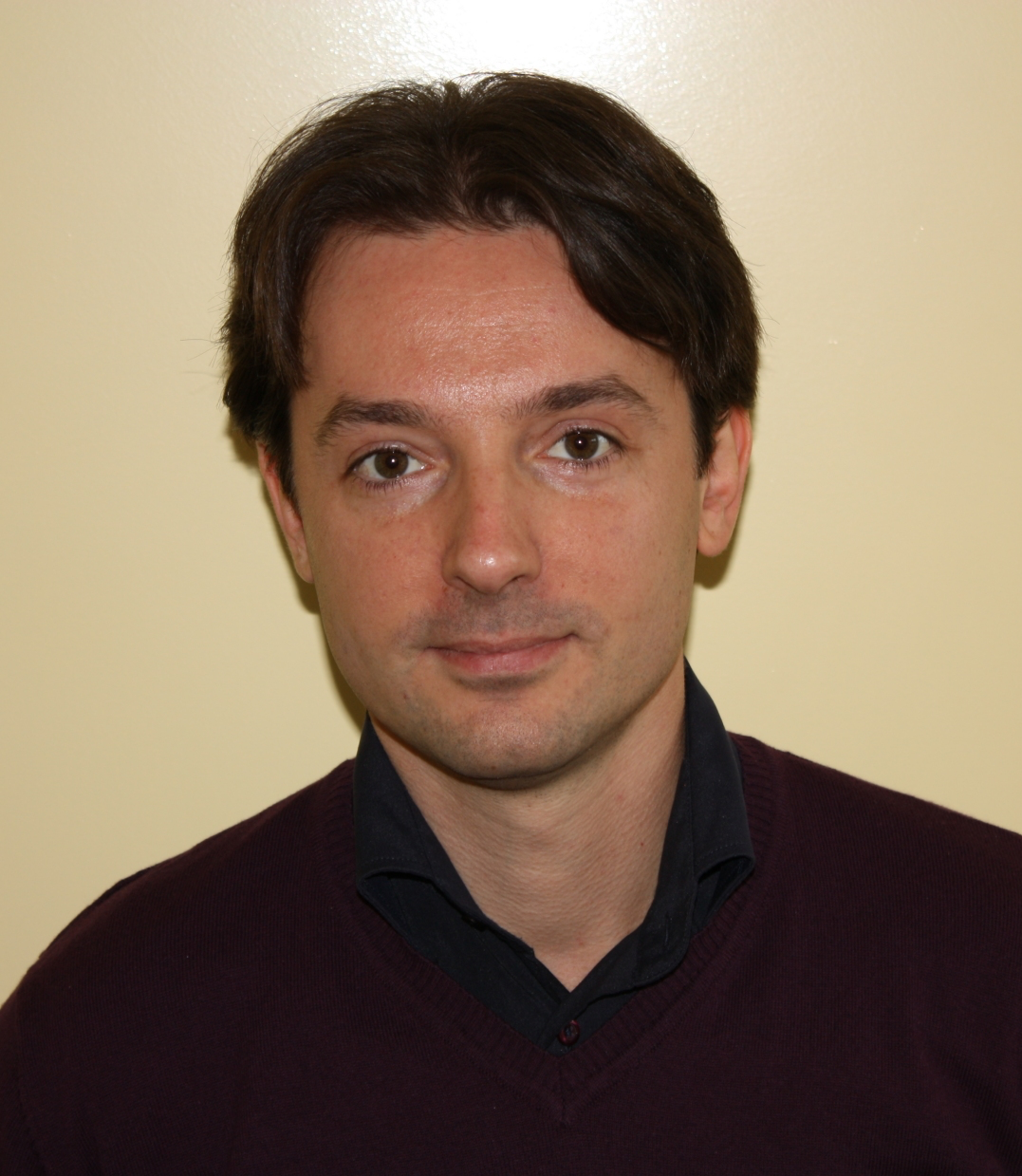}}]{Andrea Tagarelli}
 is an associate professor of computer engineering at the University of Calabria, Italy. From the same university, he received his PhD degree  in computer and systems engineering in 2006.  His research interests are in the areas of data mining, machine learning, network analysis, web and semistructured data management, information retrieval. 
 On these topics, he has co-authored about 100 peer-reviewed papers, and edited one book on XML data mining.  
  He co-organized   workshops  on data mining topics in premier conferences in the field (ACM SIGKDD, SIAM DM, PAKDD, ECML-PKDD, ECIR). He is co-program-chair of the 2018 IEEE/ACM ASONAM Conference.
 % Since 2015, he is member of the editorial board of Computational Intelligence journal and Social Network Analysis and Mining journal.  
\end{IEEEbiography}

\end{document}